\documentclass[twocolumn,showpacs,preprintnumbers,superscriptaddress,amsmath,amssymb,pra,10pt]{revtex4-1}
\usepackage{graphicx}
\usepackage{color}
\usepackage{amssymb,amsfonts,amsmath}
\usepackage{colordvi}
\usepackage{ytableau}
\usepackage{mathrsfs}
\usepackage{xspace}
\usepackage{tabularx}

\newtheorem{proposition}{Proposition}

\usepackage{bbm}
\newcommand{\unity}{\mathbbmss{1}}
\renewcommand{\o}{\myo}

\newcommand\bex{\begin{exemple} \rm}
\newcommand\eex{\end{exemple}}

\newcommand{\C}{\mathbb C}

\newcommand{\R}{\mathbb R}
\newcommand{\Z}{\mathbb Z}

\providecommand{\abs}[1]{\lvert#1\rvert}

\newcommand{\fig}[1]{Fig.~\ref{#1}}

\newcommand{\tab}[1]{Table~\ref{#1}}

\newcommand{\pa}[1]{p.~#1}
\newcommand{\eq}[1]{Eq.~\eqref{#1}}
\newcommand{\eqs}[1]{Eqs.~\eqref{#1}}
\newcommand{\pp}[1]{p.~#1}
\newcommand{\ppp}[1]{pp.~#1}
\newcommand{\secn}[1]{Sec.~#1}

\newcommand{\B}{\mathcal B} 
\renewcommand{\H}{\mathscr H} 
\newcommand{\bra}[1]{\left\langle #1\right|}
\newcommand{\ket}[1]{\left|#1\right\rangle} 

\newcommand{\su}{\mathfrak{su}}
\newcommand{\SU}{\mathrm{SU}}
\newcommand{\SO}{\mathrm{SO}}

\newcommand{\GL}{\mathrm{GL}}

\newcommand{\II}[1]{I_{#1}} 
\newcommand{\TT  }{\mathrm{T}}
\newcommand{\J}{\mathrm{J}}
\newcommand{\JJ}{\mathcal{J}}

\newcommand{\Id}{\mathrm{Id}}

\newcommand{\Tr}{\mathrm{Tr}}

\newcommand{\Y}{\mathrm{Y}} 

\newcommand{\TTab}[1]{\ytableaushort{#1}}

\newcommand{\taun}[2]{\tau_{#1}^{[#2]}}

\newcommand{\TTlb}[2]{\TT_{#1}^{\{#2\}}} 
 
\renewcommand{\tt}[3]{\TT_{#1}(\taun{#2}{#3})}

\newcommand{\tttt}[2]{\TT_{#1}(\tau^{[3]}_{#2})}

\newcommand{\tsymF}[3]{\TT_{#1}(\taun{#2}{#3})}

\newcommand{\droplet}{droplet\xspace}
\newcommand{\droplets}{droplets\xspace}

\newcommand{\lisa}{LISA\xspace}
\newcommand{\drops}{DROPS\xspace}
\newcommand{\Drops}{DROPS\xspace}
\newcommand{\arr}{\hspace{-1.1mm}\rightarrow\hspace{-1.1mm}}

\usepackage[bookmarks=false,pdfstartview={FitH}]{hyperref}

\usepackage{tikz}
\newcommand\mybox[2][]{\tikz[overlay]\node[inner sep=2pt, anchor=text, 
rectangle, draw=black,#1] {#2};\phantom{#2}}
\allowdisplaybreaks

\begin{document}
\title{Visualizing operators of coupled spin systems}

\author{Ariane Garon}
\affiliation{Department Chemie, Technische Universit{\"a}t
M{\"u}nchen, Lichtenbergstrasse 4, 85747 Garching, Germany}

\author{Robert Zeier}
\email{robert.zeier@ch.tum.de}
\affiliation{Department Chemie, Technische Universit{\"a}t
M{\"u}nchen, Lichtenbergstrasse 4, 85747 Garching, Germany}

\author{Steffen J. Glaser}
\email{steffen.glaser@tum.de}
\affiliation{Department Chemie, Technische Universit{\"a}t
M{\"u}nchen, Lichtenbergstrasse 4, 85747 Garching, Germany}

\date{January 26, 2015}
\begin{abstract}
The state of quantum systems, their energetics, and their time evolution
is modeled by abstract operators.
How can one visualize such operators for coupled spin systems? 
A general approach is presented which consists of
several shapes representing linear combinations of spherical harmonics.
It is applicable to an arbitrary number of spins and 
can be interpreted as a generalization of Wigner functions.
The corresponding visualization transforms naturally under
non-selective spin rotations as well as spin permutations.
Examples and applications are illustrated for the case of three spins~$1/2$.
\end{abstract}

\maketitle

\ytableausetup{smalltableaux,centertableaux}

\section{Introduction}

We present a technique to visualize operators acting on coupled spin systems.
Their high-dimensional structure is uniquely described 
by several shapes (cf.\ \fig{random DROPS} below),
which represent linear combinations of spherical harmonics.
Crucial features are directly observable
and transform naturally under non-selective spin rotations as well as spin permutations.
This provides a general approach to systematically analyze 
coupled
spin systems and their time evolution.
We emphasize that our approach is generally applicable and that
arbitrary operators on multi-spin systems can be visualized.
Examples applicable in research and education include density operators 
which describe the state of a quantum-mechanical system
(e.g., spin systems or quantum bits from quantum information processing),
Hamilton operators which specify energy terms, and unitary transformations 
modeling the time evolution.

Various approaches to visualize quantum systems are known.
A quantum-mechanical operator for a two-level system (such as an isolated spin $1/2$ 
particle in an external magnetic field) can always be mapped to a three-dimensional (real) vector  
as shown in the seminal work of Feynman~et~al.\@~\cite{Feynman_Vernon_57}.
This vector can represent a Bloch vector, a field vector, or
a rotation vector related to applications ranging from magnetic resonance imaging~\cite{handbook_04,EBW87} 
and spectroscopy~\cite{EBW87}  to quantum optics~\cite{SchleichBook}.

A multi-spin operator can be displayed as a bar chart of the absolute value
(or the real and imaginary parts) of its individual matrix elements.
This technique is commonly used, e.g., to present experimental results
of state tomography of a quantum system \cite{NC00}.
Alternatively, energy-level diagrams are used, e.g., in 
quantum optics and 
magnetic resonance spectroscopy. The corresponding populations
can be represented by circles on energy levels, and coherences can be depicted
by lines between energy levels \cite{SEL:1983}.
Another visualization of a density operator is based on the non-classical vector representation based on
single-transition operators \cite{EBW87,Donne_Gorenstein,Freeman97}.
Moreover, graphical shorthand notations for coupled-spin dynamics have been proposed in \cite{EggBoden90_eng}.
All these approaches are cumbersome for many
spins, in particular if the density matrix has many
non-zero entries. Frequently, non-selective spin rotations do not act naturally 
on these visualizations.

Our method surmounts these difficulties
by relying on a map between a multi-spin basis given by 
tensor operators~\cite{Racah42} (\emph{vide infra}) and multiple sets of spherical harmonics \cite{Jac99}
which are  independently plotted in different locations. Related work can be at least traced back
to Pines~et~al.~\cite{Pines} where 
(albeit without a formal map)
selected density operator terms of a spin 1 particle
and their symmetry properties
are depicted using spherical harmonics.
Further visualizations have been presented
in \cite{HalsMNMRIX,SH91}.
Dowling~et~al.~\cite{DowlingAgarwalSchleich} 
illustrated ``collections'' of (essentially non-interacting) 
two-level atoms while highlighting connections to Wigner functions 
(which will be discussed in Sec.~\ref{sect_WIgner}).
Similar figures can also be found in \cite{JHKS}.
More recently, the usefulness of visualizing single-spin systems with spherical harmonics has been 
impressively demonstrated for nuclear magnetic resonance experiments of 
quadrupolar nuclei, including the generation of multiple-quantum coherence and multiple-quantum 
filters \cite{PhilpKuchel}. However, the authors were skeptical if this approach could be generalized to coupled 
spins, see the discussion in the appendix of \cite{PhilpKuchel}.
A special class of two coupled spins
was treated in \cite{MJD}. 
Certain states of two (and three) spins could be visualized by the method
of \cite{Harland}; however in \cite{Harland} it is was also emphasized that a
general method was still missing.
We present 
in this work a versatile approach which is applicable to an arbitrary number of coupled spins.

Although our approach is completely general,
we focus in the following on the most common situation typically found in the field of  
magnetic resonance spectroscopy and
quantum information processing where all spins are distinguishable
and have spin number $1/2$ unless otherwise stated.
This article has the following structure:
First,  maps between tensor operators and 
sets of spherical harmonics are analyzed which furnishes a general
framework for our approach to visualization.
Then, the 
\lisa basis (with defined \textbf{li}nearity, \textbf{s}ubsystem, and \textbf{a}uxiliary criteria, 
such as permutation symmetry) which provides a particular
choice for this map is discussed in detail for the case of three coupled spins.
We continue with various applications. 
Afterwards, we discuss 
connections to Wigner functions
and provide the mathematical details for the \lisa basis
for an arbitrary number of spins. Alternatives to 
the \lisa basis are discussed before we conclude.
Ancillary information is collected in the Appendices \ref{App_Motivation}--\ref{App_multipole}.

\section{Visualization\label{sec_vis}}

Abstract objects such as quantum mechanical operators can be visualized by mapping them
into vivid, three-dimensional objects
(such as three-dimensional functions). For example, the state of a quantum 
mechanical two-level system can be mapped to the Bloch vector visualized as a three-dimensional arrow
\cite{Feynman_Vernon_57}. 
In order to generalize this idea, the mapping from an abstract object to its visualization should ideally 
satisfy the following essential properties: (A) An operator should be
bijectively mapped to a unique function (or object). (B) Crucial features should be directly 
visible. In our context, (B) can refer to (e.g.) observables, symmetries 
under rotations or permutations, natural transformation characteristics 
under rotations, as well as the set of involved spins.

In order to describe the mapping from operators to functions, we first recall a complete, 
orthonormal operator basis
which captures the symmetries of rotations and which is
known as irreducible tensor operators $\TT_{j}$ 
\cite{Wigner31}: \nocite{Wigner59}
The components $\TT_{jm}$ of $\TT_{j}$ with fixed rank $j$ and varying order $m \in \{-j,\ldots,j\}$  
form a basis of a space which stays invariant
under the action of the rotation group $\SO(3)$ (or any group) and which does not contain
a proper invariant subspace. 
In the following, we usually substitute the rotation group $\SO(3)$ by the
locally-isomorphic unitary group $\SU(2)$ which consists of all unitary $2{\times} 2$-matrices
of determinant one \cite{SW86}.
The tensor operators form the foundation for the theory of angular momentum
\cite{Racah42,Rose1957Book,edmonds1957Book,BrinkSatchler,BL81,Zare88,Thom94}
and are part of the standard curriculum of quantum mechanics, see, e.g., \cite{Merzbacher98}.
It can be  illuminating to note 
(as has been done by Mackey \cite{Mackey81,Mackey93}, \nocite{BL81b,Wigner93}
see also \cite{Michel})
that the tensor operators provide an explicit form for the well-established
representation theory of the Lie algebra $\su(2)$ (and more general ones), see, e.g., \cite{Sepanski07}. 

Some readers might find it convenient to have a more explicit
definition for tensor operators $\TT_{j}$ which is provided using the conditions of Racah \cite{Racah42} 
\begin{subequations}\label{racah}
\begin{align}
[\JJ_{z}, \TT_{jm}] 
=&\, m\, \TT_{jm},\\
[\JJ_{\pm}, \TT_{jm}] 
=&\, \sqrt{j(j+1)-m(m\pm1)}\, \TT_{j,m\pm1},
\end{align}
\end{subequations}
which feature the raising and lowering operators
$\JJ_{\pm}:=
\JJ_x\pm i \JJ_y$, the infinitesimal rotation operators
$\JJ_x$, $\JJ_y$, $\JJ_z$,
and the commutator $[A,B]:=AB{-}BA$.
In the case of a single-spin system with spin number $s$,
all these operators can be interpreted as $(2s{+}1){\times}(2s{+}1)$-matrices and
all possible tensors have distinct ranks $j\in\{0,1,\ldots,2s\}$. 
As an example for $s=1/2$, we obtain
the Pauli spin matrices $\JJ_x=\II{x}:= \sigma_x/2 =
\left(\begin{smallmatrix}
0 & 1\\
1 & 0
\end{smallmatrix}\right)/2$,
$\JJ_y=\II{y}:= \sigma_y/2=
\left(\begin{smallmatrix}
0 & -i\\
i & 0
\end{smallmatrix}\right)/2$,
 $\JJ_z=\II{z}:= \sigma_z/2 =
\left(\begin{smallmatrix}
1 & 0\\
0 & -1
\end{smallmatrix}\right)/2$ as well as the tensor operator components
$\TT_{0,0}=
\left(\begin{smallmatrix}
1 & 0\\
0 & 1
\end{smallmatrix}\right)/\sqrt{2}$, 
$\TT_{1,-1}=
\left(\begin{smallmatrix}
0 & 0\\
1 & 0
\end{smallmatrix}\right)$, 
$\TT_{1,0}=
\left(\begin{smallmatrix}
1 & 0\\
0 & -1
\end{smallmatrix}\right)/\sqrt{2}$, 
$\TT_{1,1}=
\left(\begin{smallmatrix}
0 & -1\\
0 & 0
\end{smallmatrix}\right)$, cf.\@~\cite{EBW87}.

Having given an operator basis by recalling tensor operators, we can complete
the discussion how to map operators to functions by deciding on a suitable set of functions.
Note that the tensor operators have been explicitly defined 
by Wigner \cite{Wigner31} and Racah \cite{Racah42}
(generalizing the vector operators discussed in \cite{CondonShortley})
to mimic the properties of spherical harmonics $\Y_{jm}$ \cite{Jac99},
which map the spherical coordinates $\theta$ and $\phi$ to a complex value
$r(\theta,\phi) \exp[i\beta(\theta,\phi)]$ with radial part $r(\theta,\phi)$ and 
phase $\beta(\theta,\phi)$. The components
$\TT_{jm}$ can consequently be mapped to 
spherical harmonics $\Y_{jm}$, see Chap.~5 of  \cite{Silver76}
or Chap.~8 of \cite{CH98}. Hence,
an operator $A$ acting on a single spin with spin number $s$ can be represented by a 
unique spherical function $f_A$ using the straightforward 
mapping 
(in this particular case $J=\{0,1,\ldots,2s\}$)
\begin{equation}
\label{direct mapping}
A=\sum_{j\in J} \sum_{m=-j}^j c_{jm}\TT  _{jm} \Leftrightarrow f_A=\sum_{j\in J} \sum_{m=-j}^j c_{jm}\Y_{jm},
\end{equation}
which translates an expansion of an operator (in terms of a tensor operator basis) into an expansion 
of a function (in terms of spherical harmonics).

In this work, we systematically generalize this approach to systems consisting of an 
arbitrary number of coupled spins. A particular focus will be the case of three coupled spins. 
In the general case of multiple spins, the set
of irreducible tensor operators contains multiple elements with the same rank $j$.
Consequently, a direct mapping as in \eqref{direct mapping} would not be bijective, 
as distinct operators would be mapped onto the same function. 
For instance, the tensor basis for a system consisting of two coupled spins 
contains three distinct tensors of 
rank $j=1$, see, e.g., \cite{EBW87}. 
The tensor operators $\TT_j$ of rank $j$ are not even uniquely determined if 
their multiplicity $n_j$ is larger than one. The corresponding subspace 
$\oplus_{p=1}^{n_j} B_{2j+1} = \unity_{n_j} \otimes B_{2j+1}$ of the tensor operator space
is decomposed into $n_j$ blocks $B_{2j+1}$ of dimension $2j{+}1$ and  
allows for transformations of the form $M\otimes \unity_{2j+1}$ (cf.\ \cite{Ledermann}),
where $M$ is a non-singular $n_j\times n_j$-matrix and
$\unity_q$ denotes an identity matrix of dimension $q$.
Therefore, the different tensor operators $\TT_j$ with identical rank $j$ can be mixed using
linear combinations.

A first idea for a generalization would be to view a multi-spin system with spin numbers equal to $s$ 
as a single spin with a higher spin $s'>s$ and visualize it using the map of  \eqref{direct mapping}. 
Even though this approach would meet the uniqueness of the map as stated in
property (A), it would destroy invariance properties under rotation and conceal
important physical features of the system contrary to our intentions in property (B).
There were even doubts if a generalization 
is possible at all \cite{PhilpKuchel}.

We present a possibility to distinguish the representations for 
the tensors $\TT  _{j}^{(\ell)}\neq\TT  _{j}^{(\ell')}$
by introducing additional labels $\ell,\ell' \in L$.
For suitably chosen labels,  the irreducible tensor operators $\TT  _{j}^{(\ell)}$ can then be
grouped  into subsets 
\begin{equation}
\label{tensor basis}
\B(\ell):=\bigcup_{j\in\J(\ell)}\TT  _{j}^{(\ell)}=\bigcup_{j\in\J(\ell)}\bigcup_{m=-j}^j \TT  _{jm}^{(\ell)}
\end{equation}
with respect to their label $\ell \in L$
such that each index set $\J(\ell)$ never contains a rank $j$ more than once.
This allows us to independently apply the approach of \eqref{direct mapping} to
each subset $\B(\ell)$. Thus, the main idea is to 
introduce multiple spherical functions $f^{(\ell)}_A$ for an operator $A$ and visualize them 
in parallel:
\begin{subequations}
\label{DROPS}
\begin{align}
\label{DROPS union}
A = \sum_{\ell\in L} A^{(\ell)} \Leftrightarrow &\bigcup_{\ell\in L} f^{(\ell)}_A \quad \text{with}\\
\label{DROPS mapping}
A^{(\ell)}=\!\!\sum_{j\in J(\ell)} \sum_{m=-j}^j \! \! c^{(\ell)}_{jm}\TT  ^{(\ell)}_{jm} \Leftrightarrow &
f^{(\ell)}_A=\!\!\sum_{j\in J(\ell)} \sum_{m=-j}^j \! \! c^{(\ell)}_{jm}\Y_{jm}.
\end{align}
\end{subequations}
In the following, such a visualization of spin operators will be denoted as \drops 
(\textbf{d}iscrete \textbf{r}epresentation of \textbf{op}erators for \textbf{s}pin systems) 
visualization 
and we will refer to each individual visualization of a spherical function $f^{(\ell)}_A$ as a \emph{\droplet}. 
It is essential that all $2j{+}1$ components $\TT^{(\ell)}_{jm}$ of a tensor operator $\TT^{(\ell)}_{j}$
are contained in the same \droplet 
in order to ensure the invariance properties under rotation due to property (B).

Finding a suitable choice for the labels $\ell \in L$  is sometimes called the 
\emph{problem of missing labels}
(see, e.g., \cite{JMPW74,Sharp75,IL95} and \pp{145} of \cite{RW10}). 
Note that labels are not restricted to numbers.
In a more general context one aims at finding a
complete set of mutually commuting operators or a set of good quantum numbers  
(see, e.g., \secn{10.4} and \pp{473} of \cite{Merzbacher98}) which 
enables the analysis of a quantum system in a complete and problem-adapted basis.

\begin{figure}[tb]
\includegraphics[]{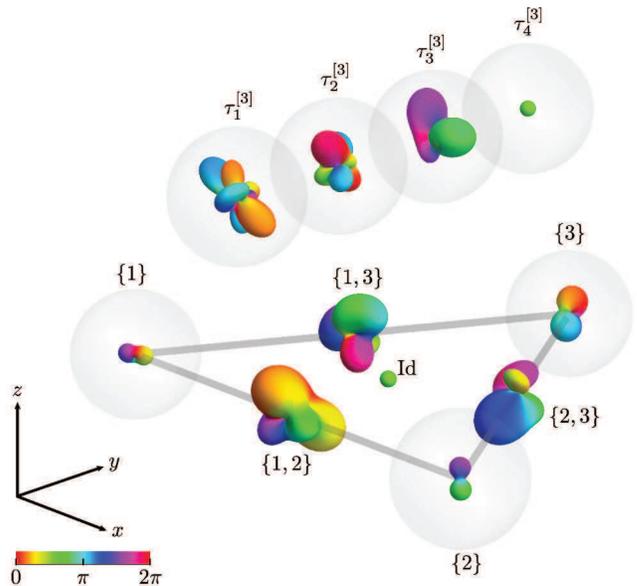}
\caption{(Color online) Visualization of an arbitrary  $8\times 8$-matrix $A$
whose complex matrix elements have been randomly chosen.
The matrix $A$ 
corresponds to an
operator acting on a 
system of three spins $1/2$.
In this particular example, $A$ is not hermitian. 
Each \droplet is associated with a specific linear combination
$f^{(\ell)}_A$ of spherical harmonics (see \eq{DROPS mapping}). 
The labels $\ell\in L$
are defined
 in \tab{tensor filter}.\label{random DROPS}}
\end{figure}

Although our choice of labels will only be presented in Sec.~\ref{Sec_lisa}, we refer the reader to 
\fig{random DROPS}
to establish ideas: Our approach is applied to
a system of three spins  
by visualizing a randomly-chosen operator $A$ 
using $|L| {=}11$ \droplets.
Here, the value $f(\theta,\phi)=r(\theta,\phi)  \exp[i\beta(\theta,\phi)]$ of a spherical function
at a point with coordinates $(\theta,\phi)$ is mapped 
to its distance $r(\theta,\phi)$ from the origin (forming the shape of the \droplet)
and its color corresponding to the phase $\beta(\theta,\phi)$
as defined by the color key in \fig{random DROPS}
(see, e.g., \cite{MJD} for an alternative technique for visualizing spherical harmonics).
We will detail further aspects
of \fig{random DROPS} in the course of 
our presentation.

Before presenting our choice of labels, we first elaborate on
how the set $L$ of labels and their quantity $\abs{L}$
is limited by what ranks $j$ (including their multiplicity) appear in a concrete system.
These limitations also directly affect how the set of irreducible tensor operators is grouped into a complete
orthonormal basis $\B = \cup_{\ell \in L} \B(\ell)$ as outlined
by \eqref{tensor basis}.
One obtains for a coupled system of three spins 
that the ranks $j\in\{0,1,2,3\}$ occur respectively with multiplicity
five, nine, five, and one (see \tab{tensor filter},
as will be explained in Secs.~\ref{Sec_lisa}
and \ref{Sec_arbitrary}). Therefore, the number $\abs{L}$ of 
labels is restricted to $9 \leq \abs{L} \leq 20$: The lower bound results from the maximal 
multiplicity of nine, and the upper bound is a consequence of the maximal number of distinguishable 
irreducible tensor operators as given by the sum of the multiplicities.
Our choice of eleven different labels (and \droplets) in \fig{random DROPS}
contains a few more labels than necessary but will yield further benefits, as explained below.

\section{\lisa tensor operator basis\label{Sec_lisa}}
Building on the discussion of how the set $L$ of labels induces a grouping of 
irreducible tensor operators and allowing for further symmetries beyond the 
ones of rotations,
the specific choice of labels for the \lisa tensor operator basis
is introduced. We proceed in three steps and sort the 
irreducible tensor operators into non-overlapping classes:  
First, we divide them with respect to the number $g$ of spins involved
(i.e.\ $g$-linearity). Second, we further split up the 
irreducible tensor operators with identical $g$ according to the set $G$ of involved spins with $\abs{G}=g$. 
Third, the symmetry types $\tau$ under permutations of the set $G$ give rise to a
decomposition of the subspace of irreducible tensor operators with identical $G$.
The permutations of a set of cardinality $g$ are known as the symmetric group $S_g$
\cite{Boerner67,Hamermesh62,Pauncz95,Sagan01}.
The third step is suppressed for $g\leq 2$ as no rank $j$ occurs more than once for
a given set $G$. In summary, a complete label is given by $(\ell)=(G,\tau)$
(if the number of spins is five or smaller as will be explained in Sec.~\ref{Sec_arbitrary}). We use the 
notations $\TT_j^{(\ell)}=\TT_j(\ell)=\TT_j^{G}(\tau)$ for a labeled irreducible tensor operator,
where both $G$ and $\tau$ can be omitted at will.
Next, we specify the labels (including the explicit form of $\tau$) for each $g$ while highlighting
the case of three spins 
(see \tab{tensor filter}).

For $g{=}0$, a single rank of zero appears in \tab{tensor filter}. 
The corresponding label is given by $\Id$ (or $\emptyset$),
and the \droplet of the single irreducible tensor component $\TT_{0,0}^{\Id}$ is plotted
in the center of the triangle in \fig{random DROPS}. 
The three linear irreducible tensor operators of rank one acting on a single spin
(i.e.\ $g{=}1$) are associated with the labels (and subsystems) $G\in \{\{1\}, \{2\}, \{3\}\}$ and 
are plotted at the vertices of the triangle in \fig{random DROPS}. 
The \droplets for bilinear tensor operators ($g{=}2$)
are plotted  at the edges of the triangle in \fig{random DROPS} and contain the ranks $j\in\{0,1,2\}$ for each
subsystem label $G\in \{\{1,2\}, \{1,3\}, \{2,3\}\}$ (see \tab{tensor filter}).
The full structure of the labeling will emerge for trilinear operators with
$g=3$ and $G=\{1,2,3\}$. Here, the ranks $j=1$ and $j=2$ occur more than once 
(see \tab{tensor filter}) and the symmetry types
\begin{equation}\label{taus}
\taun{1}{3}:=\TTab{1 2 3}\, ,\; \taun{2}{3}:=\TTab{1 2, 3}\, , \; \taun{3}{3}:=
\TTab{1 3, 2}\, ,\; \taun{4}{3}:=\TTab{1, 2, 3}
\end{equation}
will be applied for a complete labeling which
reflects the symmetries under permutations of the elements in $G$.
Each $\taun{i}{g}$ is a standard Young tableau
of size $g$ \cite{Boerner67,Hamermesh62,Pauncz95,Sagan01} 
which is a left-aligned arrangement of $g$ boxes where the number of boxes
does not increase from one row to following ones and 
where each box contains a different number from a set $G$
such that the numbers are ordered strictly
increasing from left to right and top to bottom.
The standard Young tableaux $\taun{1}{3}$ and  $\taun{4}{3}$ represent
complete symmetrization and antisymmetrization, respectively.
The four \droplets for the three-linear operators are given above the triangle in 
\fig{random DROPS}.

\begin{table}[tb]
\includegraphics[]{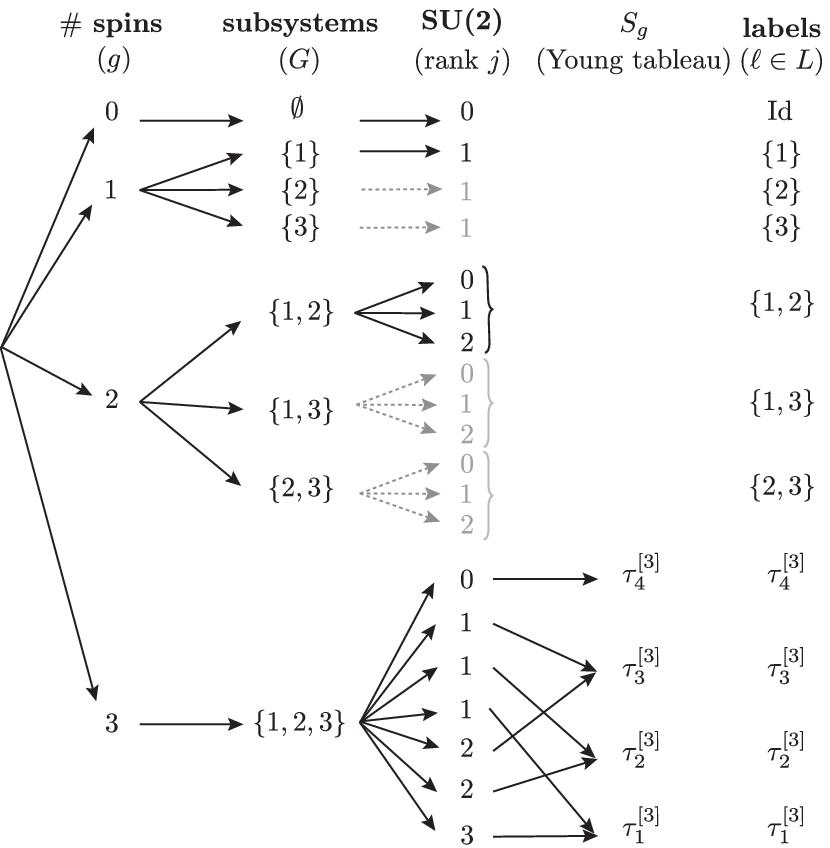}
\caption{Labels for the irreducible tensor operators 
corresponding to the \lisa basis for three coupled spins. 
Each irreducible tensor operator is first labeled by the number of spins
involved (i.e.\ its $g$-linearity) and then with the set of spins (i.e.\ subsystems) involved.
Standard Young tableaux $\taun{i}{g}$ of 
the symmetric group $S_g$ are used for $g\geq 3$
to complete the labeling process (for less than six spins). 
As some of this information is redundant, the labels are simplified
in the last column while providing a complete partitioning into \droplets.\label{tensor filter}}
\end{table}

Note that the symmetry types are trivial for $g\in\{0,1\}$.
But it is worthwhile to discuss their explicit form for $g{=}2$ 
even though they are suppressed in the labels of  \tab{tensor filter}. 
The bilinear irreducible tensor operators of 
rank zero and two are automatically symmetric under spin permutations
(i.e.\ have the symmetry type $\taun{1}{2}{:=}\TTab{2 3}$ for the case of $G{=}\{2,3\}$), 
while the case of  rank one is antisymmetric
(i.e.\ $\taun{2}{2}{:=}\TTab{2, 3}$). 
In general, different symmetry types are mapped to different \droplets.
But bilinear operators are an exception where all symmetry types are combined
in a single \droplet (see \tab{tensor filter}).

Evidently, the \lisa tensor operator basis and the corresponding
decomposition of the tensor space 
are based on methods perfected by Weyl~\cite{Weyl31,Weyl50,Weyl46}
which relate the structures of the unitary group $\SU(2)$ and 
the symmetric group $S_n$, where $n$ denotes the total number of spins.
But we symmetrize only with respect to spin permutations over tensors
with defined linearity and subsystem (see \tab{tensor filter}).
Approaches which symmetrize over all tensors of 
fixed linearity (or even over the complete set of tensors) as in 
\cite{Listerud_thesis,LGD93} are more suitable for sets of
\emph{indistinguishable} spins (refer also to the discussion in Sec.~\ref{Sec_Discussion}). 
The corresponding decomposition of the tensor 
space has been analyzed before  by \cite{Chakrabarti64,LLLN65} (see also \cite{McI60})
using different methods. We refer in this context also
to the detailed work of Temme et al.\ 
\cite{Temme90a,Temme92,Temme2000a,Temme2004,ST2008},
and references therein.

\subsection{Phase and sign}
Before we outline how to explicitly construct the irreducible tensor operators, 
we address the non-uniqueness of their phase and sign. The phase of a tensor 
component $\TT_{jm}$
is determined by observing the Condon-Shortley phase convention 
$\TT_{jm} = (-1)^{m}\,\TT_{j,-m}^{\dagger}$ \cite{CondonShortley,Schwinger}, where $A^{\dagger}$
denotes the complex conjugate and transpose of a matrix~$A$. Consequently, 
the tensor components $\TT_{jm}$ are defined up to an algebraic sign and
the ones for $m=0$ are hermitian.  Moreover, 
visualizations of hermitian matrices feature only the colors red and cyan corresponding
to phases of zero and $\pi$, respectively. More generally,
only the two colors for the phases $\gamma$ and $\gamma+\pi$ appear
if a matrix is hermitian up to a factor of $\exp(i\gamma)$. 
Although the choice of sign for each rank $j$ is completely arbitrary, 
the resulting 
visualizations can differ notably.
Our choice for the \lisa basis is motivated in Appendix~\ref{App_Motivation}.

\subsection{Iterative construction\label{construction}}
We outline the explicit construction of the \lisa tensor operator basis which is built up
iteratively from the tensor operators $\TT_0$ and $\TT_{1}$ for one spin.
For a general system consisting of $n$ spins 
with $n\leq 5$, the construction consists
of three steps: (I)~In the first step, the $g$-linear tensor operators 
of a $g$-spin system are constructed and symmetrized for all $g\in \{0,1,\ldots,n\}$ such that
they reflect both the symmetries of the unitary group $\SU(2)$ and the symmetric group $S_g$.
(II)~In the second step,
the tensor operators are multiplied with suitable phase factors in order
to comply with the phase and sign conventions detailed above.
(III)~Lastly, the tensor operators which have been constructed for a 
$g$-spin system 
are naturally embedded into the full $n$-spin system for each $g$-element subset of $\{1,\ldots,n\}$. 
The steps (II) and (III) are straightforward,
and we provide further details for the first step which will be subdivided into two parts
(Ia) and (Ib).

In part (Ia), the tensor operators $\tsymF{j}{i}{g-1}$ for $g{-}1$ spins are combined 
with the tensor operator
$\TT_1$ for one spin  to iteratively build up tensor operators $\tsymF{j}{i}{g-1}\otimes \TT_1$ 
for $g$ spins. We determine the
occurring tensor operators using
the Clebsch-Gordan decomposition
\begin{equation}
\label{CG decomposition}
\TT_j \otimes\TT  _1=\TT_{|j-1|}\oplus\cdots\oplus \TT_{j+1},
\end{equation}
and their explicit 
form is given by the 
Clebsch-Gordan coefficients \cite{Wigner59,BL81,Zare88}
for which---in the case of \eq{CG decomposition}---simple closed formulas (cf.\ \pp{635} in \cite{BL81}) and tables 
exist (see \pp{419} in
\cite{PDG12}).

After completing part (Ia) the tensor operators observe the symmetries of $\SU(2)$ but not the ones of the 
symmetric group $S_g$. In step (Ib), the symmetrization can be completed using one of two approaches.
The first approach relies on explicit projection operators
\cite{Boerner67,Hamermesh62,SancTemMNMRXIII,tung1985group,Sagan01} which project
onto subspaces with distinct symmetry type $\tau_i^{[g]}$ 
as given in \tab{tensor filter} (for details refer to Appendix~\ref{App_construction}).  In the second approach,
fractional parentage coefficients 
\cite{Racah65,EL57,Kaplan75,Silver76,Chisholm76,KJS81} 
determine how the tensor operators are recombined into their permutation-symmetrized versions.
Formulas for the fractional parentage coefficients are analyzed in the literature 
\cite{Hassitt55,Kaplan62a,Kaplan62b,Horie64}, but in this case 
one can conveniently rely on explicit tables from \cite{JvW51}.

The explicit matrix form of the \lisa basis is detailed in Appendix~\ref{App_construction}.
Before addressing the relation of the \drops visualization to Wigner functions as well as
the extension of our approach to an arbitrary number of spins, we 
provide explicit applications and examples for the \lisa basis.

\section{Visualization of typical operators in the \lisa basis}
\subsection{Cartesian product operators\label{Sec_cartesian}}
\begin{figure}[b]
\includegraphics[]{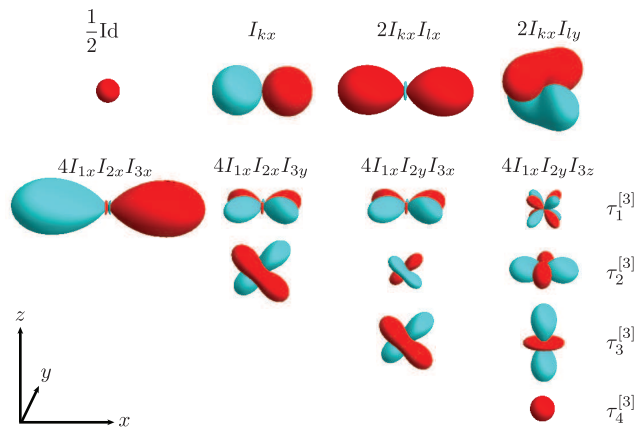}
\caption{(Color online) Examples of characteristic \droplets for Cartesian product  operators \cite{EBW87}.
The red and cyan colors refer respectively to positive and negative values of the spherical functions.
The \droplets for most of the remaining Cartesian product operators can be obtained 
by rotating the displayed ones. The \droplets for $4I_{1y}I_{2x}I_{3x}$ differ from
the ones for $4I_{1x}I_{2y}I_{3x}$ only in an inversion of color for the $\tau_3^{[3]}$-component.
\label{cartesian}}
\end{figure}

The \lisa basis is in particular suitable for visualizing
Cartesian product operators
which form a widely-used orthogonal basis in spin physics \cite{EBW87}.
For a single spin, the Cartesian product operators $\II{\eta}$ with $\eta\in\{x,y,z\}$ have been defined above.
The one-spin operators $\II{\eta}$ are embedded in an 
$n$-spin system as
$\II{k \eta} := \bigotimes_{s=1}^{n} \II{a_{s}}$ 
where $a_{s}{=}\eta$ for $s{=}k$ 
and $a_{s}{=}0$ otherwise; $\II{0}:=
\left(\begin{smallmatrix}
1 & 0\\
0 & 1
\end{smallmatrix}\right)$.
Cartesian product operators 
as
$\II{2x}$, $2\II{1z}\II{3_y}$ and 
$4\II{1x}\II{2x}\II{3y}$ have usually a prefactor of $2^{d-1}$ where $d$ denotes the number of single-spin operators.
Explicit transformations between the \lisa basis and the Cartesian product operators
are given in Appendix~\ref{transformation}.
A Cartesian product operator acts on a well-defined subsystem
and can consequently be represented with very few  \droplets.
Each of these \droplets features only the colors red and cyan (see \fig{cartesian})
as Cartesian product operators are hermitian. 
We discuss now the visualizations of the linear, bilinear, and trilinear cases (see \fig{cartesian}).

The only  \droplet for a linear Cartesian product operator $I_{k\eta}$ on spins $k$
consists of two spheres which are colored red and cyan, while the red one corresponds to a positive sign 
and points into the direction of the axis $\eta\in\{x,y,z\}$ (see \fig{cartesian}).
More generally, the axis of the \droplet for a Cartesian operator $a_x I_{kx} +a_y I_{ky} +a_z I_{kz}$ 
with $a_\eta \in \R$ is
collinear and proportional to the vector $a_x \vec{x} +a_y \vec{y} +a_z \vec{z}$.
This corresponds to a vector generalizing respectively a
(magnetic) field vector for Hamiltonians and
the Bloch vector for density matrices, which represents
the magnetization in nuclear magnetic resonance (NMR) \cite{EBW87} or 
polarization in quantum optics \cite{SchleichBook,HalsMNMRIX}.

Bilinear Cartesian operators $2I_{k\eta_1}I_{l\eta_2}$  on spins $k$ and $l$ with $\eta_1 \neq \eta_2$
such as an anti-phase coherence operator \cite{EBW87} with $\eta_1=x$ and $\eta_2=z$ induce a 
droplet with a particular shape (see $2I_{kx}I_{\ell y}$ in \fig{cartesian})  
which is not common for 
atomic or molecular orbitals.
This \droplet consists of two bean-shaped lobes 
with colors red and cyan whose major axes are orthogonal to each other. 
The major axis of the red lobe is oriented in the direction $\vec{\eta_1}+\vec{\eta_2}$, while the  axis from
the cyan lobe to the red one points in the direction $\vec{\eta_1}\times\vec{\eta_2}$ given by the right-hand 
rule. The \droplet transforms naturally under rotations, e.g., when both spins are simultaneously
rotated by $\pi$ around the $x$-axis, the anti-phase operator $2I_{kx}I_{lz}$
is mapped to $-2I_{kx}I_{lz}$ corresponding to a \droplet of the same shape but with
inverted colors.
More general shapes appear for the bilinear operators $2I_{k\eta}I_{l\eta}$
in the form of elongated shapes as well as 
antisymmetrically
elongated shapes oriented along the $\eta$-direction
for the trilinear operators $4I_{1\eta}I_{2\eta}I_{2\eta}$ (see \fig{cartesian}).

\subsection{Multiple quantum coherences\label{Sec_multiple}}
Operators $A_p$ of defined coherence order $p$ 
play an important role 
in NMR spectroscopy~\cite{EBW87}. 
They are invariant under global $z$-rotations up to a phase factor:
$$
\exp( -i\alpha\sum_{k=1}^{n}\II{kz})\, A_p\, \exp( i\alpha\sum_{k=1}^{n}\II{kz})
=
A_p \exp(-ip\alpha).
$$
This property is nicely captured in the \drops representation 
as detailed
in Appendix~\ref{App_multiple}, where
characteristic multiple-quantum terms for linear, bilinear and trilinear operators 
are displayed in the \lisa basis.

\begin{figure}[b]
\includegraphics[]{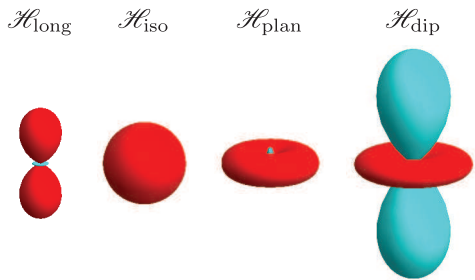}
\caption{(Color online) Non-trivial \droplets for coupling Hamiltonians: 
Ising-$\mathrm{ZZ}$ (or Heisenberg-Ising) model $\H_{\mathrm{long}}$, 
Heisenberg-$\mathrm{XXX}$ model $\H_{\text{iso}}$, 
Heisenberg-$\mathrm{XX}$ model $\H_{\mathrm{plan}}$, and
dipolar coupling $\H_{\mathrm{dip}}$  as detailed in the main  text.
\label{fig:Hamiltonians}}
\end{figure}

\subsection{Hamiltonians\label{Hamiltonians}}
Hamiltonians can also be conveniently visualized using the \drops representation.
The cases of linear and bilinear terms of the Hamiltonian mirror the properties of Cartesian 
product operators as 
discussed above. We analyze the shape of the \droplets for
bilinear coupling Hamiltonians 
$$\H_{\text{bil}}=2\pi \sum_{kl} c_{kl}(a\II{k x}\II{l x}{+}a\II{k y}\II{l y}{+}b\II{k z}\II{l z})$$
representing characteristic spin-spin interactions (see, e.g., \fig{fig:Hamiltonians}):
The cases $a=0$ and $b=1$ correspond to
the Ising-$\mathrm{ZZ}$ (or  Heisenberg-Ising) model $\H_{\mathrm{long}}$ \cite{Isi:1925,Cas:1989}, 
which is also known as
weak \cite{EBW87}  or longitudinal coupling \cite{Glaser93}
and is represented by a longitudinally elongated \droplet.
The Heisenberg-$\mathrm{XXX}$ model $\H_{\mathrm{iso}}$ with  $a=1$ and $b=1$ is also denoted
as strong or isotropic coupling \cite{EBW87} and results in an isotropic  \droplet of spherical shape.
For $a=1$ and $b=0$, we obtain the Heisenberg-$\mathrm{XX}$ model $\H_{\mathrm{plan}}$
which is also know as the planar coupling \cite{SMSE91,Glaser93}
and is represented as a planar disc-shaped \droplet in the $x$-$y$ plane.
The case $a=1$ and $b=-2$ corresponds to
a dipolar coupling $\H_{\mathrm{dip}}$.
More general coupling terms 
can also be visualized. 
Examples 
as
anisotropic Heisenberg $\mathrm{XYZ}$ or effective trilinear coupling terms
could be used in visualizations of multi-spin systems such 
as the well-known Kitaev honeycomb lattice \cite{Kitaev111}.
Hence, even for very large spin systems the \lisa basis (presented here explicitly for three spins) can be used to 
visualize Hamiltonians with at most trilinear terms.

\begin{figure}[b]
\includegraphics[]{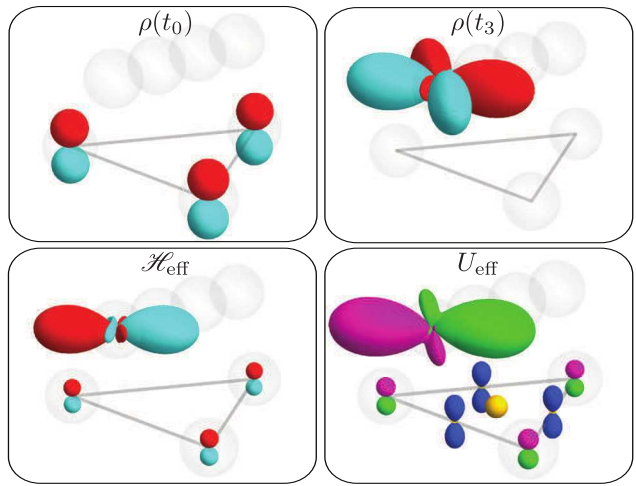}
\caption{(Color online) Example of an NMR pulse sequence creating triple-quantum coherences
which consists of a $\pi/2$ pulse (with phase $x$) followed by a delay 
and a second $\pi/2$ pulse (with phase $y$). The upper left panel shows (the traceless part of) the initial
density operator $\rho(t_0)=I_{1z}+I_{2z}+I_{3z}$. 
The final density operator $\rho(t_3)=4\II{1y}\II{2x}\II{3x}{+}4\II{1x}\II{2y}\II{3x}{+}4\II{1x}\II{2x}\II{3y}$
is given in the upper right panel.
The effective Hamiltonian $\H_{\mathrm{eff}}$ \cite{EBW87} and the effective propagator ${U}_{\mathrm{eff}}$ for 
the pulse sequence are shown in the lower left and lower right panels, respectively. \label{Short_wall}}
\end{figure}

\subsection{Time evolution} 
We provide an example visualizing
density operators, Hamiltonians, and unitary transformations
for a non-trivial
pulse sequence in NMR spectroscopy (see \fig{Short_wall}); full details are given in 
Appendix~\ref{App_time_evolution}.
The pulse sequence consists of two $\pi/2$ pulses separated by 
a delay, which is designed to excite triple-quantum coherence starting from the thermal density 
operator in the high-temperature limit \cite{EBW87}.
This highlights crucial information 
for the system and provides a better understanding of the corresponding time evolution.
Note that the unitary transformation is not hermitian and therefore requires  in general more colors.

\begin{figure}[b]
\includegraphics[]{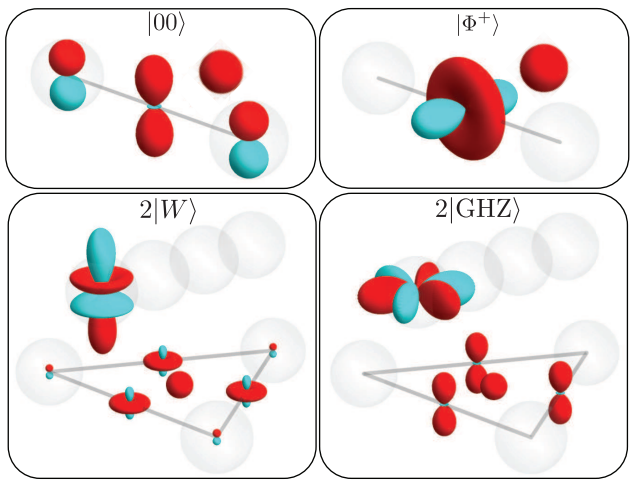}
\caption{(Color online) Visualizations of density matrices for separable and entangled pure states.
The upper panels show  
the product state $\ket{00}$ (as
a separable state) and the maximally-entangled
Bell state $\ket{\Phi^{+}}$ 
for two spins.
In the lower panels, the W state $\ket{W}$
and the 
Greenberger-Horne-Zeilinger state $\ket{\text{GHZ}}$
are shown as entangled quantum states for three spins. 
\label{Short_Entanglement}}
\end{figure}

\subsection{Pure quantum states\label{purequantum}}
In the field of quantum information  \cite{NC00}, 
(in addition to mixed states) pure states and their entanglement measures 
are of particular interest.
Density matrices for four examples of pure states are visualized in  
\fig{Short_Entanglement}: a product state  $\ket{00}$ and a maximally-entangled
Bell state  $\ket{\Phi^{+}}=(\ket{00} {+} \ket{11})/\sqrt{2}$ for two spins 
as well as the 
W state $\ket{W}=(\ket{100}{+}\ket{010}{+}\ket{001})/\sqrt{3}$ and the 
Greenberger-Horne-Zeilinger state $\ket{\text{GHZ}}=(\ket{000}{+}\ket{111})/\sqrt{2}$
for three spins (further cases are given in Appendix~\ref{App_pure_states}).
The reduced density matrix $\rho_1= \Tr_2 \rho$ of a two-spin system
is obtained by tracing over the second spin which
translates into deleting in the \lisa basis expansion
all linear terms corresponding to
the second spin and all bilinear terms.
Let $r_1$ denote the maximal radius of the droplet
for the first spin, i.e.\ the maximal
absolute value  of
the spherical function 
$f_{\rho}^{\{1\}}(\theta,\phi)$ 
(cf.\ \eq{DROPS})
for the corresponding  linear terms. Thus,
the length  of the Bloch vector 
corresponding to $\rho_1$ is given by $b_1=r_1  \sqrt{2^{n+2}\pi /3}$ for $n$-spin systems.

As an example for an entanglement measure for two spins,
consider the concurrence $C=\sqrt{1{-}b_1^2}$ \cite{Woo:1998}
which can 
be expressed as a function of $b_1$ (see also \cite{OV06}, \cite[\pa{168}]{HR06}, 
or \cite[\pp{50}]{StolzeSuter}).
Hence, it can also be obtained from the maximal radius $r_1$ as $C=\sqrt{1{-}16\pi r_1^2 /3}$.
The value of the concurrence $C$ for the pure state
$\ket{00}$ and $\ket{\Phi^{+}}$ of \fig{Short_Entanglement} is zero and one, respectively.

For the three-spin examples in \fig{Short_Entanglement},
interesting information about their bipartite entanglement 
measured by the concurrence
\cite{Woo:1998b,CKW:2000,DVC00} can be directly deduced
form the size of the \droplets corresponding to linear terms.
As the linear terms in the \lisa basis expansion are very small
for the W state $\ket{W}$,
each spin is strongly entangled with the rest of the system \cite{CKW:2000}. 
For $\ket{\text{GHZ}}$, the linear terms in the \lisa basis expansion 
are even zero implying that 
each spin is maximally entangled 
with the rest of the system \cite{Woo:1998b,CKW:2000}.

The preceding examples highlight that a \drops representation 
using the \lisa basis implicitly includes all reduced density 
matrices. A reduced density matrix is obtained by deleting
all \droplets (or terms in the \lisa basis) which correspond to
spins which have to be traced out. This should be particularly 
instrumental in visualizing quantum states while emphasizing
their symmetries and entanglement properties.

\section{Generalized Wigner representation\label{sect_WIgner}}

In this section 
we describe how the \drops representation 
can be interpreted as a generalized Wigner function.
Recall that 
the Wigner \emph{quasi}-probability distribution (or Wigner function for short) 
provides an equivalent phase-space formulation 
for the standard Hilbert-space framework of 
quantum mechanics and mimics the phase-space probability distribution in classical physics
\cite{Wey27,Wig32,Gro46,Ville48,Moy49,Cohen95,Leonhardt97,Shadows,SchleichBook,AM04,Curtright-review}.
It is formally not a probability distribution as negative values may appear.
These negative values in Wigner functions might be interpreted as signatures
of quantum effects (cf.\ \cite{Ferrie11}). However, the scope of this interpretation
is still widely discussed in the literature 
\cite{Johansen97,BW98,BW99,KZ04,RMMJ05,DMWS06,Spekkens08,MKC09,KMMR09,WB12,MO14}.
Although Wigner functions were originally developed for infinite-dimensional quantum systems with 
continuous degrees of freedom, they can  be extended to 
finite-dimensional quantum systems following the work of Stratonovich~\cite{Stratonovich},
see, e.g., \cite{VGB89,Brif97,Brif98}.
For the finite-dimensional case, a different perspective
is provided by a comprehensive theory of square-integrable functions on compact Lie groups
(e.g., functions on the sphere for $\SU(2)$)
introduced in the seminal work of Peter and Weyl~\cite{PW27} (see \cite{Sepanski07}).
The case of non-compact Lie groups 
is still an area of active research \cite{Helgason00,Varadarajan89}.
However, a relatively simple example of a non-compact Lie group is the symplectic group
 which  is widely studied in quantum optics \cite{SMD94} in the context of 
infinite-dimensional systems.

Returning to 
Wigner functions of finite-dimensional systems, 
the case of one spin (and a ``collection'' of spins)
was detailed in \cite{ACGT72,Agarwal81,DowlingAgarwalSchleich}
where each 
tensor operator is mapped to a unique (square-integrable) function on 
a sphere along the lines of \eq{direct mapping}.
In these spherical plots, 
qualitative signatures of quantum effects such as ``oscillating fringes'' and ``interference patterns''
have been analyzed in \cite{DowlingAgarwalSchleich,APS97,BC99,Harland,SFGD14}.
But similarly as for the above discussed infinite-dimensional Wigner functions the significance of these
qualitative signatures is still debated in the literature. For example, Refs.~\cite{FE08,FE09,Ferrie11} state
that quasi-probability distributions corresponding to states \emph{and} measurements
are necessary for reliably detecting quantum effects from negative values.
A different strategy for visualizing entanglement properties of quantum states
could be based on the localized information in the \lisa basis.
A first example in this direction is given in Sec.~\ref{purequantum} where pure quantum 
states are analyzed with the help of reduced density matrices.

The approach of Stratonovich \cite{Stratonovich}
defines a ``spherical phase space'' \cite{Ferrie11}.
But the general case of Wigner functions for coupled spin systems
has not been solved so far \cite{PhilpKuchel,Harland}.
Therefore, it is important to point out that the approach introduced
in this work (c.f.\ \eq{DROPS})
provides in fact a solution to this open problem.
Rather than mapping each operator $A$ to a single function on a sphere (see \eq{direct mapping}),
it is mapped to a set  $\{f^{(\ell)}_A(\theta,\phi) \text{ with } \ell\in L\}$ of functions on multiple spheres.
This set satisfies conditions generalizing
the ones of Stratonovich \cite{Stratonovich,Brif97,Brif98} and hence
can be interpreted as a generalized Wigner function:

\begin{proposition}\label{prop}
We assume 
that the \drops representation  of \eq{DROPS} observes the Condon-Shortley phase convention
and that the  functions $f^{(\ell)}_A(\theta,\phi)$ 
are correctly normalized. The following conditions are fulfilled:\\
(a) Linearity: 
$A\mapsto f^{(\ell)}_A(\theta,\phi)$ is linear for each $\ell \in L$.\\
(b) Reality: $f^{(\ell)}_{A^{\dagger}}(\theta,\phi)=[f^{(\ell)}_A(\theta,\phi)]^*$ holds for each $\ell\in L$.\\
(c) Norm: $\sum_{\ell\in L}\int_{S^2}f^{(\ell)}_A(\theta,\phi)\, f^{(\ell)}_{\Id}(\theta,\phi)d\mu=\Tr(A)$.\\
(d) Covariance: $f^{(\ell)}_{R(A)}(\theta,\phi)=f^{(\ell)}_A(R^{-1}(\theta,\phi))$ holds for 
each $\ell\in L$ and all non-selective rotations $R\in \SU(2)$,\\
(e) Trace: $\sum_{\ell\in L}\int_{S^2}f^{(\ell)}_A(\theta,\phi)\, f_B^{(\ell)}(\theta,\phi)d\mu=\Tr(AB)$.\end{proposition}

The (inversely) rotated point on the sphere has the coordinates
$(\theta',\phi'):=R^{-1}(\theta,\phi)$. The corresponding action $R(A)=UAU^{-1}$ describes a 
non-selective spin conjugation where the unitary matrix $U$ is a non-selective rotation operator 
(acting on complex column vectors, in particular on quantum-mechanical Hilbert-space vectors).
The straightforward proof of Prop.~\ref{prop} is given in Appendix~\ref{App_Prop_1}.
Based on these criteria, it is possible to describe the state of spin systems using droplets, 
i.e.\ sets of linear combinations of spherical harmonics. In particular, for a given set of droplets, 
the expectation value of an operator $A$ can be calculated based on (e), where $B$ is replaced 
by the density operator $\rho$:
$\langle A \rangle = {\rm Tr}(A \rho)$.
The relations of Prop.~\ref{prop} are obtained from the Stratonovich conditions in a straightforward fashion 
by simply applying them to each \droplet individually (for (a), (b), and (d)) or by summing over 
all the \droplets (for (c) and (e)).
Note that in contrast to the original Stratonovich conditions, in (c) the function $f^{(\ell)}_{\Id}(\theta,\phi)$
appears. 
This is a direct consequence of (e) if $B$ is replaced by the identity operator.
It ensures that unwanted contributions from functions $f^{(\ell)}_A(\theta,\phi)$ with traceless $A$ are
eliminated.
Consider for example the  traceless basis operator $A=\TT_{00}^{\{1,2\}}$  
for two spins which is mapped to $\Y_{00}(\theta,\phi):=1/\sqrt{4\pi}$ on one \droplet.
Without the presence of $f^{(\ell)}_{\text{Id}}(\theta,\phi)$, (c) would result in
$ 2\sqrt{\pi}=\int_{S^2}\Y_{00}(\theta,\phi)d\mu =\Tr(\TT_{00}^{\{1,2\}})=0$,
which is a contradiction.
For the special case of a single spin, the only droplet with rank $j=0$ corresponds to the identity operator 
and the generalized criterion (c) given above can be reduced to the original form 
$\int_{S^2}f_A(\theta,\phi)d\mu=\Tr(A)$ of Stratonovich.
As the integrals  $\int_{S^2} \Y_{jm}(\theta,\phi)d\mu$ vanish for $j>0$ anyway, 
the only terms contributing to (c) come from the cases with $j=0$ 
corresponding to the spherical harmonic $\Y_{00}(\theta,\phi)$.
In the case of the \lisa basis, the identity operator is mapped to a unique \droplet
as the basis operators are also characterized by particle number. Then, the sum in (c)  
reduces to a single term with the integral  corresponding to this particular \droplet 
while ignoring all the other ones.

\section{Arbitrary number of spins\label{Sec_arbitrary}}

We demonstrate now to what extent the construction of the \lisa basis 
(as outlined in Sec.~\ref{Sec_lisa}  and detailed in Appendix~\ref{App_construction})
is applicable to an arbitrary number of spins. It is explained that the available quantum numbers
(or the corresponding symmetries) are sufficient for labeling a tensor operator
basis of the full quantum system or that the quantum numbers can be easily extended with ad-hoc labels.
In particular, no ad-hoc labels are necessary for up to five spins.
Our analysis also identifies the inherent symmetry structure of the quantum system 
and provides specifics on the number of \droplets for the \lisa basis. Moreover,
lower and upper bounds for the number of \droplets are given for general \drops representations.

To this end, we resume discussing the choice of labels $\ell \in L$ which 
provides a partition of the irreducible tensor operators $\TT_{j}^{(\ell)}$
into subsets never containing tensor operators of rank $j$ more than once (see \eqref{tensor basis}).
This partition induces also a decomposition into different \droplets for the \drops representation. 
Recall that the number $\abs{L}$ of \droplets  for three spins 
is bounded by $9 {\leq} \abs{L} {\leq} 20$. Bounds for up to twelve spins  
are given in \tab{tab_two}.

In the particular case of the \lisa tensor operator basis,
the irreducible tensor operators $\TT_{j}^{(\ell)}$ 
of rank $j$ are divided according to the number $g$ of spins involved, the set $G$ of involved spins, 
and the symmetry type $\tau$ under permutations of the set $G$.
All possible combinations of $j$, $g$, $G$, 
and $\tau$ for up to three spins  
are shown in \tab{tensor filter}, where the notation was simplified
by suppressing some
trivial symmetry types. Below, 
we will explain the labels of \tab{tensor filter} 
for an arbitrary number of spins 
and provide a general 
method  for
computing all possible combinations of rank $j$ and symmetry type $\tau$ reflecting
both the symmetries of the unitary group $\SU(2)$ and the symmetric group $S_g$.
This also determines the number of \droplets for the \lisa basis
as shown in \tab{tab_two}.

We infer from \tab{tab_two} that the number of \droplets is significantly smaller than
the dimension $4^n$ of the operator space for $n$ spins. 
Moreover, the \lisa basis uses more \droplets than strictly necessary.
But meaningful labels are essential
as even the minimum number of \droplets grows very fast. Therefore, each \droplet
of the \lisa basis has a unique permutation symmetry type $\tau$
except for
bilinear operators,
where the two occurring symmetry types  (e.g., $\taun{1}{2}$ and $\taun{2}{2}$) are combined
into one \droplet which is possible
as no $j$-value appears more than once.
Based on the values in \tab{tab_two}, we consider
the \lisa basis 
as an efficient and informative
visualization for a moderate number of spins. 
In the following, we first present a method for computing the 
minimum and maximal number of \droplets for a \drops representation.
Secondly, we describe the explicit form of the labels for the \lisa basis and thereby determine
the corresponding number of \droplets.

\begin{table}[tb]
\caption{The minimum and maximal number of \droplets 
for a \drops representation of $n$ spins with
$n\in\{1,\ldots,12\}$
is compared to the \lisa basis. The column named multipole will be addressed in Sec.~\ref{Sec_Discussion}.
The minimum number of \droplets
divided by the dimension $4^n$ is also shown.
\label{tab_two}}
\color{black}
\begin{tabular}[t]{@{\hspace{1mm}}r@{\hspace{4mm}}r@{\hspace{2mm}}r@{\hspace{3mm}}r@{\hspace{2mm}}r@{\hspace{4mm}}r@{\hspace{1mm}}}
$n$ & \multicolumn{4}{c}{number of \droplets} & $\tfrac{\text{minimum}}{4^n}$\\
& minimum & multipole &\lisa & maximum\\
\hline\\[-3mm]
1 & 1 & 1 & 2 & 2 & 0.250\\
2 & 3 & 4 & 4 & 6 & 0.188\\
3 & 9 & 9 & 11 & 20 & 0.141\\
4 & 28 & 36 & 36 & 70 & 0.109\\
5 & 90 & 100 & 122 & 252 & 0.088\\
6 & 297 & 400 & 423 & 924 & 0.073\\
7 & 1001 & 1225 & 1486 & 3432 & 0.061\\
8 & 3640 & 4900 & 5246 & 12870 & 0.056\\  
9 & 13260 & 15876 & 18689 & 48620 & 0.051\\
10 & 48450 & 63504 & 67356 & 184756 & 0.046\\
11 & 177650 & 213444 & 244917 & 705432 & 0.042\\
12 & 653752 & 853776 & 896899 & 2704156 & 0.039\\
\end{tabular}
\end{table}

\subsection{The minimum and maximal number of \droplets}

Recall that the set of infinitesimal rotation operators (or equivalently the Lie algebra $\su(2)$) 
acts on the irreducible tensor operators $\TT_0$ and $\TT_1$ of a single spin 
via \eqref{racah}.
This means that the unitary group $\SU(2)$
(and its Lie algebra $\su(2)$) acts for a single spin 
non-trivially 
on a three-dimensional (complex) space $\C^3$ via its three-dimensional irreducible 
representation $\varphi_1$.
Here, an irreducible representation $\varphi_j$ of $\SU(2)$ denotes
in the language of representation theory 
\cite{FH91,Sepanski07,Sagan01} an action of $\SU(2)$ on the abstract space $\C^{2j+1}$ (for which
the irreducible tensor operator $\TT_j$ provides an explicit model)
and maps an element $g\in \SU(2)$ to a $(2j{+}1){\times}(2j{+}1)$ matrix $\varphi_j(g)$.

\begin{table}[b]
\caption{
We provide for each appearing $j$-value the multiplicities
$n_j$ for the $n$-linear operators in an $n$-spin system.
The multiplicity
$\bar{n}_j$ represents the number of all rank-$j$ tensors considering 
all possible subsets $G$ of the $n$ spins (including the $n$-linear operators).
The example values discussed in the text are highlighted.
\label{tab_one}}
\color{black}
\begin{tabular}{l@{\hspace{3mm}}c@{\hspace{3mm}}r}
\begin{tabular}[t]{@{\hspace{1mm}}l@{\hspace{2mm}}l@{\hspace{2mm}}r@{\hspace{2mm}}r@{\hspace{1mm}}}
$n$ & $j$ & $n_j$ & $\bar{n}_j$\\
\hline\\[-3mm]
0 & 0 & 1 & 1 \\[0.5mm]
1 & 0 & 0 & 1\\
   & 1 & 1 & 1\\[0.5mm]
2 & 0 & 1 & 2\\
   & 1 & 1 & 3\\
   & 2 & \mybox{1} & 1\\[0.5mm]
3 & 0 & 1 & 5\\
   & 1 & 3 & 9\\
   & 2 & \mybox{2} & \mybox{5}\\
   & 3 & 1 & 1\\[0.5mm]
4 & 0 & 3 & 14\\
   & 1 & 6 & 28\\
   & 2 & 6 & 20\\
   & 3 & 3 & 7\\
   & 4 & 1 & 1
\end{tabular}&
\begin{tabular}[t]{@{\hspace{1mm}}l@{\hspace{2mm}}l@{\hspace{2mm}}r@{\hspace{2mm}}r@{\hspace{1mm}}}
$n$ & $j$ & $n_j$ & $\bar{n}_j$\\
\hline\\[-3mm]
5 & 0 & 6 & 42\\
   & 1 & 15 & 90\\
   & 2 & 15 & 75\\
   & 3 & 10 & 35\\
   & 4 & 4   & 9\\
   & 5 & 1   & 1\\[0.5mm]
6 & 0 & 15 & 132\\
   & 1 & 36 & 297\\
   & 2 & 40 & 275\\
   & 3 & 29 & 154\\
   & 4 & 15 & 54\\
   & 5 & 5   & 11\\
   & 6 & 1   & 1\\[0.5mm]
7 & 0 & 36 & 429\\
   & 1 & 91 & 1001
\end{tabular}&
\begin{tabular}[t]{@{\hspace{1mm}}l@{\hspace{2mm}}l@{\hspace{2mm}}r@{\hspace{2mm}}r@{\hspace{1mm}}}
$n$ & $j$ & $n_j$ & $\bar{n}_j$\\
\hline\\[-3mm]
7  & 2 & 105 & 1001\\
   & 3 & 84 & 637\\
   & 4 & 49 & 273\\
   & 5 & 21 & 77\\
   & 6 & 6 & 13\\
   & 7 & 1 & 1\\[0.5mm]
8 & 0 & 91 & 1430\\
   & 1 & 232 & 3432\\
   & 2 & 280 & 3640\\
   & 3 & 238 & 2548\\
   & 4 & 154 & 1260\\
   & 5 & 76 & 440\\
   & 6 & 28 & 104\\
   & 7 & 7 & 15\\
   & 8 & 1 & 1
\end{tabular}\end{tabular}
\end{table}

For multiple spins, a simultaneous action of $\SU(2)$ on (e.g.) all $n$-linear operators of 
an $n$-spin system arises and the (inner) 
tensor product representation 
$\varphi_1^{\otimes n}:=\otimes^n \varphi_1=\varphi_1 {\otimes} \cdots {\otimes} \varphi_1$ 
of $\SU(2)$
naturally acts on the 
$n$-fold tensor product 
$\otimes^n \C^3:=\C^3{\otimes} \cdots {\otimes} \C^3$ of a three-dimensional (complex) space.
The tensor product representation $\varphi_1^{\otimes n}$ is known as the $n$th tensor power
of $\varphi_1$ and decomposes into a sum of representations $\varphi_j$ with multiplicities $n_j$
by means of the well-known technique of \eq{CG decomposition} 
(for general methods, cf.\ \ppp{424--429} of \cite{FH91} or \ppp{135--142} of \cite{Humphreys72}). 
The explicit values for $n_j$ in \tab{tab_one} have been computed using the computer algebra system
{\sc magma} \cite{MAGMA}.
The corresponding multiplicities $\bar{n}_j$ for
the full $n$-spin system are obtained by summing the multiplicities $n_j$ for $g$-linear operators with 
$g\in\{0,\ldots,n\}$ which have to be multiplied with the number 
$\tbinom{n}{g}{=}{n!}/[g! (n{-}g)!]$
of possible sets $G{\subseteq}\{1,\ldots,n\}$ of $\abs{G}{=}g$ spins.
For example consider the number $\bar{n}_j$  for $n=3$ spins with rank $j=2$. Here, for the set  
$G=\{1,2,3\}$ we find $n_2=2$ and for each of the sets $\{1,2\}$, $\{1,3\}$, and $\{2,3\}$ 
we have $n_2=1$ which can be inferred from the corresponding case of two spins. None of the subsets
$G\in\{ \emptyset,\{1\},\{2\},\{3\}\}$ yields an
operator with rank $j=2$ and the corresponding $n_2$-values are zero. Thus, 
$\bar{n}_j=1\times 2 + 3\times 1 + 3\times 0 + 1\times 0=5$.
For a given $n$, the minimum and maximal number of \droplets in  \tab{tab_two}
are now given by the maximum of the multiplicities $\bar{n}_j$
for all ranks $j$
and the sum $\sum_j \bar{n}_j$, respectively.

\subsection{All combinations of symmetry types}

We determine all possible combinations of rank $j$ and 
permutation symmetry type $\tau$
by refining our symmetry analysis of $g$-linear operators. 
Before,
we identified the symmetries of $g$-linear operators for rank $j$ 
which are modeled by 
a $g$-fold tensor product $\otimes^g \C^3$ and 
acted on by the unitary group $\SU(2)$.
We extend this action on $\otimes^g \C^3$ to an action of the direct 
product $\SU(2)\times S_g$, where
the symmetric group $S_g$ acts by permuting spins from a 
set $G{\subseteq}\{1,\ldots,n\}$ with $\abs{G}{=}g$. 

\begin{table}[b]
\caption{All combinations of  partitions $\lambda$ and ranks $j$ for $n$-linear operators 
in a $n$-spin system with $1\leq n \leq 8$.
The third column 
shows the number of symmetry types $\tau$ for each~$\lambda$.\label{tab_klambda}}
\color{black}
\begin{tabular}{l@{\hspace{2mm}}r}
\begin{tabular}[t]{@{\hspace{1mm}}l@{\hspace{2mm}}l@{\hspace{2mm}}r@{\hspace{5mm}}l@{\hspace{1mm}}}
$n$ & $\lambda$ & $\#\tau$ & $j$\\
\hline\\[-3mm]
1 & [1] & 1 & 1\\[0.5mm]
2 & [2]  & 1 & 0, 2\\
   & [1,1] & 1 & 1 \\[0.5mm]
3 & [3] & 1 & 1, 3\\
   & [2,1] & 2 & 1, 2\\
   & [1,1,1] & 1 & 0\\[0.5mm]
4 & [4] & 1 & 0, 2, 4\\
   & [3,1] & 3 & 1, 2, 3\\
   & [2,2] & 2 & 0, 2\\
   & [2,1,1] & 3 & 1\\[0.5mm]
5 & [5] & 1 & 1, 3, 5\\
   & [4,1] & 4 & 1, 2, 3, 4\\
   & [3,2] & 5 & 1, 2, 3\\
   & [3,1,1] & 6 & 0, 2\\
   & [2,2,1] & 5 & 1\\[0.5mm]
6 & [6] & 1 & 0, 2, 4, 6\\
   & [5,1] & 5 & 1, 2, 3, 4,\\
   & & & 5\\
   & [4,2] & 9 & 0, \underline{{\bf 2}, {\bf 2}}, 3,\\
   & & & 4\\
   & [4,1,1] & 10 & 1, 3\\
   & [3,3] & 5 & 1, 3\\
   & [3,2,1] & 16 & 1, 2\\
   & [2,2,2] & 5 & 0\\
\end{tabular}&
\begin{tabular}[t]{@{\hspace{1mm}}l@{\hspace{2mm}}l@{\hspace{2mm}}r@{\hspace{5mm}}l@{\hspace{1mm}}}
$n$ & $\lambda$ & $\#\tau$ & $j$\\
\hline\\[-3mm]
7 & [7] & 1 & 1, 3, 5, 7\\
   & [6,1] & 6 & 1, 2, 3, 4, \\
   & & & 5, 6\\
   & [5,2] & 14 & 1, 2, \underline{{\bf 3}, {\bf 3}},\\
   & & & 4, 5\\
   & [5,1,1] & 15 & 0, 2, 4\\
   & [4,3] & 14 & 1, 2, 3, 4\\
   & [4,2,1] & 35 & 1, 2, 3\\
   & [3,3,1] & 21 & 0, 2\\
   & [3,2,2] & 21 & 1\\[0.5mm]
8 & [8] & 1 & 0, 2, 4, 6\\
   & & & 8\\
   & [7,1] & 7 & 1, 2, 3, 4\\
   & & & 5, 6, 7\\
   & [6,2] & 20 & 0, \underline{{\bf 2}, {\bf 2}}, 3,\\
   & & & \underline{{\bf 4}, {\bf 4}}, 5, 6\\
   & [6,1,1] & 21 & 1, 3, 5\\
   & [5,3] & 28 & 1, 2, \underline{{\bf 3}, {\bf 3}},\\
   & & & 4, 5\\
   & [5,2,1] & 64 & 1, 2, 3, 4\\
   & [4,4] & 14 & 0, 2, 4\\
   & [4,3,1] & 70 & 1, 2, 3\\
   & [4,2,2] & 56 & 0, 2\\
   & [3,3,2] & 42 & 1\\
\end{tabular}\end{tabular}
\end{table}

The corresponding symmetry analysis for the number of spins $n\in\{1,\ldots,8\}$
is summarized in \tab{tab_klambda}, where 
we consider $g$-linear operators with $g=n$ in a $n$-spin system.
\tab{tab_klambda}
states all possible combinations of rank $j$ and partition
$\lambda$ from which
the corresponding permutation symmetry types $\tau$ can be easily determined.
A partition $\lambda:=[\lambda_1,\ldots,\lambda_{\kappa(\lambda)}]$
of length $\kappa(\lambda)$ and degree $\abs{\lambda}:=\sum_{p=1}^{\kappa(\lambda)} \lambda_p$ 
consists of positive integers $\lambda_p$ with $\lambda_p \geq \lambda_{p+1}$ 
and can be identified with a Young diagram (i.e.\ a Young tableaux $\tau$ without entries)
which is a left-aligned arrangement of 
$\abs{\lambda}$ boxes 
into $\kappa(\lambda)$ rows where the $p$th row contains $\lambda_p$ boxes (cf.\ \ppp{44--45} of \cite{FH91}).
The number of different symmetry types $\tau$ for each partition $\lambda$
is given in the third column ($\#\tau$) of \tab{tab_klambda}.
The results for $n\in\{1,2,3,4\}$ and some partial results for
$n\in\{5,6\}$ can also be found 
in Table~12 of \cite{Jahn50} (cf.\ Table~2 on \pp{294} of \cite{Kaplan75}).

Combinations of $\lambda$ and $j$ which appear more than once 
are highlighted in \tab{tab_klambda}. The case of
$\lambda=[4,2]$ and $j=2$ for $n=6$ has been known at least 
since \cite{FP37} (see also \cite{Jahn50,JvW51}).
If no combination of $\lambda$ and $j$ appears more than once (as for $1\leq n\leq 5$), the 
\lisa basis is 
uniquely defined without any additional labels.
This ensures that the visualization technique described in this work
is directly applicable for up to five spins. 
Additional labels are required in the general case, but ad-hoc labels as in \cite{FP37} 
are usually sufficient.
The resulting permutation symmetry types $\tau$ for  a partition $\lambda$ 
are given by the standard Young tableaux of shape $\lambda$ 
 \cite{Boerner67,Hamermesh62,Pauncz95,Sagan01}, i.e.\ Young diagrams of shape $\lambda$
which are filled with the numbers 
$G{\subseteq}\{1,\ldots,n\}$ from the set of the $\abs{G}{=}g$ involved spins.
The employed method for the computation of \tab{tab_klambda} is a combination of the
Schur-Weyl duality \cite{GW09} and a technique known as plethysm 
\cite{Littlewood58,Macdonald95,Wybourne70,RW10}. The results
in \tab{tab_klambda} were obtained using the computer algebra system
{\sc magma} \cite{MAGMA} and details are given in Appendix~\ref{App_table}.

We explain now how to read \tab{tab_klambda} and how to 
recover some of the labels for the three-spin case of \tab{tensor filter}.
In particular, we consider the subsystem of trilinear operators
(i.e.\ $g=3$ and $G=\{1,2,3\}$). Referring to the case of $n=3$ in \tab{tab_klambda}, 
we obtain the combinations of $j=0$ and $\lambda=[1,1,1]$, 
$j=1$ and $\lambda=[3]$,  $j=1$ and $\lambda=[2,1]$, $j=2$ and $\lambda=[2,1]$,
as well as $j=3$ and $\lambda=[3]$. Note that $\lambda=[3]$ induces
the permutation symmetry $\taun{1}{3}$ (using the notation of Eq.~\eqref{taus}). 
Similarly, one obtains $\taun{2}{3}$ and $\taun{3}{3}$ for $\lambda=[2,1]$
as well as $\taun{4}{3}$ for $\lambda=[1,1,1]$. Consequently, all the relevant 
labels in \tab{tensor filter} can be recovered.

\section{Discussion\label{Sec_Discussion}}
Before concluding, we discuss  alternative \drops visualizations which complement
the \lisa representation. A suitable choice can reflect the considered system and
application. 
One possibility arises form partitioning 
 the tensor operators from the Clebsch-Gordan decomposition of \eq{CG decomposition}
into \droplets without symmetrizing with respect to spin permutations as in the \lisa basis.
As before, one has to ensure that no rank $j$ appears more than once in any \droplet. 
Many different partitions are possible, and applying the Clebsch-Gordan decomposition
recursively could provide a natural partition.

Indistinguishable spins
utilize only a proper subspace of all tensor operators. 
The corresponding symmetry-adapted tensor basis can be obtained by symmetrizing 
tensor operators with respect to the relevant spin
permutations. This usually reduces the number of droplets. In particular, one can discard all \droplets
with incompatible symmetry types.
For example,  the Hamiltonian and the density 
operator in a three-spin system of type $I_2S$ \cite{Nielsen,EBW87}
are both invariant with respect to permutations of the first two spins.
Consequently, the  \lisa basis can be restricted to
a $40$-dimensional space consisting of the twelve irreducible tensor operators $\TT^{\emptyset}_0$,
$\tfrac12(\TTlb11{+}\TTlb12)$, $\TTlb 13$, $\TTlb 0{1,2}$, $\TTlb 2{1,2}$,
$\tfrac12(\TTlb 0{1,3}{+}\TTlb 0{2,3})$, $\tfrac12(\TTlb 1{1,3}{+}\TTlb 1{2,3})$, 
$\tfrac12(\TTlb 2{1,3}{+}\TTlb 2{2,3})$, $\tt 113$, $\tt 313$,
$\tt 123$, and $\tt 223$. This allows us to reduce the number of \droplets from eleven to seven.
For a three-spin system which is totally symmetric with respect to spin permutations, the \lisa 
basis can  be limited to a $20$-dimensional space spanned by the tensor components of the 
six irreducible tensor operators $\TT^{\emptyset}_0$,
$\tfrac13(\TTlb11{+}\TTlb12{+}\TTlb13)$, $\tfrac13(\TTlb 0{1,2}{+}\TTlb 0{1,3}{+} \TTlb 0{2,3})$,
$\tfrac13(\TTlb 2{1,2}{+}\TTlb 2{1,3}{+} \TTlb 2{2,3})$, $\tt 113$, and $\tt 313$. Thus,
one obtains four \droplets.

\begin{figure}[t]
\includegraphics[]{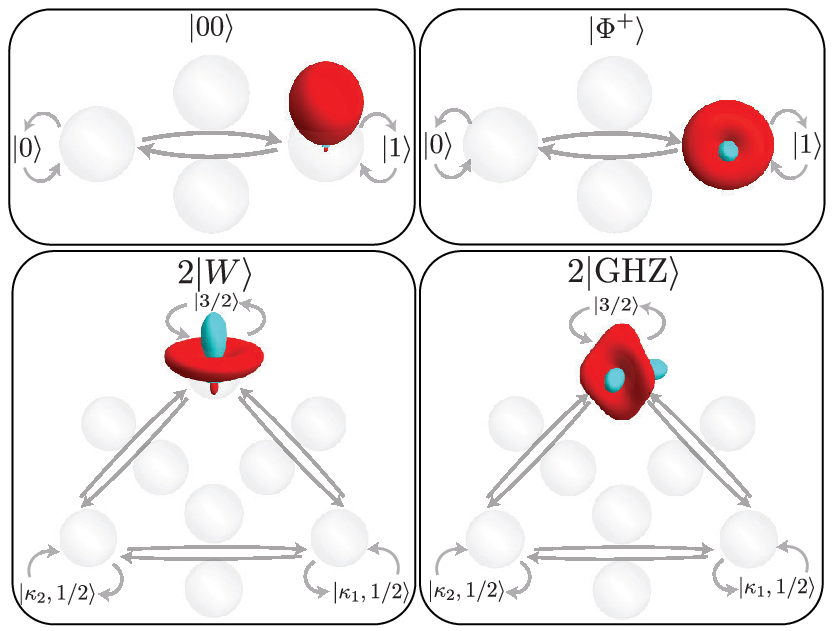}
\caption{(Color online) \Drops visualization based on multipole tensor operators
of the density matrices for the pure states of \fig{Short_Entanglement}.
\label{Short_Entanglement_Tmpo}}
\end{figure}

One further variant of the \drops visualization is based on
multipole tensor operators 
\cite{Sanct75,SancTemMNMRXIII,SanctuaryX,Sanct_MNMRXI,SanctNMRIII,SancAll89}
which reflect the state-space structure of angular momentum states
with suitable-chosen auxiliary labels. The multipole tensor operators
are defined by means of transforming one set of states 
of defined rank into another one
according to the Clebsch-Gordan decomposition
$\TT_{p} \otimes \TT_{{1}/{2}} = \TT_{\abs{p-{1}/{2}}} \otimes \cdots \otimes \TT_{p+{1}/{2}}$,
where $p\in\{0,{1}/{2},1,{3}/{2},\ldots\}$.
The multipole tensor operators differ from the tensor operators
in the \lisa basis in not having a defined particle number (i.e.\ linearity) and 
inducing a different grouping into \droplets (see Appendix~\ref{App_multipole}).
The corresponding number of \droplets is given in the column named multipole of  \tab{tab_two}
and can be computed as the square of the number of irreducible tensor operators $\TT_p$
(with multiplicity).
The multipole-based
\drops representation introduced here can be viewed as a generalization of the 
special Wigner representation introduced in
\cite{MJD} for a spin $1/2$ particle coupled to a second particle of arbitrary spin number.
Figure~\ref{Short_Entanglement_Tmpo} illustrates the \drops representation based
on multipole operators using the same pure-state examples which are shown
in \fig{Short_Entanglement} for the \lisa basis.
In contrast to \fig{Short_Entanglement}, 
in \fig{Short_Entanglement_Tmpo} all examples have 
only one single \emph{non-empty} \droplet.

\section{Conclusion}
We introduced a general approach for representing arbitrary operators by a finite set of 
functions.
Their properties make this representation particularly appealing and useful for
the visualization of important quantum mechanical concepts and properties, which are conventionally
represented by abstract operators or plain matrices.
There are many possible bases on which such a mapping between operators and sets of 
functions can be based.
Here we focused on the \lisa basis which transforms naturally under non-selective spin rotations 
as well as spin permutations and which is particularly suitable for distinguishable spins.
However, depending on the application, other bases can be more appropriate.
It is noteworthy that the \drops visualization can be seen as a generalization of the Bloch vector 
representation of simple two-level quantum systems, such as uncoupled spin 1/2 particles.
On the other hand, the \drops method can also be interpreted as a natural (albeit not obvious) 
generalization of Wigner functions that  have been studied extensively for ``collections'' of spins.
The physical interpretation of the 
\drops representation in terms of a quasi-probability distribution for experimental
observables and the related question of non-classical signatures are interesting open 
problems. These are however beyond the scope of the present work, which focusses on the theoretical 
construction and the symmetry properties of the \drops representation.

Although we considered here the relatively simple but non-trivial example of three coupled spins 1/2, 
our method can also be applied to more than three spins and is also not limited to spins 1/2. 
It is important to emphasize that general coupled spin systems constitute
complex high-dimensional quantum systems whose description cannot be expected to be both
\emph{complete} and \emph{simple}.
We outlined in the beginning of Sec.~\ref{sec_vis} the
essential properties which our visualization should satisfy. These properties can be summarized
as (A) the bijectivity of the mapping and (B) the immediate visibility of crucial features 
(such as symmetries under non-selective rotations). 
Under these assumptions,
the analysis of Sec.~\ref{Sec_arbitrary} (see \tab{tab_two}) rules out 
a complete representation of operators with only a single \droplet for spin systems with more than
one spin. In contrast to that, the visualization technique of Ref.~\cite{Harland} 
(see also \cite{ACGT72,Agarwal81,DowlingAgarwalSchleich}) 
uses only a single \droplet (even for more than one spin).  Therefore, in general, it cannot 
respect all symmetries of operators under 
non-selective rotation and be complete at the same time. The approach of \cite{Harland}
emphasizes the structure of pure quantum states and chooses quantum numbers from 
the eigenvalues of the operators 
$S^2:=S_x^2+S_y^2+S_z^2$ and $S_z$, where $S_\alpha:=(\sigma_\alpha \oplus 
\cdots \oplus \sigma_\alpha)/2$ with $\alpha\in\{x,y,z\}$.
But this captures only a subset 
of the symmetries of pure quantum states under non-selective rotations.
All these symmetries  are revealed by directly applying the Clebsch-Gordan 
decomposition of the tensor product structure. Hence, the approach of \cite{Harland} could be modified
into a \drops visualization using the multipole tensor operators (see Sec.~\ref{Sec_Discussion} 
and Appendix~\ref{App_multipole}),
which also emphasizes the structure of pure quantum states.
In summary, we believe that the analysis of coupled spin systems should be primarily guided
by finding their inherent symmetries. 
This provides a solid foundation for studying nontrivial properties of general  spin systems.

We illustrated applications which benefit from the \drops approach, such as
the visualization of mixed quantum states of spin systems. It can also used
to represent the density matrix of 
pure quantum states with or without entanglement. Section~\ref{purequantum}
contains a first step in understanding how entanglement properties of quantum states
can be visualized in the \lisa basis.
Furthermore,
the \drops representation can be applied to arbitrary operators, including
Hamilton operators and time evolution operators. 
This approach is also well suited to show the time evolution of quantum mechanical operators
as animations, rather than static figures.
The \drops visualization lends itself to building intuition about the dynamics of coupled spins
and is expected to become a valuable tool both in education and research.
Potential applications range from theoretical and experimental quantum information theory, where 
quantum bits (corresponding to spins $1/2$)
can be realized by trapped ions, quantum dots, superconducting circuits, and spin systems,
to electron and nuclear magnetic resonance applications in physics, chemistry, biology, and medicine.
A Mathematica package \cite{GG14} and a mobile application 
software \cite{ipad_app} are made available so that readers may apply the \drops mapping to new areas 
and interactively explore the 
\drops visualization of coupled spin dynamics in real time.

\begin{acknowledgments}
The authors acknowledge support from the Deutsche Forschungsgemeinschaft (DFG) via the grant
{GL}~203/7-1 and the SFB 631 as well as from the EU programmes QUAINT and SIQS.
A.G.\ was supported by NSERC (Canada).
R.Z.\ was also funded by the DFG through the grant 
{SCHU}~1374/2-1.
\end{acknowledgments}

\appendix

\section{Motivation for  the choice of signs in the \lisa basis\label{App_Motivation}}
The phase of the irreducible tensor operators
for the \lisa basis is fixed by the Condon-Shortley convention up to a sign.
The specific choice of signs made in the  \lisa basis
will be presented in Appendix~\ref{App_details_details}, 
 during the construction of the \lisa basis via projectors (see, e.g.,
\tab{sign_and_phase}).
Here, we present desirable properties which motivate this specific choice of signs:

The droplet of the identity operator 
has a positive value (i.e.\ it is red in Fig.~\ref{cartesian}).
The
droplets of the
linear Cartesian operators $\II{k\eta}$ with $\eta\in\{x,y,z\}$ for each spin $k$
are consistent with its Bloch vector representation, i.e., 
the positive lobe in the \lisa 
visualization of $\II{k\eta}$ is pointing in the 
$\eta$-direction (see Fig.~\ref{cartesian}).

\begin{figure}[t]
\includegraphics[]{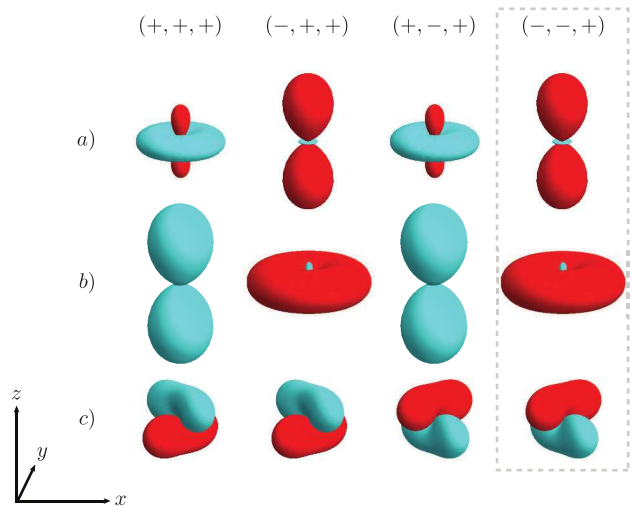}
\caption{(Color online) Different choices  $(a_0,a_1,a_2)$ for the signs of the bilinear tensors 
$\TTlb{0}{k,l}$, $\TTlb{1}{k,l}$, and $\TTlb{2}{k,l}$ with rank $0$, $1$, and $2$
result in different visualizations for the operators $2I_{kz}I_{lz}$ in~a),  
$2I_{kx}I_{lx}{+}2I_{ky}I_{ly}$ in~b), and $2I_{kx}I_{ly}$ in~c). The red and the cyan color refers to
the positive and the negative value of the \droplet function $f^{\{k,l\}}(A)$, respectively.
The signs for the \lisa basis are given in the last column
which is highlighted by a dashed rectangle. \label{different signs}} 
\end{figure}

Characteristic coupling Hamiltonians, such as longitudinal 
and planar 
couplings have elongated 
and disc-shaped droplets (see last column of \fig{different signs}).
For spins $k$ and $l$,  different shapes appear if $\TT^{\{k,l\}}_0$ and $\TT^{\{k,l\}}_2$
have different relative signs  (\fig{different signs}). 
If the sign of $\TTlb{1}{k,l}$ is negative,
the positive lobe of the 
\droplet representing $2\II{k\eta_1}\II{l\eta_2}$ with $\eta_1 \neq \eta_2$ 
(refer to case c) in \fig{different signs}) is displaced in the
$\vec\eta_3$-direction with respect to the center of the \droplet, where
 $\vec\eta_3=\vec\eta_1\times\vec\eta_2$ is given by the right hand rule.
The \droplet for the 
fully symmetric Cartesian tensor $4\II{1\eta}\II{2\eta}\II{3\eta}$ has an elongated shape
and its positive lobe  points in the $\eta$ direction 
(see Fig.~\ref{cartesian}).

\section{Construction of the \lisa basis via projectors\label{App_construction}} 
We provide here the details for the iterative construction
of the \lisa basis which
is directly applicable to general spin systems with 
$n\leq 5$ spins. We determine 
the basis elements by applying explicit projection operators. The presentation
aims at providing all necessary details for replicating our approach. 
To this end, we specify also all applied conventions which regrettably vary significantly in the literature.

\subsection{The symmetric group and the standard Young tableaux}
We start by recalling some basic notation,
see, e.g., \cite{Boerner67,Hamermesh62,Pauncz95,Sagan01}. 
The $g!$ permutations of a 
set $G=\{1,\ldots,g\}$ of cardinality $g$ 
are known as the symmetric group $S_g$.
They form a finite group whose group multiplication
is defined by the
composition $(\sigma_2\sigma_1)(p):=(\sigma_2 \circ \sigma_1)(p)=\sigma_2(\sigma_1(p))$
for $\sigma_1,\sigma_2\in S_g$ and $p\in G$.
An element
$\sigma\in S_g$ maps $p\in G$ to $\sigma(p)\in G$ such
that  $\sigma(p_1)\neq \sigma(p_2)$ for $p_1\neq p_2$.
Permutations can be compactly specified 
as products of disjoint cycles.
A cycle $c:=(c_1\cdots c_{\abs{c}} )\equiv(c_1,\ldots,c_{\abs{c}})$ 
of length
$\abs{c}$ (where $c_p\in G$ and $c_p\neq c_q$ for $p\neq q$)
represents a permutation $\sigma\in S_g$ where $\sigma(c_p)=c_{p{+}1}$ for 
$p<\abs{c}$ and $\sigma(c_{\abs{c}})=c_1$.
Two cycles $c$ and $\tilde{c}$ are disjoint
if $c_{p}\neq \tilde{c}_{q}$ for all $p$ and $q$.
A transposition $(p,q)$ with $p\neq q$ is a cycle of length two
which permutes $p$ and $q$.
Note that the symmetric group acts naturally on the set of 
Cartesian product operators (for notation refer to Sec.~\ref{Sec_cartesian})
by permuting particles (and their labels), e.g., $\sigma(\II{1x}\II{2y}\II{3z}) 
= \II{\sigma(1)x}\II{\sigma(2)y}\II{\sigma(3)z}
= \II{3x}\II{1y}\II{2z} = \II{1y}\II{2z}\II{3x}$ for
$\sigma=(132)$. 
This is equivalent to setting $\sigma(v_{1}\otimes \cdots \otimes v_{g}):=
v_{\sigma^{-1}(1)}\otimes \cdots \otimes v_{\sigma^{-1}(g)}$ for $g$ spins with
$v_p \in \{\II{x},\II{y},\II{z}\}$.

Recall that
a standard Young tableau
of size $g$ 
is a left-aligned arrangement of $g$ boxes where the number of boxes
does not increase from one row to following ones and 
where each box contains a different number from the set $G$
such that the numbers are ordered strictly
increasing from left to right and top to bottom.
For every standard Young tableau $\tau$, one introduces
a partition (i.e.\ its shape) 
$\lambda=\lambda(\tau)=[\lambda_1,\ldots,\lambda_{\kappa(\lambda)}]$
of length $\kappa(\lambda)$ where the positive integers $\lambda_p$
with $\lambda_p \geq \lambda_{p+1}$ are equal to the number of boxes
in row $p$ of $\tau$ and $\kappa(\lambda)$ agrees with the number of rows of $\tau$.
The number of columns of $\tau$ is equal to $\lambda_1$.
Also, the filling pattern $w(\tau)=[w(\tau)_1,\ldots,w(\tau)_g]$ of $\tau$ consists of the entries of $\tau$ 
which are strung together from left to right in each row and
from top to bottom for all rows. It is convenient to introduce a total order on the set of 
standard Young tableaux of 
size $g$
where $\tau<\tilde{\tau}$ if either $\lambda(\tau)>\lambda(\tilde{\tau})$ or if $\lambda(\tau)
=\lambda(\tilde{\tau})$ and $w(\tau)<w(\tilde{\tau})$. Here, we imply that
$a=[a_1,\ldots,a_{\abs{a}}] < b=[b_1,\ldots,b_{\abs{b}}]$ if
there exists an index $q$ such that $a_p=b_p$ for $p<q$ 
and $a_q<b_q$ while setting $a_p:=0$ for $p>\abs{a}$ and $b_p:=0$ for $p>\abs{b}$.
This total order for standard Young tableaux 
is reflected by the subscript $p$ in the notation $\taun pg$
introduced in Eq.~\eqref{taus}, e.g.,
\begin{equation}
\label{ordered_tau}
\taun{1}{3}=\TTab{1 2 3}\, <\, \taun{2}{3}=\TTab{1 2, 3}\, <\, \taun{3}{3}=
\TTab{1 3, 2}\, <\, \taun{4}{3}=\TTab{1, 2, 3}.
\end{equation}

\subsection{Young symmetrizers and projectors}
We detail now the construction for Young symmetrizers and  the corresponding projectors
with symmetry type $\tau$
\cite{Boerner67,Hamermesh62,James78,JK81,Sagan01,GW09,CSST10,
Corio66,SancTemMNMRXIII,tung1985group}.
These objects are elements of the group algebra $\R[S_g]$ of $S_g$
which consists of all (formal) real linear combinations of permutations 
$\sigma\in S_g$.
This means that every element $x\in \R[S_g]$
can be decomposed as $x=\sum_{\sigma\in S_g}x_{\sigma}\sigma $ 
with coefficients $x_{\sigma}\in\R$. Given two elements $ x, y\in \R[S_g]$, the sum
in  $\R[S_g]$ is naturally defined as $ x+ y = 
\sum_{\sigma \in S_g} (x_\sigma+y_\sigma)\sigma$ and the product is given by $ x y = 
\sum_{\sigma,\tilde{\sigma} \in S_g} (x_\sigma y_{\tilde{\sigma}})(\sigma \tilde{\sigma})$.

Given a standard Young tableau $\tau$, let $R(\tau,p)$ [resp.\ $C(\tau,q)$]
denote the set of all entries in the $p$th row [resp.\ $q$th column] of $\tau$.
Let $S_M$ denote the permutations of a set $M$
and let us introduce the permutations $S_{R(\tau,p)}$
of the elements $R(\tau,p)$ in the $p$th row of the standard Young tableau $\tau$
where $1\leq p \leq \kappa(\tau)$. In the example
of $\tau=\taun{3}{3}$, one obtains $R(\tau,1)=\{1,3\}$ and $R(\tau,2)=\{2\}$
as well as $S_{R(\tau,1)}=\{e,(1,3)\}$ and $S_{R(\tau,2)}=\{e\}$
where $e$ denotes the identity permutation. The set 
$R(\tau)$ of row-wise permutations is now defined as 
$R(\tau)=S_{R(\tau,1)} \times \cdots
\times S_{R(\tau,{\kappa(\tau)})}:=\{\prod^{\kappa(\tau)}_{p=1} \sigma_{p}\, 
\text{ for all possibilities of }\, \sigma_{p}\in S_{R(\tau,p)}\}$. Note that the order 
in the product is irrelevant as the different permutations act on non-overlapping subsets
of $G=\{1,\ldots,g\}$. For $\tau=\taun{3}{3}$, we have $R(\tau)=\{e,(1,3)\}$.
Similarly, we obtain the set of column-wise permutations
$C(\tau):=S_{C(\tau,1)} \times \cdots
\times S_{C(\tau,\lambda_1)}$ where $\lambda=\lambda(\tau)$. One introduces 
$H_\tau:=\sum_{\sigma \in R(\tau)} \sigma$ and 
$V_\tau:=\sum_{\sigma \in C(\tau)} (-1)^{\abs{\sigma}}\, \sigma$
where $\abs{\sigma}$ denotes the minimal number 
of transpositions $(p,q)$ necessary to write $\sigma$ as product thereof.
Finally, we can write the Young symmetrizer for $\tau$ as
the product $e_\tau:=f_\tau\, H_\tau V_\tau$ where the normalization factor $f_\tau \in \R$
is equal to the number of standard Young tableaux of shape $\tau$ divided by $g!$ which
ensures that $e_\tau e_\tau =e_\tau$.
For instance, the Young symmetrizers for the standard Young tableaux of \eq{ordered_tau} are
\begin{subequations}
\label{symmetrizer}
\begin{align}
e^{[3]}_1 &= e_{\tau_1^{[3]}} {=} \tfrac{1}{6} [e{+}(12){+}(13){+}(23){+}(123){+}(132)], \\[-1.5mm]
e^{[3]}_2 &= e_{\tau_2^{[3]}} {=} \tfrac{1}{3} [e{+}(12){-}(13){-}(132)], \\[-1.5mm]
e^{[3]}_3 &= e_{\tau_3^{[3]}} {=} \tfrac{1}{3} [e{-}(12){+}(13){-}(123)], \label{Young_ex} \\[-1.5mm]
e^{[3]}_4 &= e_{\tau_4^{[3]}} {=} \tfrac{1}{6} [e{-}(12){-}(13){-}(23){+}(123){+}(132)], 
\end{align}
\end{subequations}
where $e$ denotes the identity in $S_g$ and
one directly verifies that $e^{[3]}_{1}+e^{[3]}_{2}+e^{[3]}_{3}+e^{[3]}_{4}=e$.
The Young symmetrizer can be computed (e.g.) for \eq{Young_ex} by multiplying
$f_\tau = 1/3$, $H_{\tau} =e+(13)$, and $V_{\tau} =e-(12)$ where $\tau = \taun33$.

We determine now projectors $P^{[3]}_p$ (i.e.\ $P^{[3]}_p P^{[3]}_p= P^{[3]}_p$) 
which can be interpreted as orthogonalized versions of 
the operators $e^{[3]}_p$ in \eq{symmetrizer}. These projectors $P_\tau=P^{[3]}_p$ 
model the symmetries for a standard Young tableau $\tau=\taun{p}{3}$ and
will be used to identify
 tensor operators $\TT_j$ which are left invariant  under
their action, i.e.\ $P_\tau \TT_{jm}=\TT_{jm}$ for all $m\in\{-j,\ldots,j\}$.
Our approach for determining the projectors $P^{[3]}_p$ is similar to
the one detailed on \ppp{114--124} of \cite{James78} and the formulas can be inferred 
from the matrices of Young's seminormal or orthogonal representation
\cite{Boerner67,Hamermesh62,JK81,CSST10}.
But we stress that the employed formulas differ widely in the literature due to
varying conventions. Note that we obtain a basis of projection operators
which differs from the so-called seminormal basis (cf.\ \ppp{109--114} of \cite{JK81}).

Let us consider the ordered sequence
$\taun {r}g,\ldots,\taun {s}g$
of all standard Young tableaux of fixed shape $\lambda$.
In particular, one has $r=s=1$ for $\lambda=[3]$, $r=2$ and $s=3$ for $\lambda=[2,1]$,
as well as $r=s=4$ for $\lambda=[1,1,1]$ (cf.\ \eq{ordered_tau}).
One defines the projection operators  $P^{[g]}_{p}$
recursively using the operators $e^{[g]}_{p}$:
For $p=r$, one has $P^{[g]}_{p}:=e^{[g]}_{p}$.
It immediately follows that 
\begin{equation}
\label{proj_one}
P^{[3]}_1=e^{[3]}_1, P^{[3]}_2=e^{[3]}_2,  \text{ and } P^{[3]}_4=e^{[3]}_4.
\end{equation}
For  $r<p\leq s$, there exists $q\in\{r,\ldots,p{-}1\}$ such that $\taun {q}g$  differs from $\taun pg$ only
by the position of two boxes $\TTab{{a}}$ and $\TTab{b}$ with consecutive labels $a$ and $b:=a{+}1$.
Let $d\in\Z$ denote the signed axial distance from the box
$\TTab{{a}}$ to $\TTab{{b}}$ in $\taun qg$, i.e., the number of  steps 
from $\TTab{{a}}$ to $\TTab{{b}}$
while counting steps down or to the left positively and
steps up or to the right negatively. After these preparations, we set
\begin{equation}
\label{projector_new}
P^{[g]}_p := f \left[d\, (ab)+  e\right] P^{[g]}_{q}\in\R[S_g],
\end{equation}
where the scalar factor  $f\in\R$ is chosen such that $P^{[g]}_pP^{[g]}_p=P^{[g]}_p$.
For $\taun{3}{3}$, one obtains $q=2$, $a=2$, $b=3$, $d=2$, and \eq{projector_new}
implies that
\begin{align}
P^{[3]}_3 &= [2(23) + e]\, P^{[3]}_2 =  [2(23) + e]\, e^{[3]}_2 \nonumber \\ 
&=
[e{-}(12){-}(13){+}2(23){-}2(123){+}(132)]/3. \label{proj_two}
\end{align}

\subsection{Details of the iterative construction\label{App_details_details}}
The iterative construction of  the \lisa basis 
is now described and exemplified for the example of $n=3$ spins.
In the Sec.~\ref{construction}, this construction was divided into three steps:
(I) We start by building and symmetrizing $g$-linear tensor operators
of a $g$-spin system for each $g\in\{1,\ldots,n\}$ which
consists in applying the Clebsch-Gordan decomposition and the  just-introduced 
projection operators.
(II) Next, the tensor operators will be phase and sign corrected.
(III) Lastly, the tensor operators are embedded into the 
full $n$-spin system for each $g$-element subset of $\{1,\ldots,n\}$.

In step (I), we begin by specifying the form of the $g$-linear tensor operators for
a $g$-spin system in the particular simple cases of $g\in\{0,1\}$.
The symmetries with respect to particle permutations
are trivial in both of these cases. The corresponding symmetry types are given by the 
standard Young tableaux $\taun 10$ and $\taun 11=\TTab{1}$ where the first one is empty
and the second one consists of a single box. 
For $g=0$, we only have the tensor operator $\TT_0(\taun 10):=\TT_0$ whose 
only tensor component is given by $\TT_{0,0} =
\left(\begin{smallmatrix}
1 & 0\\
0 & 1
\end{smallmatrix}\right)/\sqrt{2}$.
There is one single linear tensor operator $\TT_1(\taun 11):=\TT_{1}$
for $g=1$
whose components are $\TT_{1,-1}=
\left(\begin{smallmatrix}
0 & 0\\
1 & 0
\end{smallmatrix}\right)$, 
$\TT_{1,0}=
\left(\begin{smallmatrix}
1 & 0\\
0 & -1
\end{smallmatrix}\right)/\sqrt{2}$, and
$\TT_{1,1}=
\left(\begin{smallmatrix}
0 & -1\\
0 & 0
\end{smallmatrix}\right)$.

For $g\geq 2$, the tensor operators $\tsymF{j}{p}{g-1}=\TT^{[g{-}1]}_j$ for $g{-}1$ spins
are combined 
with the tensor operator
$\TT_1$ for one spin in order to iteratively build up the tensor operators 
$\TT^{[g]}_{p}$ with $p\in\{|j{-}1|,\ldots,j{+}1\}$ for $g$ spins using the Clebsch-Gordan decomposition 
$\TT^{[g{-}1]}_j \otimes\TT  _1=\TT^{[g]}_{|j{-}1|}\oplus\cdots\oplus \TT^{[g]}_{j{+}1}$. 
The corresponding tensor components $\TT^{[g]}_{j,m}$ with $m\in\{-j,\ldots,j\}$
are determined by the Clebsch-Gordan coefficients \cite{Wigner59,BL81,Zare88}.
The three bilinear tensor operators $\TT^{[2]}_0$, $\TT^{[2]}_1$, and $\TT^{[2]}_2$
for $g=2$ are obtained via $\TT_1 \otimes \TT_1 = {\TT^{[2]}_0} \oplus {\TT^{[2]}_1} \oplus {\TT^{[2]}_2}$
and their components are given by (see, e.g., \pp{419} in
\cite{PDG12})
\begin{align*}
\TT^{[2]}_{0,0} & = \tfrac{1}{\sqrt{3}} [\TT_{1,-1} {\otimes} \TT_{1,1} - \TT_{1,0} {\otimes} \TT_{1,0 } + 
\TT_{1,1} {\otimes} \TT_{1,-1}],\\
\TT^{[2]}_{1,-1} & = \tfrac{1}{\sqrt{2}} [- \TT_{1,-1} {\otimes} \TT_{1,0} + \TT_{1,0} {\otimes} \TT_{1,-1} ],\\
\TT^{[2]}_{1,0} & = \tfrac{1}{\sqrt{2}} [- \TT_{1,-1} {\otimes} \TT_{1,1} + \TT_{1,1} {\otimes} \TT_{1,-1} ],\\
\TT^{[2]}_{1,1} & = \tfrac{1}{\sqrt{2}} [- \TT_{1,0} {\otimes} \TT_{1,1} + \TT_{1,1} {\otimes} \TT_{1,0} ],\\
\TT^{[2]}_{2,-2} & = \TT_{1,-1} {\otimes} \TT_{1,-1},\\
\TT^{[2]}_{2,-1} & = \tfrac{1}{\sqrt{2}} [\TT_{1,-1} {\otimes} \TT_{1,0} + \TT_{1,0} {\otimes} \TT_{1,-1} ],\\
\TT^{[2]}_{2,0} & = \tfrac{1}{\sqrt{6}} [\TT_{1,-1} {\otimes} \TT_{1,1} + 2\, \TT_{1,0} {\otimes} \TT_{1,0 } +
 \TT_{1,1} {\otimes} \TT_{1,-1}],\\
\TT^{[2]}_{2,1} & = \tfrac{1}{\sqrt{2}} [\TT_{1,0} {\otimes} \TT_{1,1} + \TT_{1,1} {\otimes} \TT_{1,0} ],\\
\TT^{[2]}_{2,2} & = \TT_{1,1} {\otimes} \TT_{1,1}.
\end{align*}
The projectors for $g=2$ are
\begin{align*}
P^{[2]}_1 &= e^{[2]}_{1} =  e_{\tau_1^{[2]}} =[e+(12)]/{2}\,\text{ where }\, \taun{1}{2} := \TTab{1 2} \,\text{ and }\\
P^{[2]}_2 &= e^{[2]}_{2} =  e_{\tau_2^{[2]}} =[e-(12)]/{2}\,\text{ where }\, \taun{2}{2} := \TTab{1,2} \,.
\end{align*}
It is obvious that $P^{[2]}_1 \TT^{[2]}_{0,0} = \TT^{[2]}_{0,0}$ and $P^{[2]}_1 \TT^{[2]}_{2,m} = \TT^{[2]}_{2,m}$
for $m \in \{-2,\ldots,2\}$. Moreover, $P^{[2]}_2 \TT^{[2]}_{1,m} = \TT^{[2]}_{1,m}$ for $m \in \{-1,0,1\}$.
Thus, both $\TT^{[2]}_0$ and $\TT^{[2]}_2$ have symmetry type $\taun{1}{2}$
and  $\TT^{[2]}_1$ has symmetry type $\taun{2}{2}$
(cf.\ Table~\ref{tab_klambda}). We set 
\begin{equation*}
\TT  _0(\taun 12):=\TT^{[2]}_0,\,
\TT  _1(\taun 22):=\TT^{[2]}_1,\,
\TT  _2(\taun 12):=\TT^{[2]}_2,
\end{equation*}
but use both variants synonymously.

\begin{table*}[t]
\begin{tabular}{@{\hspace{1mm}}l@{\hspace{1mm}}|@{\hspace{1mm}}r@{\hspace{4mm}}r@{\hspace{4mm}}r@{\hspace{1mm}}r@{\hspace{1mm}}r@{\hspace{4mm}}r@{\hspace{1mm}}r@{\hspace{1mm}}r@{\hspace{1mm}}r@{\hspace{1mm}}r@{\hspace{1mm}}r@{\hspace{1mm}}r@{\hspace{1mm}}}
before & $\TT_0(\taun 00)$ &
$\TT_1(\taun 11)$ &
$-\TT_0(\taun 12)$ &
$-i \TT_1(\taun 22)$ &
$\TT_2(\taun 12)$ &
$i\TT_0(\taun 43)$ &
$-\TT_1(\taun 13)$ &
$\TT_1(\taun 23)$ &
$\TT_1(\taun 33)$ &
$i \TT_2(\taun 23)$ &
$i \TT_2(\taun 33)$ &
$\TT_3(\taun 13)$\\[0.5mm]\hline\\[-2.75mm]
after & $\TT_0(\taun 00)$ &
$\TT_1(\taun 11)$ &
$\TT_0(\taun 12)$ &
$\TT_1(\taun 22)$ &
$\TT_2(\taun 12)$ &
$\TT_0(\taun 43)$ &
$\TT_1(\taun 13)$ &
$\TT_1(\taun 23)$ &
$\TT_1(\taun 33)$ &
$\TT_2(\taun 23)$ &
$\TT_2(\taun 33)$ &
$\TT_3(\taun 13)$
\end{tabular}
\caption{Transformation in step (II) to ensure the correct phases and signs.\label{sign_and_phase}}
\end{table*}

For  $g=3$, the Clebsch-Gordan decomposition results in
various trilinear tensors and the multiplicities of their ranks 
agree with Tables~\ref{tab_one} and \ref{tab_klambda}:
\begin{subequations}
\begin{align}
\TT_0(\taun 12)\otimes\TT^{[1]}_1 &= \TT^{[2]}_0\otimes\TT^{[1]}_1 =  \TT_1' \label{GC1}\\
\TT_1(\taun 22)\otimes\TT^{[1]}_1 &= \TT^{[2]}_1\otimes\TT^{[1]}_1 = \TT_0'' \oplus 
\TT_1'' \oplus \TT_2'' \label{GC2}\\
\TT_2(\taun 12)\otimes\TT^{[1]}_1 &= \TT^{[2]}_2\otimes\TT^{[1]}_1 = \TT_1''' \oplus \TT_2''' \oplus \TT_3'''.
\label{GC3}
\end{align}
\end{subequations}
Referring again to \pp{419} in \cite{PDG12},
the corresponding components for \eq{GC1} are  given by
\begin{align*}
\TT_{1,m}' &= \TT^{[2]}_{0,0} {\otimes}\TT^{[1]}_{1,m} \; \text{ for } \; m\in\{-1,0,1\},\\
\intertext{and for \eq{GC2} one obtains}
 \TT_{0,0}'' &= \tfrac{1}{\sqrt{3}} [\TT^{[2]}_{1,-1} {\otimes}\TT^{[1]}_{1,1} - \TT^{[2]}_{1,0} {\otimes}\TT^{[1]}_{1,0} + 
 \TT^{[2]}_{1,1} {\otimes}\TT^{[1]}_{1,-1}]\\ 
\TT_{1,-1}'' &= \tfrac{1}{\sqrt{2}} [-\TT^{[2]}_{1,-1} {\otimes}\TT^{[1]}_{1,0} + \TT^{[2]}_{1,0} {\otimes}\TT^{[1]}_{1,-1}]\\ 
\TT_{1,0}'' &= \tfrac{1}{\sqrt{2}} [-\TT^{[2]}_{1,-1} {\otimes}\TT^{[1]}_{1,1} + \TT^{[2]}_{1,1} {\otimes}\TT^{[1]}_{1,-1}]\\ 
\TT_{1,1}'' &= \tfrac{1}{\sqrt{2}} [-\TT^{[2]}_{1,0} {\otimes}\TT^{[1]}_{1,1} + \TT^{[2]}_{1,1} {\otimes}\TT^{[1]}_{1,0}]\\ 
\TT_{2,-2}'' &= \TT^{[2]}_{1,-1} {\otimes}\TT^{[1]}_{1,-1}\\
\TT_{2,-1}'' &= \tfrac{1}{\sqrt{2}} [\TT^{[2]}_{1,-1} {\otimes}\TT^{[1]}_{1,0} + \TT^{[2]}_{1,0} {\otimes}\TT^{[1]}_{1,-1}]\\
\TT_{2,0}'' &= \tfrac{1}{\sqrt{6}} [\TT^{[2]}_{1,-1} {\otimes}\TT^{[1]}_{1,1} + 2\, \TT^{[2]}_{1,0} {\otimes}\TT^{[1]}_{1,0} + 
\TT^{[2]}_{1,1} {\otimes}\TT^{[1]}_{1,-1}]\\ 
\TT_{2,1}'' &= \tfrac{1}{\sqrt{2}} [\TT^{[2]}_{1,0} {\otimes}\TT^{[1]}_{1,1} + \TT^{[2]}_{1,1} {\otimes}\TT^{[1]}_{1,0}]\\
\TT_{2,2}'' &= \TT^{[2]}_{1,1} {\otimes}\TT^{[1]}_{1,1}.\\
\intertext{The case of \eq{GC3} results in}
\TT_{1,-1}''' &= \tfrac{1}{\sqrt{10}} [\sqrt{6} \TT^{[2]}_{2,-2} {\otimes}\TT^{[1]}_{1,1} 
{-} \sqrt{3} \TT^{[2]}_{2,-1} {\otimes}\TT^{[1]}_{1,0}\\ 
&\phantom{= \tfrac{1}{\sqrt{10}} [\hspace{1mm}}{+} \TT^{[2]}_{2,0} {\otimes}\TT^{[1]}_{1,-1}]\\ 
\TT_{1,0}''' &= \tfrac{1}{\sqrt{10}} [\sqrt{3} \TT^{[2]}_{2,-1} {\otimes}\TT^{[1]}_{1,1} 
{-} 2 \TT^{[2]}_{2,0} {\otimes}\TT^{[1]}_{1,0}\\
&\phantom{= \tfrac{1}{\sqrt{10}} [\hspace{1mm}}{+}\sqrt{3} \TT^{[2]}_{2,1} {\otimes}\TT^{[1]}_{1,-1}]\\ 
\TT_{1,1}''' &= \tfrac{1}{\sqrt{10}} [\TT^{[2]}_{2,0} {\otimes}\TT^{[1]}_{1,1} {-} \sqrt{3}\, 
\TT^{[2]}_{2,1} {\otimes}\TT^{[1]}_{1,0} 
{+}\sqrt{6} \TT^{[2]}_{2,2} {\otimes}\TT^{[1]}_{1,-1}]
\intertext{for $j=1$,}
\TT_{2,-2}''' &= \tfrac{1}{\sqrt{3}} [-\sqrt{2} \TT^{[2]}_{2,-2} {\otimes}\TT^{[1]}_{1,0} 
{+} \TT^{[2]}_{2,-1} {\otimes}\TT^{[1]}_{1,-1}]\\
\TT_{2,-1}''' &= \tfrac{1}{\sqrt{6}} [-\sqrt{2} \TT^{[2]}_{2,-2} {\otimes}\TT^{[1]}_{1,1} 
{-} \TT^{[2]}_{2,-1} {\otimes}\TT^{[1]}_{1,0} \\
&\phantom{= \tfrac{1}{\sqrt{6}} [\hspace{1mm}}{+} \sqrt{3} \TT^{[2]}_{2,0} {\otimes}\TT^{[1]}_{1,-1}]\\
\TT_{2,0}''' &= \tfrac{1}{\sqrt{2}} [- \TT^{[2]}_{2,-1} {\otimes}\TT^{[1]}_{1,1} 
{+} \TT^{[2]}_{2,1} {\otimes}\TT^{[1]}_{1,-1}]\\
\TT_{2,1}''' &= \tfrac{1}{\sqrt{6}} [-\sqrt{3} \TT^{[2]}_{2,0} {\otimes}\TT^{[1]}_{1,1} 
{+} \TT^{[2]}_{2,1} {\otimes}\TT^{[1]}_{1,0} 
{+}\sqrt{2} \TT^{[2]}_{2,2} {\otimes}\TT^{[1]}_{1,-1}]\\
\TT_{2,2}''' &= \tfrac{1}{\sqrt{3}} [- \TT^{[2]}_{2,1} {\otimes}\TT^{[1]}_{1,1} 
{+} \sqrt{2} \TT^{[2]}_{2,2} {\otimes}\TT^{[1]}_{1,0}]
\intertext{for $j=2$, and}
\TT_{3,-3}''' &= \TT^{[2]}_{2,-2} {\otimes}\TT^{[1]}_{1,-1}\\
\TT_{3,-2}''' &= \tfrac{1}{\sqrt{3}} [ \TT^{[2]}_{2,-2} {\otimes}\TT^{[1]}_{1,0} 
{+} \sqrt{2} \TT^{[2]}_{2,-1} {\otimes}\TT^{[1]}_{1,-1}]\\
\TT_{3,-1}''' &= \tfrac{1}{\sqrt{15}} [ \TT^{[2]}_{2,-2} {\otimes}\TT^{[1]}_{1,1} 
{+} \sqrt{8} \TT^{[2]}_{2,-1} {\otimes}\TT^{[1]}_{1,0}\\
&\phantom{= \tfrac{1}{\sqrt{15}} [\hspace{1mm}}{+} \sqrt{6} \TT^{[2]}_{2,0} {\otimes}\TT^{[1]}_{1,-1}]\\
\TT_{3,0}''' &=\tfrac{1}{\sqrt{5}} [ \TT^{[2]}_{2,-1} {\otimes}\TT^{[1]}_{1,1} 
{+} \sqrt{3} \TT^{[2]}_{2,0} {\otimes}\TT^{[1]}_{1,0} 
{+} \TT^{[2]}_{2,1} {\otimes}\TT^{[1]}_{1,-1}]\\
\TT_{3,1}''' &= \tfrac{1}{\sqrt{15}} [\sqrt{6} \TT^{[2]}_{2,0} {\otimes}\TT^{[1]}_{1,1} 
{+} \sqrt{8} \TT^{[2]}_{2,1} {\otimes}\TT^{[1]}_{1,0} 
{+} \TT^{[2]}_{2,2} {\otimes}\TT^{[1]}_{1,-1}]\\
\TT_{3,2}''' &= \tfrac{1}{\sqrt{3}} [\sqrt{2} \TT^{[2]}_{2,1} {\otimes}\TT^{[1]}_{1,1} 
{+}  \TT^{[2]}_{2,2} {\otimes}\TT^{[1]}_{1,0}]\\
\TT_{3,3}''' &= \TT^{[2]}_{2,2} {\otimes}\TT^{[1]}_{1,1}\quad
\text{for the rank of $j=3$.}
\end{align*}

We apply now the projectors
of \eqs{proj_one} and \eqref{proj_two} in order
to obtain the permutation-symmetrized versions of the tensor operators.
These permutation-symmetrized versions can be obtained by tedious but straightforward linear algebra.
For example, the zero-rank tensor operator $\TT_{0,0}''$ is unchanged by the action of $P^{[3]}_{4}$, i.e.\
$\TT_{0,0}''    = P^{[3]}_{4} \TT_{0,0}''$. It also holds that 
$P^{[3]}_{1} \TT_{0,0}'' = P^{[3]}_{2} \TT_{0,0}'' = P^{[3]}_{3} \TT_{0,0}''=0$.
Consequently, $\TT_{0,0}''$ is completely antisymmetric which also
follows from the expansion
\begin{align*}
\sqrt{6} \TT_{0,0}'' =
& + 
 \TT^{[1]}_{1,-1} {\otimes} \TT^{[1]}_{1,1} {\otimes} \TT^{[1]}_{1,0}
+ \TT^{[1]}_{1,0} {\otimes} \TT^{[1]}_{1,-1} {\otimes} \TT^{[1]}_{1,1}\phantom{.}\\
&+ \TT^{[1]}_{1,1} {\otimes} \TT^{[1]}_{1,0} {\otimes} \TT^{[1]}_{1,-1}
- \TT^{[1]}_{1,-1} {\otimes} \TT^{[1]}_{1,0} {\otimes} \TT^{[1]}_{1,1}\phantom{.}\\
&- \TT^{[1]}_{1,0} {\otimes} \TT^{[1]}_{1,1} {\otimes} \TT^{[1]}_{1,-1}
- \TT^{[1]}_{1,1} {\otimes} \TT^{[1]}_{1,-1} {\otimes} \TT^{[1]}_{1,0}.
\end{align*}
Thus, the  permutation-symmetrized version of $\TT_{0,0}''$ is
\begin{align}
\TT_{0,0}(\taun 43) &:=  \TT_{0,0}'' \quad (= P^{[3]}_{4} \TT_{0,0}''). \nonumber \\
\intertext{These computations can be in general quite unwieldy but are easily automated.
The permutation-symmetrized tensor components of rank one are computed as (where $m\in\{-1,0,1\}$)}
\TT_{1,m}(\taun 13) &:= \tfrac{3}{\sqrt{5}} \TT_{1,m}' + \tfrac{6}{5} \TT_{1,m}''' 
\quad (=  \tfrac{3}{\sqrt{5}} P^{[3]}_{1} \TT_{1,m}') \label{nontriv_a}\\
\TT_{1,m}(\taun 23) &:= \tfrac{3}{2} \TT_{1,m}' - \tfrac{3\sqrt{5}}{4} \TT_{1,m}''' 
\quad (= \tfrac{3}{2} P^{[3]}_{2} \TT_{1,m}') \label{nontriv_b}\\
\TT_{1,m}(\taun 33) &:= \TT_{1,m}'' 
\quad (=  \tfrac{\sqrt{3}}{2} P^{[3]}_{3} \TT_{1,m}'). \nonumber  \\
\intertext{The  rank-two case results in (where $m\in \{-2,\ldots,2\}$)}
\TT_{2,m}(\taun 23) &:= \TT_{2,m}''' 
\quad (= \sqrt{3} P^{[3]}_{2} \TT_{2,m}'')  \nonumber \\
\TT_{2,m}(\taun 33) &:= \TT_{2,m}'' 
\quad (= P^{[3]}_{3} \TT_{2,m}''), \nonumber \\
\intertext{and one obtains for the case of rank three that}
\TT_{3,m}(\taun 13) &:=  \TT_{3,m}'''    
\quad (= P^{[3]}_{1} \TT_{3,m}'''), \nonumber
\end{align}
where $m\in\{-3,\ldots,3\}$.
All tensor components $\TT_{j,m}(\tau)$ have been normalized 
such that $\Tr[\TT^{\phantom{\dagger}}_{j,m}(\tau)\,  \TT_{j,m}^\dagger(\tau)]=1$, where
$\Tr(A) =\sum_{p} A_{pp}$ denotes the trace of a matrix $A$.
Note that non-trivial recombinations occur for trilinear tensor operators only in 
\eqs{nontriv_a} and~\eqref{nontriv_b}.

In step (II), we apply the transformations of \tab{sign_and_phase}.
This ensures that the phases and signs of the tensor operators are set 
according to the  conventions discussed and motivated before.
We emphasize that the transformations of \tab{sign_and_phase}
lead only to the correct form of the tensor operators if one has executed step (I) 
exactly as described.
By abuse of notation, the tensor operators after the transformation
are denoted by the same symbol $\TT_j(\tau)$.
In the following, we assume that the tensor operators $\TT_j(\tau)$ have been phase
and sign corrected.

Our construction is now completed by step (III) where each $g$-linear tensor operator
of a $g$-spin system is embedded into $n$-spin systems for $n\geq g$.
In particular, we detail the three-spin case with $n=3$. The zero-linear tensor component
$\TT_{0,0}(\taun 10)=\TT_{0,0}=
\left(\begin{smallmatrix}
1 & 0\\
0 & 1
\end{smallmatrix}\right)/\sqrt{2}$ (i.e.\ $g=0$) is embedded as
\begin{equation*}
\TT^{\emptyset}_0(\taun 10):= (\tfrac{1}{\sqrt{2}})^n \unity_{2^n}
= \TT_{0,0}^{\otimes n} = \TT_{0,0} \otimes \cdots \otimes \TT_{0,0},
\end{equation*}
where $\unity_q$ denotes the $q\times q$-dimensional identity matrix.
The embedding of linear tensor components $\TT_1(\taun 11)$ (i.e.\ $g=1$) 
is given for $n=3$ as follows  (where $m \in\{-1,0,1\}$)
\begin{align*}
\TT^{\{1\}}_{1,m}(\taun 11) &= \TT_{1,m}(\taun 11)\otimes  \TT_{0,0} \otimes  \TT_{0,0}\\
\TT^{\{2\}}_{1,m}(\taun 11) &= \TT_{0,0} \otimes\TT_{1,m}(\taun 11)  \otimes   \TT_{0,0}\\
\TT^{\{3\}}_{1,m}(\taun 11) &= \TT_{0,0}  \otimes   \TT_{0,0} \otimes\TT_{1,m}(\taun 11).
\end{align*}
More generally, one applies for $g\geq 1$ a transposition $(1k)$ to $\TT^{\{1\}}_{1,m}(\taun 11)$
which permutes the first
and the $k$th particle:
\begin{align*}
\TT^{\{1\}}_{1,m}(\taun 11)&:=\TT_{1,m}(\taun 11){\otimes}  \TT_{0,0}^{\otimes n-1}\\
&\phantom{:}= \TT_1(\taun 11){\otimes} \TT_{0,0} {\otimes} \cdots {\otimes} \TT_{0,0}\\
\TT^{\{k\}}_{1,m}(\taun 11)&:= (1k)\, \TT^{\{1\}}_{1,m}(\taun 11) \;\text{ for } k\neq 1.
\end{align*}
This technique can be easily generalized to bilinear tensor components
$\TT_{j,m}(\taun p2)$ with $m\in\{-j,\ldots,j\}$ (and beyond) which are embedded
into a system of $n\geq 2$ spins:
\begin{align*}
\TT^{\{1,2\}}_{j,m}(\taun p2)&:=\TT_{j,m}(\taun p2) \otimes  \TT_{0,0}^{\otimes n-2}\\
\TT^{\{1,l\}}_{j,m}(\taun p2)&:= (2l)\, \TT^{\{1,2\}}_{j,m}(\taun p2) \;\text{ for } l> 2\\
\TT^{\{k,l\}}_{j,m}(\taun p2)&:= (1k) (2l)\, \TT^{\{1,2\}}_{j,m}(\taun p2) \;\text{ for } l > k
> 1.
\end{align*}
Finally, trilinear tensor components are embedded into a three-spin system
by (where $m\in\{-j,\ldots,j\}$)
\begin{equation*}
\TT^{\{1,2,3\}}_{j,m}(\taun p3):=\TT_{j,m}(\taun p3).
\end{equation*}

\begin{table*}[tb]
\footnotesize
$
\begin{array}{r@{\hspace{1mm}}l}
\tttt{1,-1} {1}     =&  
\frac{2}{\sqrt{15}} \left[
3I_{xxx}
{-}3iI_{yyy} 
{-}i(I_{xxy}{+}I_{xyx}{+}I_{yxx})
{+} (I_{xyy}{+}I_{yxy}{+}I_{yyx})
{+}(I_{xzz}{+}I_{zxz}{+}I_{zzx})
{-}i(I_{yzz}{+}I_{zyz}{+}I_{zzy})\right]\\
\tttt{1,0} {1}      
=& {\frac{\sqrt{8}}{\sqrt{15}}} \left[(I_{xxz}{+}I_{xzx}{+}I_{zxx}){+}(I_{yyz}{+}I_{yzy}{+}I_{zyy}){+}3 I_{zzz}\right] \\
\tttt{1,1} {1}   =& {-}\frac{2}{\sqrt{15}} \left[
i(I_{xxy}{+}I_{xyx}{+}I_{yxx}) 
{+}(I_{xyy}{+}I_{yxy}{+}I_{yyx})
{+}(I_{xzz}{+}I_{zxz}{+}I_{zzx})
{+}i(I_{yzz}{+} I_{zyz}{+}I_{zzy})
{+}3I_{xxx}{+}3iI_{yyy}
\right] \\
\tttt{3,-3} {1}    =& \left[
(I_{xxx}{+}i I_{yyy}){-}i(I_{xxy}{+}I_{xyx}{+}I_{yxx}){-}(I_{xyy}{+}I_{yxy}{+}I_{yyx})
\right]\\
\tttt{3,-2} {1}    =& {\frac{\sqrt{2}}{\sqrt{3}}} \left[
(I_{xxz}{+} I_{xzx}{+} I_{zxx}){-}(I_{yyz}{+}I_{yzy}{+}I_{zyy})
{-}i(I_{xyz}{+}I_{xzy}{+}I_{yxz}{+}I_{yzx}{+}I_{zxy}{+}I_{zyx})\right] \\
\tttt{3,-1} {1}   =& \frac{1}{\sqrt{15}}\left[
-3(I_{xxx}{-} iI_{yyy})
{+}i(I_{xxy}{+}I_{xyx}{+}I_{yxx})
{-}(I_{xyy}{+}I_{yxy}{+}I_{yyx})
{+}4(I_{xzz}{+} I_{zxz}{+}I_{zzx})
{-}4 i(I_{yzz}{+} I_{zyz}{+} I_{zzy})
   \right] \\
\tttt{3,0} {1}     =& -\frac{2}{\sqrt{5}} \left[
(I_{xxz}{+}I_{xzx}{+}I_{zxx}){+}(I_{yyz}{+}I_{yzy}{+}I_{zyy}){-}2 I_{zzz}
\right] \\
\tttt{3,1} {1}     =& \frac{1}{\sqrt{15}}\left[ 
3(I_{xxx}{+}iI_{yyy})
{+}i(I_{xxy}{+}I_{xyx}{+}I_{yxx})
{+}(I_{xyy}{+}I_{yxy}{+}I_{yyx})
{-}4(I_{xzz}{+} I_{zxz}{+} I_{zzx})
{-}4i(I_{yzz}{+}I_{zyz}{+}I_{zzy})
   \right]\\
\tttt{3,2} {1}     =& {\frac{\sqrt{2}}{\sqrt{3}}} \left[
(I_{xxz}{+} I_{xzx}{+} I_{zxx}){-}(I_{yyz}{+}I_{yzy}{+}I_{zyy}){+}i(I_{xyz}
{+}I_{xzy}{+}I_{yxz}{+}I_{yzx}{+}I_{zxy}{+}I_{zyx})\right] \\
\tttt{3,3} {1}     =&  \left[
(-I_{xxx}{+}i I_{yyy}){-}i(I_{xxy}{+}I_{xyx}{+}I_{yxx}){+}(I_{xyy}{+}I_{yxy}{+}I_{yyx})\right]\\[1.5mm]
\tttt{1,-1} {2}  =& \frac{1}{\sqrt{3}}\left[-i(I_{yxx}{+}I_{xyx}{-}2I_{xxy}){-}i(I_{yzz}{+} I_{zyz}{-}2I_{zzy}){+}(I_{xyy}{+}
I_{yxy}{-}2 I_{yyx}){+}(I_{xzz}{+}I_{zxz}{-}2 I_{zzx})\right] \\
\tttt{1,0} {2}   =& {\frac{\sqrt{2}}{\sqrt{3}}} \left[-2(I_{xxz}{+}I_{yyz}){+}(I_{zxx}{+}I_{xzx}){+}(I_{zyy}{+}I_{yzy})\right] \\
\tttt{1,1} {2}   =& \frac{1}{\sqrt{3}}\left[(-I_{xyy}{-}I_{yxy}{+}2 I_{yyx}){+}(-I_{xzz}
{-}I_{zxz}{+}2 I_{zzx}){+}i(-I_{yxx}{-}I_{xyx}
{+}2I_{xxy}){+}i(-I_{yzz}{-} I_{zyz}{+}2 I_{zzy})\right] \\
\tttt{2,-2} {2} =& \frac{1}{\sqrt{3}}\left[(I_{yzx}{+}I_{zyx}){+}(I_{xzy}{+}I_{zxy}){-}2(I_{xyz}{+}I_{yxz}) {-}(2 i I_{xxz}
{-}i I_{xzx}{-}i I_{zxx}){+}(2 i I_{yyz}{-}i I_{yzy}{-}i I_{zyy})\right] \\
\tttt{2,-1} {2} =& \frac{1}{\sqrt{3}}\left[-(2 I_{xxy}{-}I_{xyx}{-}I_{yxx}){-}i(2 I_{yyx}{-}I_{yxy}{-}I_{xyy}){+}(2iI_{zzx}
{-}iI_{zxz}{-}iI_{xzz}){+}(2 I_{zzy}{-}I_{zyz}{-}I_{yzz})\right] \\
\tttt{2,0} {2}  =& \sqrt{2} \left[(I_{yzx}{+}I_{zyx}){-}(I_{xzy}{+}I_{zxy})\right] \\
\tttt{2,1} {2} =& \frac{1}{\sqrt{3}}\left[
(2 I_{xxy}{-}I_{xyx}{-}I_{yxx}){-}(2 I_{zzy}{-}I_{zyz}{-}I_{yzz}){+}i(2I_{zzx}
{-}I_{zxz}{-}I_{xzz}){-}i(2I_{yyx}{-}I_{yxy}{-}I_{xyy})
\right] \\
\tttt{2,2} {2}  =& \frac{1}{\sqrt{3}}\left[
-(2 I_{xyz}{-}I_{xzy}{-}I_{zxy}){-}(2 I_{yxz}{-}I_{yzx}{-}I_{zyx})
{+}i(2I_{xxz}{-}I_{xzx}{-}I_{zxx}){-}i(2I_{yyz}{-}I_{yzy}{-}I_{zyy})
\right] \\[1.5mm]
\tttt{1,-1} {3} =& \left[(I_{xyy}{-}I_{yxy}){+}(I_{xzz}{-}I_{zxz}){-}i(I_{yxx}{-}I_{xyx}){-}i(I_{yzz}{-} I_{zyz})\right] \\
\tttt{1,0}{3} =& \sqrt{2} \left[(I_{zxx}{-}I_{xzx}){+}(I_{zyy}{-}I_{yzy})\right] \\
\tttt{1,1} {3}  =& -[(I_{xyy}{-}I_{yxy}){+}(I_{xzz}{-}I_{zxz}){+}i(I_{yxx}{-} I_{xyx}){+}i (I_{yzz}{-} I_{zyz})] \\
\tttt{2,-2} {3} =& \left[(I_{zxy}{-}I_{xzy}){+}(I_{zyx}{-}I_{yzx}){+}i(I_{zxx}{-}I_{xzx}){+}i(I_{yzy}{-}I_{zyy})\right] \\
\tttt{2,-1} {3} =& \left[(I_{yxx}{-}I_{xyx}){+}i(I_{xyy}{-}I_{yxy}){+}i(I_{zxz}{-}I_{xzz}){+}(I_{zyz}{-}I_{yzz})\right]\\
\tttt{2,0} {3}  =& {\frac{\sqrt{2}}{\sqrt{3}}} \left[-(2 I_{xyz}{+}I_{xzy}{-}I_{zxy}){+} (2I_{yxz}{+}I_{yzx}{-}I_{zyx})\right] \\
\tttt{2,1} {3}  =&  \left[(I_{xyx}{-}I_{yxx}){+} (I_{yzz}{-} I_{zyz}){+}i(I_{xyy}{-}I_{yxy}){+}i(I_{zxz}{-}I_{xzz})\right] \\
\tttt{2,2} {3}  =& \left[
(I_{zxy}{-}I_{xzy}){+}(I_{zyx}{-}I_{yzx}){+}i(I_{xzx}{-}I_{zxx}){+}i(I_{zyy}{-}I_{yzy})\right] \\[1.5mm]
\tttt{0,0} {4} =&  \frac{2}{\sqrt{3}} \left[(I_{xyz}{-}I_{xzy}{-}I_{yxz}{+}I_{yzx}{+}I_{zxy}{-}I_{zyx})\right]
\end{array}
$
\caption{Decomposition of the trilinear \lisa tensor operators in terms of the 
Cartesian-product operators $\II{abc}=\II {1a}\II {2b}\II {3c}$. This form is valid for 
a three-spin system, for a general system with $n\geq 3$ spins
one has to multiply $\TT_{j,m}(\tau)$ with $(1/\sqrt{2})^{n-3}$.\label{Tril I->T}}
\end{table*}

\subsection{From the \lisa basis to Cartesian product operators\label{transformation}}
Before we close this section, our computations 
are summarized by providing explicit basis transformations from 
Cartesian product operators (for definitions refer to 
Sec.~\ref{Sec_cartesian})
to  the \lisa basis and vice versa.
The basis transformations for the linear 
tensor components
of an $n$-spin system are
\begin{align*}
\left(\begin{smallmatrix}
\TTlb {1,-1}{k}\\
\TTlb {1,0}k \\
\TTlb {1,1}k \\
\end{smallmatrix}\right) 
&= (\tfrac{1}{\sqrt{2}})^{n-1} 
\begin{pmatrix}
 \phantom{-}1 & -i & 0 \\
 \phantom{-}0 & \phantom{-}0 & {\sqrt{2}} \\
 -1 & -i & 0
\end{pmatrix}
\begin{pmatrix}
\II{kx}\\
\II{ky}\\
\II{kz}\\
\end{pmatrix}\\
\begin{pmatrix}
\II{kx}\\
\II{ky}\\
\II{kz}\\
\end{pmatrix}
&= (\sqrt{2})^{n-1}
\begin{pmatrix}
 {1}/{2} & 0 & -{1}/{2} \\
 {i}/{2} & 0 & \phantom{-}{i}/{2} \\
 0 & {1}/{\sqrt{2}} & \phantom{-}0
\end{pmatrix}
\left(\begin{smallmatrix}
\TTlb {1,-1}{k}\\
\TTlb {1,0}k \\
\TTlb {1,1}k \\
\end{smallmatrix}\right).
\end{align*}
Similarly, the transformations in the bilinear case are
\begin{gather*}
\left(\hspace{-0.75mm}\begin{smallmatrix}
\TTlb{0,0}{k,\ell }\\
\TTlb{1,-1}{k,\ell }\\
\TTlb{1,0}{k,\ell }\\
\TTlb{1,1}{k,\ell }\\
\TTlb{2,-2}{k,\ell }\\
\TTlb{2,-1}{k,\ell }\\
\TTlb{2,0}{k,\ell }\\
\TTlb{2,1}{k,\ell }\\
\TTlb{2,2}{k,\ell }\\
\end{smallmatrix}\hspace{-0.75mm}\right) = A_n \hspace{-0.75mm} \left(\hspace{-0.75mm}\begin{smallmatrix}
2\II{k x}\II{\ell x}\\
2\II{k x}\II{\ell y}\\
2\II{k x}\II{\ell z}\\
2\II{k y}\II{\ell x}\\
2\II{k y}\II{\ell y}\\
2\II{k y}\II{\ell z}\\
2\II{k z}\II{\ell x}\\
2\II{k z}\II{\ell y}\\
2\II{k z}\II{\ell z}\\
\end{smallmatrix}\hspace{-0.75mm}\right) \text{and}
\left(\hspace{-0.75mm}\begin{smallmatrix}
2\II{kx}\II{\ell x}\\
2\II{kx}\II{\ell y}\\
2\II{kx}\II{\ell z}\\
2\II{ky}\II{\ell x}\\
2\II{ky}\II{\ell y}\\
2\II{ky}\II{\ell z}\\
2\II{kz}\II{\ell x}\\
2\II{kz}\II{\ell y}\\
2\II{kz}\II{\ell z}\\
\end{smallmatrix}\hspace{-0.75mm}\right)
= B_n \hspace{-0.75mm}
\left(\hspace{-0.75mm}\begin{smallmatrix}
\TTlb{0,0}{k,\ell }\\
\TTlb{1,-1}{k,\ell }\\
\TTlb{1,0}{k,\ell }\\
\TTlb{1,1}{k,\ell }\\
\TTlb{2,-2}{k,\ell }\\
\TTlb{2,-1}{k,\ell }\\
\TTlb{2,0}{k,\ell }\\
\TTlb{2,1}{k,\ell }\\
\TTlb{2,2}{k,\ell }\\
\end{smallmatrix}\hspace{-0.75mm}\right),\\
\text{with }
A_n = 
(\tfrac{1}{\sqrt{2}})^{n{-}2}
\left(\hspace{-0.5mm}
\begin{smallmatrix}
\tfrac{1}{\sqrt{3}} & 0 & 0 & 0 & \tfrac{1}{\sqrt{3}} & 0 & 0 & 0 & \tfrac{1}{\sqrt{3}} \\
0 & 0 & \tfrac{i}{2} & 0 & 0 & \tfrac{1}{2} & \tfrac{-i}{2} & \tfrac{-1}{2} & 0 \\
0 & \tfrac{1}{\sqrt{2}} & 0 & \tfrac{-1}{\sqrt{2}} & 0 & 0 & 0 & 0 & 0 \\
0 & 0 & \tfrac{i}{2} & 0 & 0 & \tfrac{-1}{2} & \tfrac{-i}{2} & \tfrac{1}{2} & 0 \\
\tfrac{1}{2} & \tfrac{-i}{2} & 0 & \tfrac{-i}{2} & \tfrac{-1}{2} & 0 & 0 & 0 & 0 \\
0 & 0 & \tfrac{1}{2} & 0 & 0 & \tfrac{-i}{2} & \tfrac{1}{2} & \tfrac{-i}{2} & 0 \\
\tfrac{-1}{\sqrt{6}} & 0 & 0 & 0 & \tfrac{-1}{\sqrt{6}} & 0 & 0 & 0 & \tfrac{2}{\sqrt{6}} \\
0 & 0 & \tfrac{-1}{2} & 0 & 0 & \tfrac{-i}{2} & \tfrac{-1}{2} & \tfrac{-i}{2} & 0 \\
\tfrac{1}{2} & \tfrac{i}{2} & 0 & \tfrac{i}{2} & \tfrac{-1}{2} & 0 & 0 & 0 & 0
\end{smallmatrix}\hspace{-0.5mm}
\right)\\
\text{and } B_n =
(\sqrt{2})^{n{-}2}
\left(\hspace{-0.5mm}
\begin{smallmatrix}
\tfrac{1}{\sqrt{3}} & 0 & 0 & 0 & \tfrac{1}{2} & 0 & \tfrac{-1}{\sqrt{6}} & 0 & \tfrac{1}{2} \\
0 & 0 & \tfrac{1}{\sqrt{2}} & 0 & \tfrac{i}{2} & 0 & 0 & 0 & \tfrac{-i}{2} \\
0 & \tfrac{-i}{2} & 0 & \tfrac{-i}{2} & 0 & \tfrac{1}{2} & 0 & \tfrac{-1}{2} & 0 \\
0 & 0 & \tfrac{-1}{\sqrt{2}} & 0 & \tfrac{i}{2} & 0 & 0 & 0 & \tfrac{-i}{2} \\
\tfrac{1}{\sqrt{3}} & 0 & 0 & 0 & \tfrac{-1}{2} & 0 & \tfrac{-1}{\sqrt{6}} & 0 & \tfrac{-1}{2} \\
0 & \tfrac{1}{2} & 0 & \tfrac{-1}{2} & 0 & \tfrac{i}{2} & 0 & \tfrac{i}{2} & 0 \\
0 & \tfrac{i}{2} & 0 & \tfrac{i}{2} & 0 & \tfrac{1}{2} & 0 & \tfrac{-1}{2} & 0 \\
0 & \tfrac{-1}{2} & 0 & \tfrac{1}{2} & 0 & \tfrac{i}{2} & 0 & \tfrac{i}{2} & 0 \\
\tfrac{1}{\sqrt{3}} & 0 & 0 & 0 & 0 & 0 & \tfrac{\sqrt{2}}{\sqrt{3}} & 0 & 0
\end{smallmatrix}\hspace{-0.5mm}
\right).
\end{gather*}
The decomposition of trilinear \lisa tensor 
components in terms of the  trilinear Cartesian-product basis
is detailed in \tab{Tril I->T}, while the other direction can be found
in  \tab{Tril T->I}; observe the shorthand $\II{abc}:=\II {1a}\II {2b}\II {3c}$.

\begin{table*}[tb]
\footnotesize
$
\begin{array}{r@{\hspace{1mm}}l}
4I_{xxx} =& 
\frac{1}{10} [
2 \sqrt{15} [\TT_{1,-1}(\tau^{[3]}_1){-}\TT_{1,1}(\tau^{[3]}_1)]
{-}\sqrt{15} [\TT_{3,-1}(\tau^{[3]}_1){-}\TT_{3,1}(\tau^{[3]}_1)]
{+}5[\TT_{3,-3}(\tau^{[3]}_1){-}\TT_{3,3}(\tau^{[3]}_1)]
]\\
4I_{xxy} =& \frac{1}{30} [
2 i \sqrt{15} [\TT_{1,-1}(\tau^{[3]}_1){+} \TT_{1,1}(\tau^{[3]}_1)]
{-}i \sqrt{15} [\TT_{3,-1}(\tau^{[3]}_1){+}\TT_{3,1}(\tau^{[3]}_1)]
{+}15 i [\TT_{3,-3}(\tau^{[3]}_1){+}\TT_{3,3}(\tau^{[3]}_1)]\\&
-10 i \sqrt{3} [\TT_{1,-1}(\tau^{[3]}_2){+}\TT_{1,1}(\tau^{[3]}_2)]
{-}10 \sqrt{3} [\TT_{2,-1}(\tau^{[3]}_2){-}\TT_{2,1}(\tau^{[3]}_2)]
]\\
4I_{xxz} =& 
{\frac{\sqrt{2}}{\sqrt{15}}} \TT_{1,0}(\tau^{[3]}_1)
{-}\frac{1}{\sqrt{5}}\TT_{3,0}(\tau^{[3]}_1)
{+}\frac{1}{\sqrt{6}}[\TT_{3,-2}(\tau^{[3]}_1){+}\TT_{3,2}(\tau^{[3]}_1)]
{-}{\frac{\sqrt{2}}{\sqrt{3}}}\TT_{1,0}(\tau^{[3]}_2)
{+}\frac{i}{\sqrt3}[\TT_{2,-2}(\tau^{[3]}_2){-} \TT_{2,2}(\tau^{[3]}_2)]
\\
4I_{xyx} =& 
\frac{1}{30} [
2 i \sqrt{15} [\TT_{1,-1}(\tau^{[3]}_1){+}\TT_{1,1}(\tau^{[3]}_1)]
{-}i \sqrt{15} [\TT_{3,-1}(\tau^{[3]}_1){+}\TT_{3,1}(\tau^{[3]}_1)]
{+}15 i [\TT_{3,-3}(\tau^{[3]}_1){+}\TT_{3,3}(\tau^{[3]}_1)]\\&
{+}5 i \sqrt{3} [\TT_{1,-1}(\tau^{[3]}_2){+}\TT_{1,1}(\tau^{[3]}_2)]
{+}5 \sqrt{3} [\TT_{2,-1}(\tau^{[3]}_2){-}\TT_{2,1}(\tau^{[3]}_2)]
{-}15 i [\TT_{1,-1}(\tau^{[3]}_3){+}\TT_{1,1}(\tau^{[3]}_3)]
{-}15 [\TT_{2,-1}(\tau^{[3]}_3){-}\TT_{2,1}(\tau^{[3]}_3)]
]\\
4I_{xyy} =& \frac{1}{30} [
2 \sqrt{15} [\TT_{1,-1}(\tau^{[3]}_1){-}\TT_{1,1}(\tau^{[3]}_1)]
{-}\sqrt{15} [\TT_{3,-1}(\tau^{[3]}_1){-}\TT_{3,1}(\tau^{[3]}_1)]
{-}15 [\TT_{3,-3}(\tau^{[3]}_1){-}\TT_{3,3}(\tau^{[3]}_1)]\\&
{+}5 \sqrt{3} [\TT_{1,-1}(\tau^{[3]}_2){-}\TT_{1,1}(\tau^{[3]}_2)]
-5 i \sqrt{3} [\TT_{2,-1}(\tau^{[3]}_2){+}\TT_{2,1}(\tau^{[3]}_2)]
{+}15 [\TT_{1,-1}(\tau^{[3]}_3){-}\TT_{1,1}(\tau^{[3]}_3)]
{-}15 i[\TT_{2,-1}(\tau^{[3]}_3){+}\TT_{2,1}(\tau^{[3]}_3)]
]\\
4I_{xyz} =& \frac{1}{2\sqrt{3}}[
i \sqrt{2} [\TT_{3,-2}(\tau^{[3]}_1){-}\TT_{3,2}(\tau^{[3]}_1)]
{-}2 \sqrt{2} \TT_{2,0}(\tau^{[3]}_3)
{-}2 [\TT_{2,-2}(\tau^{[3]}_2){+}\TT_{2,2}(\tau^{[3]}_2)]
{+}2 \TT_{0,0}(\tau^{[3]}_4)]\\
4I_{xzx} =& \frac{1}{30} [
2 \sqrt{30} [\TT_{1,0}(\tau^{[3]}_1)]
{-}6 \sqrt{5} [\TT_{3,0}(\tau^{[3]}_1)]
{+}5 \sqrt{6}[\TT_{3,-2}(\tau^{[3]}_1){+}\TT_{3,2}(\tau^{[3]}_1)]\\&
{+}5 \sqrt{6} \TT_{1,0}(\tau^{[3]}_2)
{-}5 i \sqrt{3} [\TT_{2,-2}(\tau^{[3]}_2){-}\TT_{2,2}(\tau^{[3]}_2)]
{-}15 \sqrt{2}\TT_{1,0}(\tau^{[3]}_3)
{+}15 i [\TT_{2,-2}(\tau^{[3]}_3){-}\TT_{2,2}(\tau^{[3]}_3)]
]\\
4I_{xzy} =& \frac{1}{6} [
i \sqrt{6}[\TT_{3,-2}(\tau^{[3]}_1){-}\TT_{3,2}(\tau^{[3]}_1)]
{-}3 \sqrt{2} \TT_{2,0}(\tau^{[3]}_2)
{+}\sqrt{3} [\TT_{2,-2}(\tau^{[3]}_2){+}\TT_{2,2}(\tau^{[3]}_2)]\\&
{-}3 [\TT_{2,-2}(\tau^{[3]}_3){+}\TT_{2,2}(\tau^{[3]}_3)]
{-}\sqrt{6}\TT_{2,0}(\tau^{[3]}_3)
{-}2 \sqrt{3} \TT_{0,0}(\tau^{[3]}_4)
]\\
4I_{xzz} =& \frac{1}{30} [
2 \sqrt{15} [\TT_{1,-1}(\tau^{[3]}_1){-}\TT_{1,1}(\tau^{[3]}_1)]
{+}4 \sqrt{15} [\TT_{3,-1}(\tau^{[3]}_1){-}\TT_{3,1}(\tau^{[3]}_1)]
{+}5 \sqrt{3} [\TT_{1,-1}(\tau^{[3]}_2){-}\TT_{1,1}(\tau^{[3]}_2)]\\&
{+}5 i \sqrt{3} [\TT_{2,-1}(\tau^{[3]}_2){+}\TT_{2,1}(\tau^{[3]}_2)]
{+}15 [\TT_{1,-1}(\tau^{[3]}_3){-}\TT_{1,1}(\tau^{[3]}_3)]
{+}15 i[\TT_{2,-1}(\tau^{[3]}_3){+}\TT_{2,1}(\tau^{[3]}_3)]
]\\
4I_{yxx} =& \frac{1}{30} [
2 i \sqrt{15} [\TT_{1,-1}(\tau^{[3]}_1){+}\TT_{1,1}(\tau^{[3]}_1)]
{-}i \sqrt{15} [\TT_{3,-1}(\tau^{[3]}_1){+}\TT_{3,1}(\tau^{[3]}_1)]
{+}15 i [\TT_{3,-3}(\tau^{[3]}_1){+}\TT_{3,3}(\tau^{[3]}_1)]\\&
{+}5 i \sqrt{3} [\TT_{1,-1}(\tau^{[3]}_2){+}\TT_{1,1}(\tau^{[3]}_2)]
{+}5 \sqrt{3} [\TT_{2,-1}(\tau^{[3]}_2){-}\TT_{2,1}(\tau^{[3]}_2)]
{+}15 i[\TT_{1,-1}(\tau^{[3]}_3){+}\TT_{1,1}(\tau^{[3]}_3)]
{+}15 [\TT_{2,-1}(\tau^{[3]}_3){-}\TT_{2,1}(\tau^{[3]}_3)]
]\\
4I_{yxy} =&\frac{1}{30} [
2 \sqrt{15} [\TT_{1,-1}(\tau^{[3]}_1){-}\TT_{1,1}(\tau^{[3]}_1)]
{-}\sqrt{15} [\TT_{3,-1}(\tau^{[3]}_1){-}\TT_{3,1}(\tau^{[3]}_1)]
{-}15 [\TT_{3,-3}(\tau^{[3]}_1){-}\TT_{3,3}(\tau^{[3]}_1)]\\&
{+}5 \sqrt{3} [\TT_{1,-1}(\tau^{[3]}_2){-}\TT_{1,1}(\tau^{[3]}_2)]
{-}5 i \sqrt{3} [\TT_{2,-1}(\tau^{[3]}_2){+}\TT_{2,1}(\tau^{[3]}_2)]
{-}15[\TT_{1,-1}(\tau^{[3]}_3){-}\TT_{1,1}(\tau^{[3]}_3)]
{+}15 i[\TT_{2,-1}(\tau^{[3]}_3){+}\TT_{2,1}(\tau^{[3]}_3)]
]\\
4I_{yxz} =& -\frac{1}{2\sqrt3}[
{-}i \sqrt{2} [\TT_{3,-2}(\tau^{[3]}_1){-}\TT_{3,2}(\tau^{[3]}_1)]
{+}2 [\TT_{2,-2}(\tau^{[3]}_2){+}\TT_{2,2}(\tau^{[3]}_2)]
{-}2 \sqrt{2} \TT_{2,0}(\tau^{[3]}_3)
{+}2 \TT_{0,0}(\tau^{[3]}_4)
]\\
4I_{yyx} =& \frac{1}{30} [
2 \sqrt{15} [\TT_{1,-1}(\tau^{[3]}_1){-}\TT_{1,1}(\tau^{[3]}_1)]
{-}\sqrt{15}[\TT_{3,-1}(\tau^{[3]}_1){-}\TT_{3,1}(\tau^{[3]}_1)]
{-}15 [\TT_{3,-3}(\tau^{[3]}_1){-}\TT_{3,3}(\tau^{[3]}_1)]\\&
{-}10 \sqrt{3} [\TT_{1,-1}(\tau^{[3]}_2){-}\TT_{1,1}(\tau^{[3]}_2)]
{+}10 i \sqrt{3} [\TT_{2,-1}(\tau^{[3]}_2){+}\TT_{2,1}(\tau^{[3]}_2)]
]\\
4I_{yyy} =& \frac{1}{10} i [
2 \sqrt{15} [\TT_{1,-1}(\tau^{[3]}_1){+}\TT_{1,1}(\tau^{[3]}_1)]
{-}\sqrt{15} [\TT_{3,-1}(\tau^{[3]}_1){+}\TT_{3,1}(\tau^{[3]}_1)]
{-}5[\TT_{3,-3}(\tau^{[3]}_1){+}\TT_{3,3}(\tau^{[3]}_1)]
]\\
4I_{yyz} =& \frac{1}{30} [
2 \sqrt{30}\TT_{1,0}(\tau^{[3]}_1)
{-}6 \sqrt{5} \TT_{3,0}(\tau^{[3]}_1)
{-}5 \sqrt{6} [\TT_{3,-2}(\tau^{[3]}_1){+}\TT_{3,2}(\tau^{[3]}_1)]
{-}10 \sqrt{6} \TT_{1,0}(\tau^{[3]}_2)
{-}10 i \sqrt{3}[\TT_{2,-2}(\tau^{[3]}_2){-}\TT_{2,2}(\tau^{[3]}_2)]
]\\
4I_{yzx} =& \frac{1}{6} [
i \sqrt{6}[\TT_{3,-2}(\tau^{[3]}_1){-}\TT_{3,2}(\tau^{[3]}_1)]
{+}3 \sqrt{2}\TT_{2,0}(\tau^{[3]}_2)
{+}\sqrt{3} [\TT_{2,-2}(\tau^{[3]}_2){+}\TT_{2,2}(\tau^{[3]}_2)]
{+}\sqrt{6}\TT_{2,0}(\tau^{[3]}_3)\\&
{-}3 [\TT_{2,-2}(\tau^{[3]}_3){+}\TT_{2,2}(\tau^{[3]}_3)]
{+}2 \sqrt{3} \TT_{0,0}(\tau^{[3]}_4)
]\\
4I_{yzy} =& \frac{1}{30} [
2 \sqrt{30} \TT_{1,0}(\tau^{[3]}_1)
{-}6 \sqrt{5} \TT_{3,0}(\tau^{[3]}_1)
{-}5 \sqrt{6}[\TT_{3,-2}(\tau^{[3]}_1){+}\TT_{3,2}(\tau^{[3]}_1)]
{+}5 \sqrt{6} \TT_{1,0}(\tau^{[3]}_2)\\&
{+}5 i \sqrt{3} [\TT_{2,-2}(\tau^{[3]}_2){-}\TT_{2,2}(\tau^{[3]}_2)]
{-}15 \sqrt{2}\TT_{1,0}(\tau^{[3]}_3)
{-}15 i [\TT_{2,-2}(\tau^{[3]}_3){-}\TT_{2,2}(\tau^{[3]}_3)]
]\\
4I_{yzz} =& \frac{1}{30} [
2 i \sqrt{15} [\TT_{1,-1}(\tau^{[3]}_1){+}\TT_{1,1}(\tau^{[3]}_1)]
{+}4 i \sqrt{15} [\TT_{3,-1}(\tau^{[3]}_1){+}\TT_{3,1}(\tau^{[3]}_1)]
{+}5 i \sqrt{3} [\TT_{1,-1}(\tau^{[3]}_2){+}\TT_{1,1}(\tau^{[3]}_2)]\\&
{-}5 \sqrt{3} [\TT_{2,-1}(\tau^{[3]}_2){-}\TT_{2,1}(\tau^{[3]}_2)]
{+}15 i[\TT_{1,-1}(\tau^{[3]}_3){+}\TT_{1,1}(\tau^{[3]}_3)]
{-}15 [\TT_{2,-1}(\tau^{[3]}_3){-}\TT_{2,1}(\tau^{[3]}_3)]
]\\
4I_{zxx} =& \frac{1}{30} [
2 \sqrt{30} \TT_{1,0}(\tau^{[3]}_1)
{-}6 \sqrt{5} \TT_{3,0}(\tau^{[3]}_1)
{+}5 \sqrt{6}[\TT_{3,-2}(\tau^{[3]}_1){+}\TT_{3,2}(\tau^{[3]}_1)]
{+}5 \sqrt{6} \TT_{1,0}(\tau^{[3]}_2)
{-}5 i \sqrt{3} [\TT_{2,-2}(\tau^{[3]}_2){-}\TT_{2,2}(\tau^{[3]}_2)]\\&
{+}15 \sqrt{2}\TT_{1,0}(\tau^{[3]}_3)
{-}15 i [\TT_{2,-2}(\tau^{[3]}_3){-}\TT_{2,2}(\tau^{[3]}_3)]
]\\
4I_{zxy} =& \frac{1}{6} [
{+}i \sqrt{6}[\TT_{3,-2}(\tau^{[3]}_1){-}\TT_{3,2}(\tau^{[3]}_1)]
{+}\sqrt{3} [\TT_{2,-2}(\tau^{[3]}_2){+}\TT_{2,2}(\tau^{[3]}_2)]
{-}3 \sqrt{2} \TT_{2,0}(\tau^{[3]}_2)
{+}3 [\TT_{2,-2}(\tau^{[3]}_3){+}\TT_{2,2}(\tau^{[3]}_3)]\\&
{+}\sqrt{6} \TT_{2,0}(\tau^{[3]}_3)
{+}2 \sqrt{3} \TT_{0,0}(\tau^{[3]}_4)
]\\
4I_{zxz} =& \frac{1}{30} [
2 \sqrt{15} [\TT_{1,-1}(\tau^{[3]}_1){-}\TT_{1,1}(\tau^{[3]}_1)]
{+}4 \sqrt{15} [\TT_{3,-1}(\tau^{[3]}_1){-}\TT_{3,1}(\tau^{[3]}_1)]
{+}5 \sqrt{3} [\TT_{1,-1}(\tau^{[3]}_2){-}\TT_{1,1}(\tau^{[3]}_2)]\\&
{+}5 i \sqrt{3} [\TT_{2,-1}(\tau^{[3]}_2){+}\TT_{2,1}(\tau^{[3]}_2)]
{-}15[\TT_{1,-1}(\tau^{[3]}_3){-}\TT_{1,1}(\tau^{[3]}_3)]
{-}15 i[\TT_{2,-1}(\tau^{[3]}_3){+}\TT_{2,1}(\tau^{[3]}_3)]
]\\
4I_{zyx} =& \frac{1}{6} [
{+}i \sqrt{6}[\TT_{3,-2}(\tau^{[3]}_1){-}\TT_{3,2}(\tau^{[3]}_1)]
{+}\sqrt{3} [\TT_{2,-2}(\tau^{[3]}_2){+}\TT_{2,2}(\tau^{[3]}_2)]
{+}3 \sqrt{2} \TT_{2,0}(\tau^{[3]}_2)
{+}3 [\TT_{2,-2}(\tau^{[3]}_3){+}\TT_{2,2}(\tau^{[3]}_3)]\\&
{-}\sqrt{6} \TT_{2,0}(\tau^{[3]}_3)
{-}2 \sqrt{3} \TT_{0,0}(\tau^{[3]}_4)
]\\
4I_{zyy} =& \frac{1}{30} [
2 \sqrt{30} \TT_{1,0}(\tau^{[3]}_1)
{-}6 \sqrt{5} \TT_{3,0}(\tau^{[3]}_1)
{-}5 \sqrt{6}[\TT_{3,-2}(\tau^{[3]}_1){+}\TT_{3,2}(\tau^{[3]}_1)]
{+}5 \sqrt{6} \TT_{1,0}(\tau^{[3]}_2)
{+}5 i \sqrt{3} [\TT_{2,-2}(\tau^{[3]}_2){-}\TT_{2,2}(\tau^{[3]}_2)]\\&
{+}15 \sqrt{2}\TT_{1,0}(\tau^{[3]}_3)
{+}15 i [\TT_{2,-2}(\tau^{[3]}_3){-}\TT_{2,2}(\tau^{[3]}_3)]
]\\
4I_{zyz} =& \frac{1}{30} [
2 i \sqrt{15} [\TT_{1,-1}(\tau^{[3]}_1){+}\TT_{1,1}(\tau^{[3]}_1)]
{+}4 i \sqrt{15} [\TT_{3,-1}(\tau^{[3]}_1){+}\TT_{3,1}(\tau^{[3]}_1)]
{+}5 i \sqrt{3} [\TT_{1,-1}(\tau^{[3]}_2){+}\TT_{1,1}(\tau^{[3]}_2)]\\&
{-}5 \sqrt{3} [\TT_{2,-1}(\tau^{[3]}_2){-}\TT_{2,1}(\tau^{[3]}_2)]
{-}15 i[\TT_{1,-1}(\tau^{[3]}_3){+}\TT_{1,1}(\tau^{[3]}_3)]
{+}15 [\TT_{2,-1}(\tau^{[3]}_3){-}\TT_{2,1}(\tau^{[3]}_3)]
]\\
4I_{zzx} =& \frac{1}{5\sqrt3}[
\sqrt{5} [\TT_{1,-1}(\tau^{[3]}_1){-}\TT_{1,1}(\tau^{[3]}_1)]
{+}2 \sqrt{5} [\TT_{3,-1}(\tau^{[3]}_1){-}\TT_{3,1}(\tau^{[3]}_1)]
{-}5 [\TT_{1,-1}(\tau^{[3]}_2){-}\TT_{1,1}(\tau^{[3]}_2)]
{-}5 i [\TT_{2,-1}(\tau^{[3]}_2){+}\TT_{2,1}(\tau^{[3]}_2)]
]\\
4I_{zzy} =& \frac{1}{5\sqrt3}[
i \sqrt{5} [\TT_{1,-1}(\tau^{[3]}_1){+}\TT_{1,1}(\tau^{[3]}_1)]
{+}2 i \sqrt{5} [\TT_{3,-1}(\tau^{[3]}_1){+}\TT_{3,1}(\tau^{[3]}_1)]
{-}5 i [\TT_{1,-1}(\tau^{[3]}_2){+}\TT_{1,1}(\tau^{[3]}_2)]
{+}5 [\TT_{2,-1}(\tau^{[3]}_2){-}\TT_{2,1}(\tau^{[3]}_2)]]\\
4I_{zzz} =& \frac{1}{\sqrt5}[
\sqrt{6} \TT_{1,0}(\tau^{[3]}_1)
{+}2 \TT _{3,0}(\tau^{[3]}_1)]\\
\end{array}$
\caption{Decomposition of trilinear Cartesian-product operators $\II{abc}=\II {1a}\II {2b}\II {3c}$ 
in terms of \lisa tensor operators. This form is valid for a three-spin system, 
for a general system with $n\geq 3$ spins
one has to multiply $4\II{abc}$ with $(\sqrt{2})^{n-3}$. \label{Tril T->I}}
\end{table*}

\section{Visualization of typical operators in the \lisa basis}

\subsection{Visualizations of multiple quantum coherences\label{App_multiple}}
\begin{figure*}[t]
\includegraphics{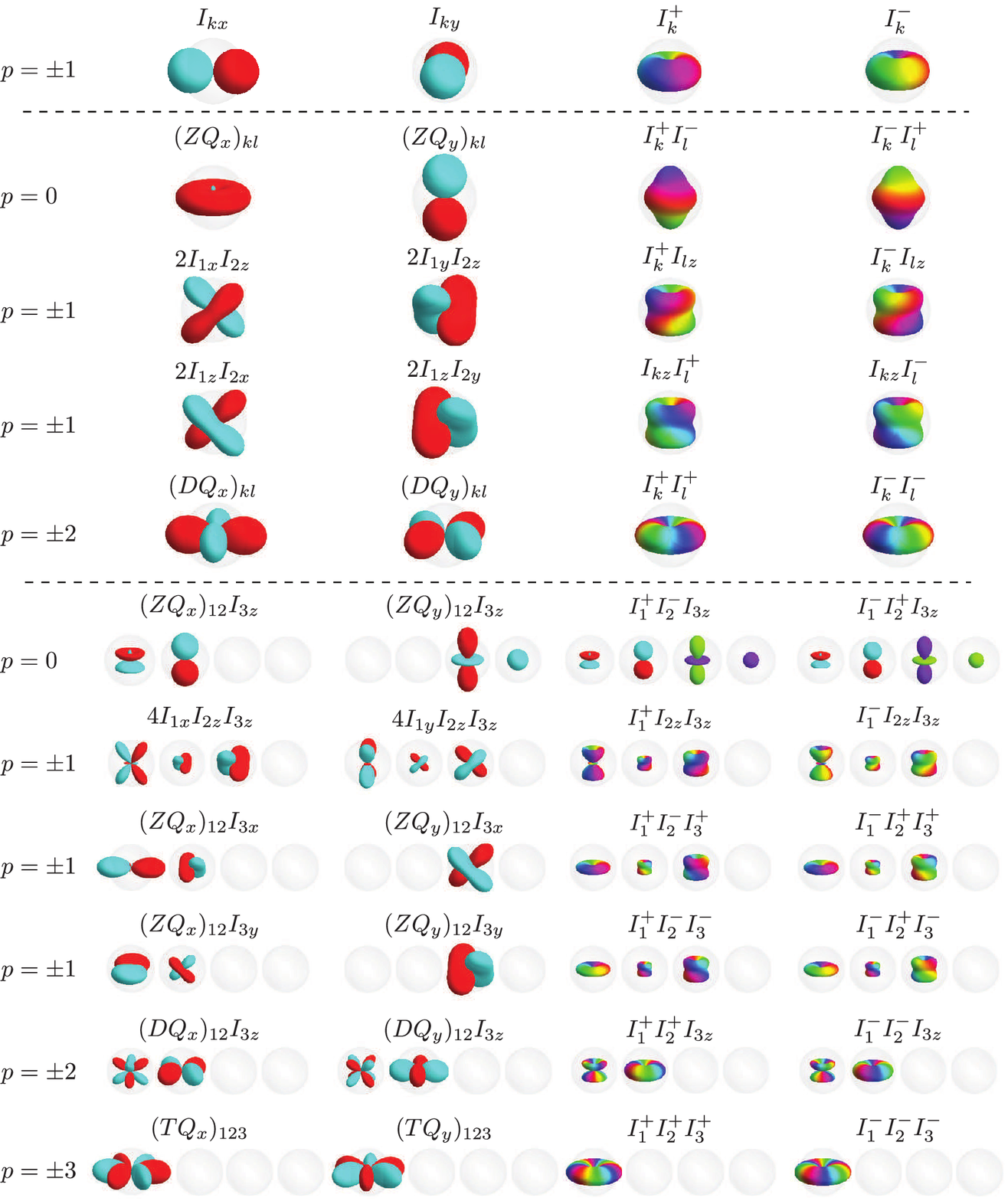}
\caption{(Color online) \Drops visualization in the \lisa basis of characteristic multiple quantum coherences
for three coupled spins. 
The operators are classified according to their linearity and their coherence order $p\in\Z$.
For visualization purposes, the \droplets 
which are always empty are not displayed.
The four trilinear \droplets are ordered from left to right according to 
the symmetry
of
each \droplet, i.e. from 
$\taun{1}{3}$ to $\taun{4}{3}$ (see \eq{ordered_tau}).
The above pictures correspond to the \drops visualization of the tensors after normalization, where
the \droplets for the trilinear operators are depicted in a smaller size.
For definitions of the zero, double, and triple quantum operators $(ZQ_\eta)_{kl}$, $(DQ_\eta)_{kl}$, and
$(TQ_\eta)_{kl}$ with $\eta\in\{x,y\}$ refer to \cite{Keeler2010,EBW87}.
\label{Quantum operators}}
\end{figure*}

Different examples of multiple-quantum coherences are given in \fig{Quantum operators}. 
Operators with defined and unique coherence order $p$ are visualized in the third and 
fourth column of  \fig{Quantum operators} 
for $p \ge 0$ and $p \le 0$, respectively.
The operators displayed in the first and second column of \fig{Quantum operators} 
correspond to mixtures of multiple quantum orders $\pm p$.

As discussed in  Sec.~\ref{Sec_multiple},  an operator $A_p$ has a well-defined coherence order $p$
if a rotation around the z axis by an arbitrary angle $\alpha$ reproduces the operator $A_p$ 
up to an additional phase factor $\exp(-ip\alpha)$:
$$
\exp( -i\alpha\sum_{k=1}^{n}\II{kz})\, A_p\, \exp( i\alpha\sum_{k=1}^{n}\II{kz})
=
A_p \exp(-ip\alpha).
$$
Similarly, a droplet representing a function $f_A^{(\ell)}$ (c.f.\ Eq.~\eqref{DROPS}) corresponds to an operator 
term $A^{(\ell)}$ with well-defined coherence order $p$, if a rotation around the $z$-axis by an 
arbitrary angle $\alpha$ reproduces the droplet (and the function $f_A^{(\ell)}$) up to an 
additional phase factor $\exp(-ip\alpha)$. 
Hence, a droplet with coherence order $p$ as well as the corresponding function 
$f_A^{(\ell)}(\theta,\phi)=r_A^{(\ell)}(\theta,\phi)  \exp[i\varphi_A^{(\ell)}(\theta,\phi)]$ 
are transformed by a $z$-rotation with angle $\alpha$ to  
$\tilde{f}^{(\ell)}_A(\theta,\phi)=r_A^{(\ell)}(\theta,\phi)  \exp[i\tilde{\varphi}^{(\ell)}_A(\theta,\phi)]$ 
with $\tilde{\varphi}^{(\ell)}_A(\theta,\phi)=\varphi_A^{(\ell)}(\theta,\phi) - p \alpha$.
In order to illustrate this point, consider how 
the operator $I_k^+$  in \fig{Quantum operators} with $p=+1$ changes under a 
$z$-rotation with angle $\alpha=\pi/2$.
A $z$-rotation of the corresponding droplet only changes its  color (representing the phase of 
the function $f_A^{(\ell)}(\theta,\phi)$) but not its shape. For example, for the azimuthal angle 
$\phi=0$, the droplet is red (corresponding to a phase $\varphi_A^{(\ell)}(\theta,0)= 0$). After the 
rotation by  $\alpha=\pi/2$, the droplet has turned dark blue at the azimuthal angle $\phi=0$ 
(corresponding to a phase $\tilde{\varphi}_A^{(\ell)}(\theta,0)= - \pi/2$), which is exactly what is 
expected from the general formula given above:
$\tilde{\varphi}^{(\ell)}_A(\theta,\phi)=\varphi_A^{(\ell)}(\theta,\phi) - p \alpha=0 - (+1) \pi/2=- \pi/2$ 
for $p=+1$ and $\alpha=\pi/2$.

\begin{figure}[tb]
\includegraphics[]{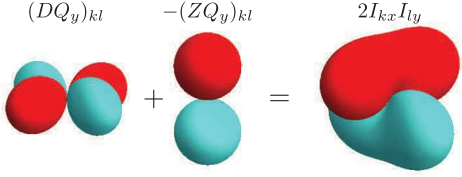}
\caption{(Color online) The bilinear Cartesian product operator $2\II{kx}\II{ly}$ is 
represented as a linear combination of
double-  and zero-quantum operators $(DQ_y)_{kl}$ and 
$(ZQ_y)_{kl}$. 
\label{IxSy}}
\end{figure}

Based on these properties, the droplets of an operator with \emph{unique} coherence order $p$ 
can be easily recognized using the following criteria:
(1) Disregarding the color, the \emph{shape} of the droplet is rotationally invariant under rotations 
around the $z$-axis, 
i.e.\ the shape is not changed by a $z$-rotation (c.f.\ third and fourth column in \fig{Quantum operators}).
(2) The coherence order $p$ of a droplet can be identified based on its \emph{color}:
(2a)~A droplet  of coherence order $p=0$ does not change its color if it is rotated by an 
arbitrary angle $\alpha$ around the $z$-axis, as illustrated by the operators 
$I_k^+ I_l^-$, $I_k^- I_l^+$, $I_1^+ I_2^- I_{3z}$, and $I_k^- I_l^+ I_{3z}$ in \fig{Quantum operators}.
(2b)~A droplet  of unique coherence order $p \ne 0$  (where $p$ is a non-zero integer with 
either positive or negative sign) is \emph{rainbow-colored}. For positive coherence order ($p>0$),  
the colors change from red to yellow to green to blue when moving counter-clockwise around the 
$z$-axis. For negative coherence order, the colors change in the opposite direction.
(2c)~For a non-zero unique coherence order $p$, the absolute value $\abs{p}$ of the coherence 
order of a droplet is reflected by the \emph{number} of rainbows encountered when moving once 
around the $z$-axis. 
(2d)~A droplet of unique coherence order $p\ne 0$ is invariant under a rotation by integer 
multiples of  $2\pi/\abs{p}$ around the $z$-axis. This is illustrated for characteristic operators with 
unique coherence orders in the third column and fourth column of  \fig{Quantum operators}.
(2e)~Even if operators do not contain a unique coherence order $p$, but a mixture of coherence 
orders $\pm p$ with $\abs{p} \ne 0$, it is still true that the corresponding droplets do not change 
their appearance if they (or the corresponding operators) are rotated by integer multiples of  
$2 \pi/\abs{p}$ around the $z$-axis. This is illustrated by the examples in the first and 
second column of \fig{Quantum operators}.
All tensor operators $\TT_{jm}$ have the unique coherence order $p=m$.
Moreover, the bean-shaped droplet of the Cartesian product operator 
$2I_{kx}I_{ly}$ (c.f.\ \fig{IxSy}) is an example of an operator that is composed of terms 
with \emph{different} coherence orders $-2$, 0, and $+2$.
The operator is a superposition of the double-quantum operator 
$(DQ_y)_{kl}=\II{kx}\II{ly}+\II{ky}\II{lx}$ with rank $j=2$ and quantum orders $p=\pm 2$ 
and the zero-quantum operator $(ZQ_y)_{kl}=-\II{kx}\II{ly}+\II{ky}\II{lx}$ with 
rank $j=1$ and quantum order $p=0$.

\subsection{Extended NMR example\label{App_time_evolution}}
\begin{figure*}[t]
\includegraphics{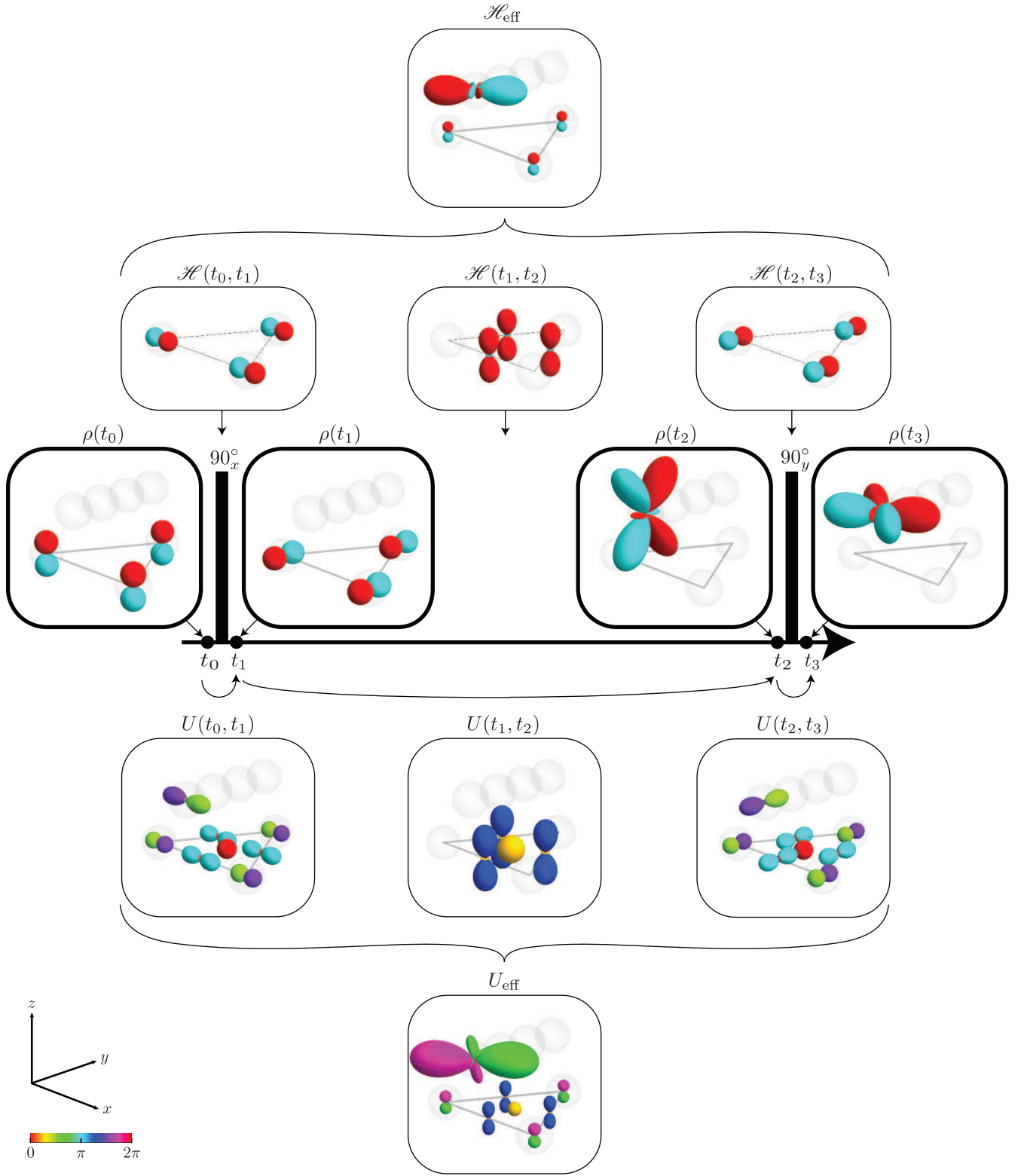}
\caption{(Color online) Experimental NMR pulse sequence to create triple-quantum coherences starting from the 
thermal equilibrium density operator in the high temperature limit \cite{EBW87}.
The pulse sequence consists of a $90^\circ$ pulse (with phase $x$) followed by a delay ($t_2-t_1$) 
and a second $90^\circ$ pulse (with phase $y$).
The density operators $\rho(t_i)$  are depicted in the middle row. 
The Hamiltonians $\H(t_i,t_{i+1})$ (which are re-scaled for better visibility)
are given in the second row 
and the effective Hamiltonian \cite{EBW87} of the experiment is shown at the top.
In the fourth row, \drops representations of the propagators 
associated to the individual time steps are displayed. The effective propagator 
is visualized at the bottom.
\label{wall}}
\end{figure*}

The example shown in \fig{wall} represents a common experiment in NMR spectroscopy
which is designed to create triple-quantum coherences from the polarization of three coupled 
spins \cite{EBW87}.
The system consists of three  spins 
in the weak-coupling limit (i.e.\ longitudinal or Ising-type coupling; see
Fig.~\ref{fig:Hamiltonians})
with identical  coupling
constants $J_{12}=J_{13}=J_{23}=10\, \text{Hz}$.
The building blocks of the experiment are visualized in \fig{wall} in the \lisa 
basis.
A triple-quantum coherence consists of combinations of tensor operators of rank $j=3$ and 
order $m=\pm3$.
At the initial time $t_0$, the system is in thermal equilibrium, which corresponds in the high-temperature limit 
to the density matrix 
$$\rho(t_0)=\II{1x}+\II{2x}+\II{3x}=\sqrt{2}(\TT_{10}^{\{1\}}+\TT_{10}^{\{2\}}+\TT_{10}^{\{3\}})$$ 
(where for simplicity only the traceless part of the density operator is considered here).

A first $90^\circ$ pulse (with phase $x$) is applied to the system with an amplitude of $10\, \text{kHz}$ for a time 
$t_1-t_0=25\, \mu\text{s}$ and flips the three magnetization vectors into the transverse plane. 
The corresponding linear control Hamiltonian is $$\H(t_0,t_1)=2\pi\, 10\, \text{kHz}\ (\II{1x}+\II{2x}+\II{3x})$$ 
and the density operator of the system at time $t_1$ is 
\begin{align*}
\rho(t_1)&={-}\II{1y}{-}\II{2y}{-}\II{3y}\\
&=-i[\TT_{1,-1}^{\{1\}}{+}\TT_{1,1}^{\{1\}}{+}\TT_{1,-1}^{\{2\}}{+}\TT_{1,1}^{\{2\}}
{+}\TT_{1,-1}^{\{3\}}{+}\TT_{1,1}^{\{3\}}].
\end{align*}
The next step consists of letting the bilinear coupling Hamiltonian act on the system in order to 
create trilinear terms 
in the density operator.
The coupling Hamiltonian is applied for a time $t_2-t_1=50\, \text{ms}$ and has the 
form 
$$\H(t_1,t_2)=2\pi\, 10\, \text{Hz}\ (\II{1z}\II{2z}+\II{1z}\II{3z}+\II{2z}\II{3z}).$$
At time $t_2$, the systems is in the state 
\begin{align*}
\rho(t_2)&=4\II{1y}\II{2z}\II{3z}+4\II{1z}\II{2y}\II{3z}+4\II{1z}\II{2z}\II{3y}\\
&\cong
+\, 0.78\, [\TT_{1,-1}(\taun13){+}\TT_{1,1}(\taun13)]\\
&\phantom{\cong\hspace{1mm}} + 1.55i\, [\TT_{3,-1}(\taun13){+}\TT_{3,1}(\taun13)].
\end{align*}
Finally, a second $90^\circ$ pulse (with phase $y$) is applied in order to create  terms of 
order $m=\pm3$.
The corresponding linear control Hamiltonian 
$$\H(t_2,t_3)=2\pi\, 10\, \text{kHz}\ (\II{1y}+\II{2y}+\II{3y})$$ 
is applied for a time $t_3-t_2=t_1-t_0.$
At time $t_3$, the density operator of the system is (where $\II{abc}:=\II {1a}\II {2b}\II {3c}$)
\begin{align*}
\rho(t_3)&=4\II{yxx}+4\II{xyx}+4\II{xxy}\\
&\cong +\, 0.78\, [\TT_{1,-1}(\taun13){+}\TT_{1,1}(\taun13)]\\
&\phantom{\cong\hspace{1mm}} -0.39i\, [\TT_{3,-1}(\taun13){+}\TT_{3,1}(\taun13)]\\
&\phantom{\cong\hspace{1mm}} +1.5i\, [\TT_{3,-3}(\taun13){+}\TT_{3,3}(\taun13)].
\end{align*}
At this point, the desired triple-quantum coherence term $\TT  _{3,-3}(\taun13)+\TT  _{3,3}(\taun13)$ 
has been created. Note that the shape of the \droplet corresponding to the term $\tau_1^{[3]}$ of $\rho(t_3)$
in \fig{wall} also exhibits the partial content of the triple-quantum coherence 
$(TQ_{y})_{123}$ (see \fig{Quantum operators}).
The remaining undesired terms of the density operator can be removed
by applying a triple quantum filter \cite{EBW87} to $\rho(t_3)$ (not shown for simplicity). 
The density operators $\rho(t_i)$ 
are depicted in the middle row of \fig{wall}.

The Hamiltonians $\H(t_i,t_{i+1})$ (which are re-scaled for better visibility) are shown in the second row and 
the effective Hamiltonian \cite{EBW87}
\begin{align*}
\H_{\text{eff}}\cong&-18.1\,\text{Hz}(\II{1z}{+}\II{2z}{+}\II{3z})\\
&-24.2\,\text{Hz}(I_{xxx}{+}I_{yyy}{+}I_{zzz})\\
&-72.5\,\text{Hz}(I_{xxy}{+}I_{xyx}{+}I_{yxx} {+} I_{yyx}{+}I_{yxy}{+}I_{xyy} )
\end{align*} 
of the experiment is given at the top.
In the fourth row, the \drops representations of the propagators ($\unity_q$ 
denotes the $q\times q$-dimensional identity matrix) 
\begin{align*}
U(t_0,t_1)\cong&+0.35\,\unity_8-0.71i(\II{1x}{+}\II{2x}{+}\II{3x})\\
&-1.41(\II{1x}\II{2x}{+}\II{1x}\II{3x}{+}\II{2x}\II{3x})\\
&+2.83\II{1x}\II{2x}\II{3x},\\
U(t_1,t_2)\cong&+0.35(1{+}i)\,\unity_8\\
&-1.41(1{+}i)(\II{1z}\II{2z}{+}\II{1z}\II{3z}{+}\II{2z}\II{3z}),\; \text{and}\\
U(t_2,t_3)\cong&+0.35\,\unity_8-0.71i(\II{1y}{+}\II{2y}{+}\II{3y})\\
&-1.41(\II{1y}\II{2y}{+}\II{1y}\II{3y}{+}\II{2y}\II{3y})\\
&+2.83\II{1y}\II{2y}\II{3y},
\end{align*}
are shown.
The overall effective propagator 
\begin{gather*}
U_{\text{eff}}\cong+0.18(1{+}i)\,~\unity_8-0.35(1{+}i)(\II{1z}{+}\II{2z}{+}\II{3z})\\
- 0.71(1{+}i)(\II{1z}\II{2z}{+}\II{1z}\II{3z}{+}\II{2z}\II{3z})
-1.41(1{-}i)\times\\(\II {xxx}{+}\II {yyy} {-} \II {zzz}
{+}\II {xxy}{+}\II {xyx} {+} \II {yxx} 
{+}\II {yyx}{+} \II {yxy}{+}\II {xyy} )
\end{gather*}
of the pulse sequence is given
at the bottom of \fig{wall}.

\begin{figure}[t]
\includegraphics[]{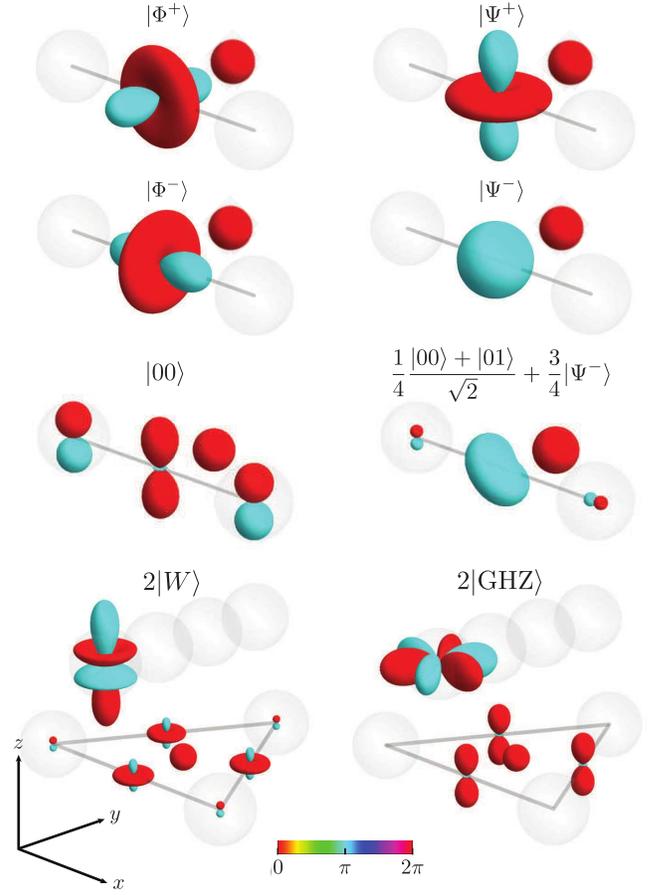}
\caption{(Color online) \lisa
representation of 
the density matrix for
characteristic entangled and separable pure states, for details refer to the text. Compare 
also to \fig{Entanglement_multipole} below.
\label{Entanglement}}
\end{figure}

\subsection{Pure quantum states\label{App_pure_states}}
The examples of \fig{Entanglement} display 
density matrices for entangled pure states
of two spins and three spins in 
the \drops representation corresponding to
the \lisa basis. 
The four Bell states $\ket{\Phi^{\pm}}=(\ket{00} {\pm} \ket{11})/\sqrt{2}$ and 
$\ket{\Psi^\pm}=(\ket{01}{\pm}\ket{10})/\sqrt{2}$ correspond to maximally 
entangled states of a two-spin system.
For comparison, a separable state ($\ket{00}$) and a partially entangled state 
$[(\ket{00}{+}\ket{01})/{\sqrt{2}}+3\ket{\Psi^-}]/4$ are depicted in the third row
of \fig{Entanglement}.
The last row of \fig{Entanglement} shows 
the W state $\ket{W}=(\ket{100}{+}\ket{010}{+}\ket{001})/\sqrt{3}$ and the 
Greenberger-Horne-Zeilinger state $\ket{\text{GHZ}}=(\ket{000}{+}\ket{111})/\sqrt{2}$,
which are both entangled quantum states of three spins.

\section{Wigner representation: Proof of Proposition~\ref{prop}\label{App_Prop_1}}
Property (a) is a direct consequence of the definition of the \drops mapping.
Proving property (b), we deduce directly from Eq.~\eqref{DROPS mapping} that 
\begin{gather*}
f^{(\ell)*}_A=\!\!\sum_{j\in J(\ell)} \sum_{m=-j}^j \! \! c^{(\ell)*}_{jm}\Y_{jm}^*,\;
A^{(\ell)\dagger}=\!\!\sum_{j\in J(\ell)} \sum_{m=-j}^j \! \! c^{(\ell)*}_{jm}\TT  ^{(\ell)\dagger}_{jm}.
\end{gather*}
Consequently, (b) follows from the Condon-Shortley phase convention 
$\TT_{jm} = (-1)^{m}\,\TT_{j,-m}^{\dagger}$ and a similar relation $\Y_{jm} = (-1)^{m}\,\Y_{j,-m}^{*}$ 
for spherical harmonics.
Property (c) is a special case of (e), which we prove below. 
Property (d) uses the fact that irreducible tensor operators and spherical harmonics are 
an explicit form of
irreducible representations for $\SU(2)$. These properties are preserved by the \drops mapping,
i.e., the components $\TT_{jm}$ with $j\leq m \leq j$ corresponding to the same tensor 
define an invariant subspace under rotations. 
Furthermore, all of these components are part of the {\it same} \droplet.
The proof of property (e) requires a more detailed analysis.
We expand both sides of (e) in order to show that they are equal.
To simplify the notation, the dependence of the spherical harmonics on 
the variables $\theta$ and $\phi$ is suppressed.

The left-hand side of the equality is expanded as
$\sum_{\ell\in L}\int_{S^2}f^{(\ell)}_A(\theta,\phi){f_B^{(\ell)}}(\theta,\phi)d\mu$, which is equal to
$$\sum_{\ell\in L}\int_{S^2}\sum_{jm}\Tr[\TT_{jm}^{\dagger(\ell)}A]\Y_{jm}\cdot\sum_{j'm'}\Tr[{\TT_{j'm'}^{\dagger(\ell)}}B]\Y_{j'm'}d\mu.$$
Applying the orthonormality of the spherical harmonics  
as well as
the property 
$\Y_{j,m}=(-1)^{m}\Y_{j,-m}^{*}$, we obtain the expression
$\sum_{\ell\in L}\sum_{jm}\Tr[\TT_{jm}^{\dagger(\ell)}A]\cdot\Tr[{\TT_{j,-m}^{\dagger(\ell)}}B](-1)^{m}$.
Applying $\TT_{jm}=(-1)^{m}\TT_{j,-m}^{\dagger}$, this
can be further simplified to 
$\sum_{\ell\in L}\sum_{jm}\Tr[\TT_{jm}^{\dagger(\ell)}A]\cdot\Tr[{\TT_{jm}^{(\ell)}}B]$.

On the right-hand side, we have $\Tr(AB) =$
$$
\Tr\Big(\sum_{\ell\in L}A^{(\ell)}\sum_{\ell'\in L}B^{(\ell')}\Big)
=\sum_{\ell\in L}\sum_{\ell'\in L}\Tr(A^{(\ell)}B^{(\ell')})$$
by the linearity of the trace, which can be further transformed to
$\sum_{\ell\in L}\Tr(A^{(\ell)}B^{(\ell)})$ relying on
the orthogonality of the operator spaces for 
different \droplets.
We use the decompositions of the $A^{(\ell)}$ and $B^{(\ell)}$
and obtain the formula
\begin{gather*}\sum_{\ell\in L}\Tr\Big(\sum_{jm}\Tr[\TT_{jm}^{\dagger (\ell)}A]
\TT_{jm}^{(\ell)}\cdot\sum_{j'm'}\Tr[\TT_{j'm'}^{\dagger (\ell)}B]\TT_{j'm'}^{(\ell)}\Big)\\
=\sum_{\ell\in L}\sum_{jm}\sum_{j'm'}\Tr[\TT_{jm}^{\dagger (\ell)}A]
\Tr[\TT_{j'm'}^{\dagger (\ell)}B]\Tr[\TT_{jm}^{(\ell)}\TT_{j'm'}^{(\ell)}].
\end{gather*}
Using the 
normalization and $\TT_{j,m}=(-1)^{m}\TT_{j,-m}^{\dagger}$, it simplifies
to $\sum_{\ell\in L}\sum_{jm}\Tr[\TT_{jm}^{\dagger (\ell)}A]\Tr[\TT_{j,-m}^{\dagger (\ell)}B](-1)^{m}
=\sum_{\ell\in L}\sum_{jm}\Tr[\TT_{jm}^{\dagger (\ell)}A]\Tr[\TT_{jm}^{ (\ell)}B]$.
This shows that the left-hand side is identical to the right-hand side.

\begin{table}[b]
\includegraphics[]{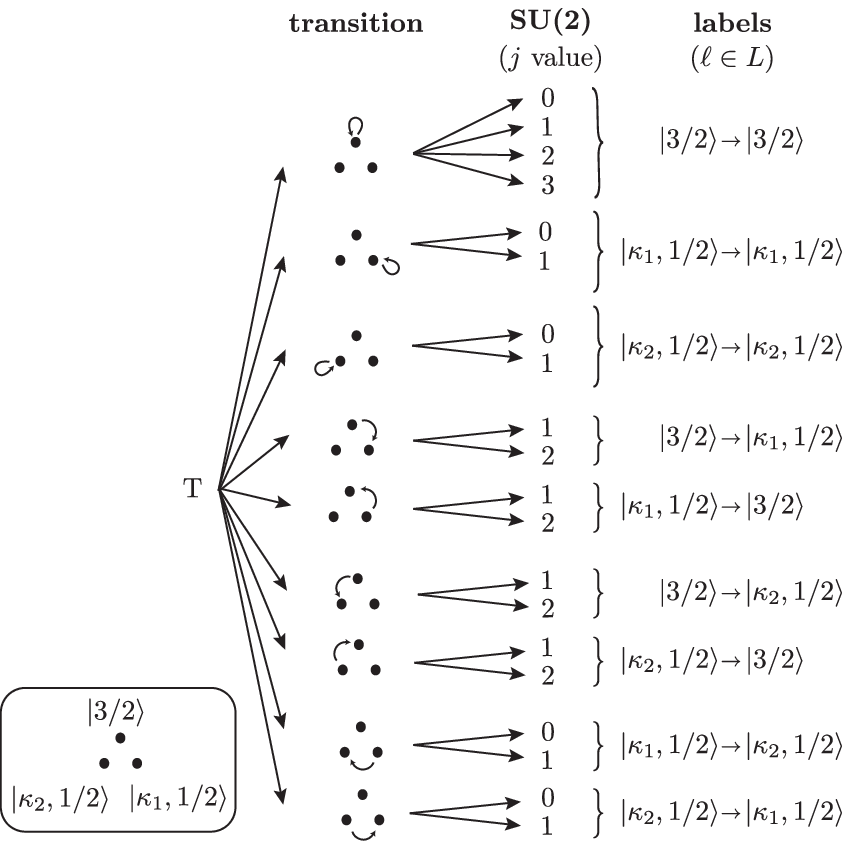}
\caption{Labels for the \drops representation 
based on multipole operators.\label{multipole pic}}
\end{table}

\section{Details for the computation of 
Table~\ref{tab_klambda}\label{App_table}}
Here, we detail the computations summarized in
Table~\ref{tab_klambda}.
These are determined in two steps.
First, the symmetries of the direct product
$\GL(d,\C)\times S_g$ acting on $\otimes^g \C^d$ (notably for $d{=}3$) are identified, 
where $\GL(d,\C)$ denotes the general linear group 
of complex $d{\times} d$-matrices with non-zero determinant.
The irreducible representations $\vartheta_{d,\lambda}$ of $\GL(d,\C)$
are labeled by partitions $\lambda$ 
with at most $d$ parts (i.e.\ $\kappa(\lambda) \leq d$), cf.\ \ppp{231--237} of \cite{FH91},
and irreducible representations $\mu_\lambda$ of the symmetric group $S_g$
are indexed with partitions $\lambda$ of degree $\abs{\lambda}=g$,
cf.\ \ppp{44--46} of \cite{FH91}.
Using these notations,
the Schur-Weyl duality (see \pp{389} of \cite{GW09})
describes how the action of $\GL(d,\C)\times S_g$ decomposes $g$-tensors
$\otimes^g \C^d$ into a multiplicity-free  sum (i.e.\ each $\lambda$ occurs only once)
\begin{equation}\label{SchurWeyl}
\bigotimes^g \C^d \cong 
\bigoplus_{\lambda \in \mathrm{Par}(g,d)}
\vartheta_{d,\lambda}\otimes \mu_{\lambda},
\end{equation}
where $\mathrm{Par}(g,d)$ denotes the set of
partitions with degree $g$ and with at most $d$ parts (i.e.\ $\kappa(\lambda) \leq d$).

\begin{figure}[b]
\includegraphics[]{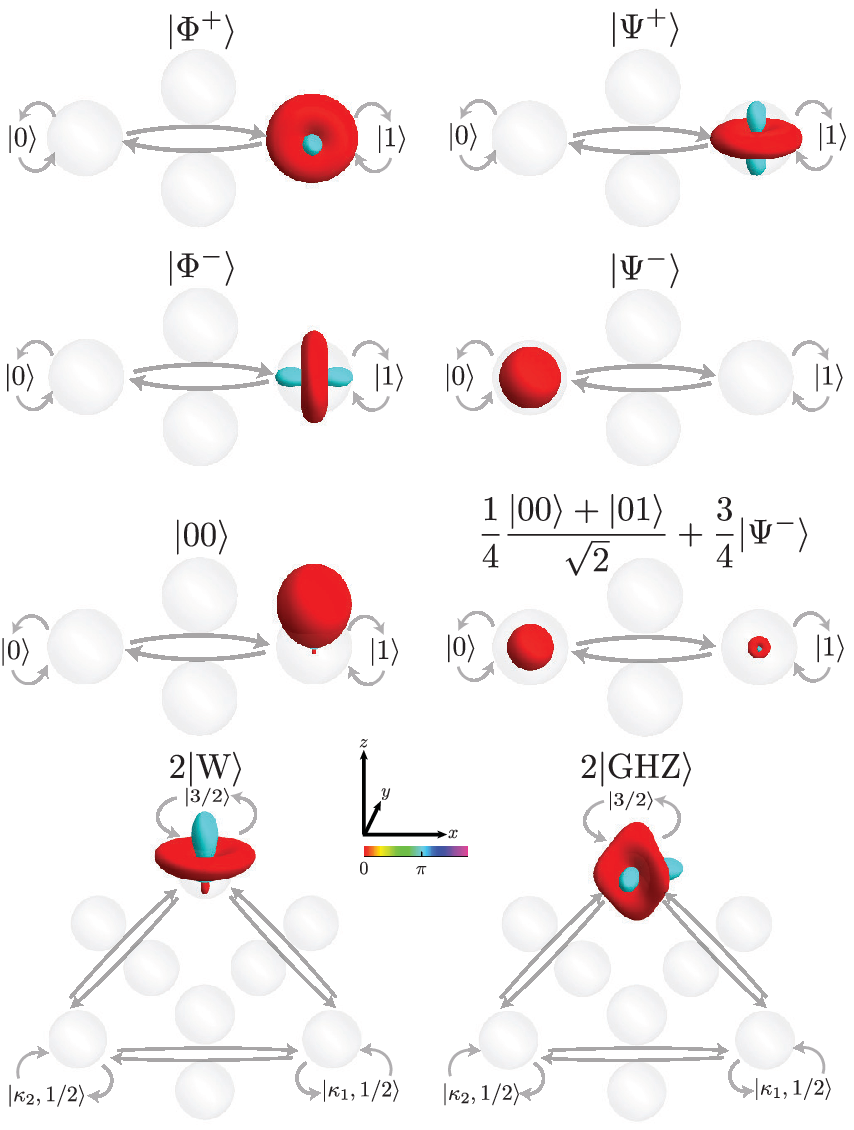}
\caption{(Color online) \Drops visualization in the multipole tensor basis
for the examples of \fig{Entanglement}.
\label{Entanglement_multipole}}
\end{figure}

\begin{table*}[t] 
\begin{tabular}{r@{\hspace{2mm}}c@{\hspace{2mm}}c@{\hspace{2mm}}l}
\begin{tabular}[t]{@{}r@{\hspace{1mm}}c@{\hspace{1mm}}l@{\hspace{3mm}}l@{}l@{}l@{}l}
\multicolumn{3}{c}{transition} & \multicolumn{4}{c}{rank}\\\hline
$\ket{3/2}$&$\arr$&$\ket{3/2}$ & 0,&1,&2,&3\\
$\ket{\kappa_1,1/2}$&$\arr$&$\ket{\kappa_1,1/2}$ & 0,&1&&\\
$\ket{\kappa_2,1/2}$&$\arr$&$\ket{\kappa_2,1/2}$ & 0,&1&&
\end{tabular}
&
\begin{tabular}[t]{@{}r@{\hspace{1mm}}c@{\hspace{1mm}}l@{\hspace{3mm}}l@{}l@{}l@{}l}
\multicolumn{3}{c}{transition} & \multicolumn{4}{c}{rank}\\\hline
$\ket{3/2}$&$\arr$&$\ket{\kappa_1,1/2}$ & &1,&2&\\
$\ket{\kappa_1,1/2}$&$\arr$&$\ket{3/2}$ & &1,&2&
\end{tabular}
&
\begin{tabular}[t]{@{}r@{\hspace{1mm}}c@{\hspace{1mm}}l@{\hspace{3mm}}l@{}l@{}l@{}l}
\multicolumn{3}{c}{transition} & \multicolumn{4}{c}{rank}\\\hline
$\ket{3/2}$&$\arr$&$\ket{\kappa_2,1/2}$ & &1,&2&\\
$\ket{\kappa_2,1/2}$&$\arr$&$\ket{3/2}$ & &1,&2&
\end{tabular}
&
\begin{tabular}[t]{@{}r@{\hspace{1mm}}c@{\hspace{1mm}}l@{\hspace{3mm}}l@{}l@{}l@{}l}
\multicolumn{3}{c}{transition} & \multicolumn{4}{c}{rank}\\\hline
$\ket{\kappa_1,1/2}$&$\arr$&$\ket{\kappa_2,1/2}$ & 0,&1&&\\
$\ket{\kappa_2,1/2}$&$\arr$&$\ket{\kappa_1,1/2}$ & 0,&1&&
\end{tabular}
\end{tabular}
\caption{All compatible combinations of transitions and ranks for multipole tensor operators\label{multipole_table}}
\end{table*}

\begin{table*}[t]
\footnotesize
$
\begin{array}{@{}r@{\hspace{1mm}}l}
T_0(\ket{3/2}\arr \ket{3/2}) &=\tfrac{1}{\sqrt{6}} [\TTlb{0}{1,2}{+}\TTlb{0}{1,3}
{+}\TTlb{0}{2,3}{+}\sqrt{3}\TT^{\emptyset}_0(\taun 10)]\\ 
T_1(\ket{3/2}\arr \ket{3/2}) &=\tfrac{\sqrt{10}}{6} [ \TTlb{1}{1}{+} \TTlb{1}{2}
{+} \TTlb{1}{3} {-}\tfrac{\sqrt{3}}{\sqrt{5}}\TT_1(\taun 13)] \\
T_2(\ket{3/2}\arr \ket{3/2}) &=  \tfrac{1}{\sqrt{3}}[\TTlb2{12}{+}\TTlb2{13}
{+}\TTlb2{23} ]\\
T_3(\ket{3/2}\arr \ket{3/2}) &= \TT_3(\taun 12)\\[0mm]
T_0(\ket{\kappa_1,1/2}\arr \ket{\kappa_1,1/2}) &=\tfrac{\sqrt{3}}{6} [\TTlb{0}{1,2}
{-}2\TTlb{0}{1,3}{-}2\TTlb{0}{2,3}{+}\tfrac{3}{\sqrt{3}}\TT^{\emptyset}_0(\taun 10)]\\ 
T_1(\ket{\kappa_1,1/2}\arr \ket{\kappa_1,1/2}) &=\tfrac16 [2\TTlb11{+}2\TTlb 12{-}\TTlb13
{-}\sqrt{15}\TT_1(\taun13){-}2\sqrt{3}\TT_1(\taun23) ] 
\\
T_0(\ket{\kappa_2,1/2}\arr \ket{\kappa_2,1/2}) &= \tfrac12[-\sqrt{3}\TTlb0{1,2}
{+} \TT^{\emptyset}_0(\taun 10)]\\
T_1(\ket{\kappa_2,1/2}\arr \ket{\kappa_2,1/2}) &= \tfrac16[3\TTlb13 {-} \sqrt{15} \tt113
{+}2\sqrt{3}\tt123]\\[0mm]
T_1(\ket{3/2}\arr \ket{\kappa_1,1/2}) &= \tfrac16 [ \sqrt{2}(-\TTlb11
{-}\TTlb12{+}2\TTlb13){+}3i(\TTlb1{1,3}{+}\TTlb1{2,3}){-}\sqrt{6}\tt123]\\
T_2(\ket{3/2}\arr \ket{\kappa_1,1/2}) &= \tfrac16 [\sqrt{3}(-2\TTlb2{1,2}{+}\TTlb2{1,3}
{+}\TTlb2{2,3}){+}3i\sqrt{2}\tt223]\\[0mm]
T_1(\ket{\kappa_1,1/2}\arr \ket{3/2}) &=\tfrac16 [ \sqrt{2}(\TTlb11
{+}\TTlb12-2\TTlb13){+}3i(\TTlb1{1,3}{+}\TTlb1{2,3}){+}\sqrt{6}\tt123]\\
T_2(\ket{\kappa_1,1/2}\arr \ket{3/2}) &= \tfrac16 [\sqrt{3}(2\TTlb2{1,2}
{-}\TTlb2{1,3}{-}\TTlb2{2,3}){+}3i\sqrt{2}\tt223]\\[0mm]
T_1(\ket{3/2}\arr \ket{\kappa_2,1/2}) &=\tfrac1{2\sqrt{3}}  [ \sqrt{2}(-\TTlb11
{+}\TTlb12){+}i(2\TTlb1{1,2}{+}\TTlb1{1,3}{-}\TTlb1{2,3}){-}\sqrt{2}\tt133]\\
T_2(\ket{3/2}\arr \ket{\kappa_2,1/2}) &=\tfrac12 [ -\TTlb2{1,3}{+}\TTlb2{2,3}{+}i\sqrt{2}\tt233]\\[0mm]
T_1(\ket{\kappa_2,1/2}\arr \ket{3/2}) &=\tfrac1{2\sqrt{3}}  [ \sqrt{2}(\TTlb11
{-}\TTlb12){+}i(2\TTlb1{1,2}{+}\TTlb1{1,3}{-}\TTlb1{2,3}){+}\sqrt{2}\tt133]\\
T_2(\ket{\kappa_2,1/2}\arr \ket{3/2}) &=\tfrac12 [\TTlb2{1,3}{-}\TTlb2{2,3}{+}i\sqrt{2}\tt233]\\[0mm]
T_0(\ket{\kappa_1,1/2}\arr \ket{\kappa_2,1/2}) &=\tfrac12 [ -\TTlb0{1,3}{+}\TTlb0{2,3}{-}i\sqrt{2}\tt143)]\\
T_1(\ket{\kappa_1,1/2}\arr \ket{\kappa_2,1/2}) &=\tfrac1{2\sqrt{3}}  [(-\TTlb11{+}\TTlb12)
{+}i\sqrt{2}(\TTlb1{1,2}{-}\TTlb1{1,3}{+}\TTlb1{2,3}){+}2\tt133]\\[0mm]
T_0(\ket{\kappa_2,1/2}\arr \ket{\kappa_1,1/2}) &=\tfrac12 [ -\TTlb0{1,3}{+}\TTlb0{2,3}{+}i\sqrt{2}\tt143]\\
T_1(\ket{\kappa_2,1/2}\arr \ket{\kappa_1,1/2}) &=\tfrac1{2\sqrt{3}}  [(-\TTlb11{+}\TTlb12)
{-}i\sqrt{2}(\TTlb1{1,2}{-}\TTlb1{1,3}{+}\TTlb1{2,3}){+}2\tt133]
\end{array}
$
\caption{Decomposition of the multipole tensor operators for a system of three spins in 
terms of the \lisa tensors.\label{Tmpo dec}}
\end{table*}

Second, these results will be
traced back to the intended group $\SU(2)\times S_g$. 
Here, $d{=}3$ since $\SU(2)$ (and its Lie algebra $\su(2)$)
acts for a single spin on the three-dimensional space $\C^3$ as
explained in Sec.~\ref{Sec_arbitrary}.
We substitute $\vartheta_{3,\lambda}$ in  \eq{SchurWeyl} by the composition
$\vartheta_{3,\lambda}\circ \varphi_{1}$ to include the action of $\SU(2)$ on $\C^3$.
The decomposition of $\vartheta_{3,\lambda}\circ \varphi_{1}$ as a representation
of $\SU(2)$ is determined by applying the well-established technique of the plethysm \cite{Littlewood58}, 
see also 
\cite{Macdonald95,Wybourne70,RW10}. The explicit computations for the plethysms have been performed 
using the computer algebra system
{\sc magma} \cite{MAGMA}, but computations could in this particular case also rely on
a generating function, see \pp{178} of \cite{PS79}. The combination of both steps leads to the results of 
Table~\ref{tab_klambda}. The resulting symmetry types $\tau$ for  a partition $\lambda$ 
are given by the standard Young tableaux of shape $\lambda$ 
 \cite{Boerner67,Hamermesh62,Pauncz95,Sagan01}, i.e.\ Young diagrams of shape $\lambda$
which are filled with the numbers 
$G{\subseteq}\{1,\ldots,n\}$ from the set of the $\abs{G}{=}g$ involved spins.
The number of symmetry types $\tau$ 
(with fixed shape $\lambda$ and set $G$)
is equal to the dimension of the irreducible representation $\mu_\lambda$ of $S_g$ 
(refer to the third column ($\#\tau$) of Table~\ref{tab_klambda}).

\section{\Drops representation based on multipole tensors\label{App_multipole}}

Multipole tensor operators 
\cite{Sanct75,SancTemMNMRXIII,SanctuaryX,Sanct_MNMRXI,SanctNMRIII,SancAll89} 
are defined by building on a state-space basis
of the quantum system which reflects
its angular momentum properties
$\{\ket{\kappa,j,m}\}$, where $\kappa\in\{\kappa_1,\ldots,\kappa_f\}$ are suitable-chosen 
auxiliary labels distinguishing angular momentum states with identical rank $j$ 
(and order $m$). 
Given $\kappa$ and $j$, the symbol $\ket{\kappa,j}:=\{\ket{\kappa,j,m} \,\text{with}\, {-}j{\leq} m {\leq} j\}$ 
denotes the ordered  set of states with angular momentum $j$ and auxiliary label $\kappa$.
Moreover, the state space of coupled spins has a basis $B = \bigcup_{\kappa,j}\ket{\kappa,j}$.
Multipole tensor operators $T_{jm}(\ket{\kappa_p,j_1}\arr\ket{\kappa_q,j_2})$
can now be defined by transforming
the states $\ket{\kappa_p,j_1}$ into $\ket{\kappa_q,j_2}$ according to the Clebsch-Gordan decomposition
\begin{align}
&T_{jm}\left(\ket{\kappa_p,j_1}\arr \ket{\kappa_q,j_2}\right)
:=  \label{Tmpo} \\
&\sqrt{\tfrac{2j{+}1}{2j_2{+}1}}\hspace{-6mm} \sum_{m_1\in\{-j_1,\ldots,j_1\}} \hspace{-6mm} 
\langle j_2, m_2|j,m|j_1,m_1\rangle 
\ket{\kappa_q,j_2,m_2}\bra{\kappa_p,j_1,m_1}, \nonumber
\end{align}
where one assumes that $m_2=m_1+m$ and where $\langle j_2,m_2|j,m|j_1,m_1\rangle$ 
denotes  the Clebsch-Gordan coefficient.
The matrix $\ket{\kappa_q,j_2,m_2}\bra{\kappa_p,j_1,m_1}$ contains only a single nonzero entry
corresponding to the state transition $\ket{\kappa_p,j_1,m_1}\arr\ket{\kappa_1,j_2,m_2}$.
All multipole tensor operators transforming $\ket{\kappa_1,j_1}$ into $\ket{\kappa_2,j_2}$ have 
integer ranks $j$ running from $|j_1-j_2|$ to $j_1+j_2$
and are grouped into a single \droplet in the associated  \drops representation.
The corresponding labels $\ell \in L$ of Eq.~\eqref{DROPS}
have the form $\ell=\ket{\kappa_1,j_1}\arr \ket{\kappa_2,j_2}$ (see, e.g., \tab{multipole pic}).


The angular momentum states are constructed by recursively coupling subsystems with an additional particle. 
For the case of one spin-$1/2$ particle, the angular momentum state basis is given by
$B_1:=\ket{1/2}:=\{\ket{1/2,1/2},\ket{1/2,-1/2}\}$.
For two coupled spin-$1/2$ particles, we use the Clebsch-Gordan decomposition of
$B_1\otimes \ket{1/2}$ and build the 
basis $B_2:=\ket{0}\cup \ket{1}$ consisting of
singlet and triplet states (see, e.g., \ppp{430--431} of \cite{Merzbacher98}).
No auxiliary labels $\kappa$ are necessary as each rank appears only once in $B_2$.
For three coupled spin-$1/2$ particles, one obtains
the basis $B_3:= \ket{3/2} \cup\ket{\kappa_1,1/2}\cup\ket{\kappa_2,1/2}$
(see \tab{multipole pic}). The  auxiliary  labels $\kappa_1:=1$ and $\kappa_2:=0$ 
of the state refer
to the parent rank of the element in $B_2$ involved in its generation. This construction
may be  in general  unwieldy, but it is always possible as $\SU(2)$ is a simply reducible 
group (see, e.g., \cite{Wigner65}), i.e., each irreducible representation in the Clebsch-Gordan 
decomposition of two irreducible representations of $\su(2)$ appears only once.
All compatible combinations of the nine possible transitions $\ket{j_1,\kappa_1}\arr \ket{j_2,\kappa_2}$
with ranks $j$ are given 
in \tab{multipole_table}.
The multipole tensor operators differ from the tensor operators
in the \lisa basis in not having a defined particle number. 
The explicit decomposition
of multipole tensor operators in the \lisa basis can be found in \tab{Tmpo dec}.

Examples for the \drops representation based on multipole tensor operators 
are illustrated in \fig{Entanglement_multipole} where
the density matrices of different pure states are given
(cf.\ \fig{Entanglement}), some of which are entangled and most of them have 
only one single \emph{non-empty} \droplet. Interestingly, both $\ket{\text{W}}$ and $\ket{\text{GHZ}}$
exhibit a symmetry under $z$-rotations, i.e.\  they are 
respectively rotationally invariant
or invariant under rotations of $2\pi/3$, see also \fig{Entanglement}.

%

\makeatletter
  \close@column@grid
  \clearpage
  \twocolumngrid
\makeatother

\end{document}